\documentclass[useAMS,usenatbib]{mn2e}

\usepackage[dvips]{graphicx}   % This is necessary to be able to include graphic
\usepackage{times}

\usepackage{epsfig}
\usepackage{xspace}
\usepackage{amsmath,amssymb}
\usepackage{latexsym}
\usepackage{graphicx}% Include figure files
\usepackage{dcolumn}% Align table columns on decimal point
\usepackage{bm}% bold math
\usepackage{epstopdf}
\def\reff@jnl#1{{\rm#1\/}}
\def\aj{\reff@jnl{AJ}}                  % Astronomical Journal
\def\araa{\reff@jnl{ARA\&A}}            % Annual Review of Astron and Astrophys
\def\apj{\reff@jnl{ApJ}}                % Astrophysical Journal
\def\apjl{\reff@jnl{ApJ}}               % Astrophysical Journal, Letters
\def\apjs{\reff@jnl{ApJS}}              % Astrophysical Journal, Supplement
\def\ao{\reff@jnl{Appl.Optics}}         % Applied Optics
\def\apss{\reff@jnl{Ap\&SS}}            % Astrophysics and Space Science
\def\aap{\reff@jnl{A\&A}}               % Astronomy and Astrophysics
\def\aapr{\reff@jnl{A\&A~Rev.}}         % Astronomy and Astrophysics Reviews
\def\aaps{\reff@jnl{A\&AS}}             % Astronomy and Astrophysics, Supplement
\def\azh{\reff@jnl{AZh}}                % Astronomicheskii Zhurnal
\def\baas{\reff@jnl{BAAS}}              % Bulletin of the AAS
\def\jrasc{\reff@jnl{JRASC}}            % Journal of the RAS of Canada
\def\memras{\reff@jnl{MmRAS}}           % Memoirs of the RAS
\def\mnras{\reff@jnl{MNRAS}}            % Monthly Notices of the RAS
\def\pra{\reff@jnl{Phys.Rev.A}}         % Physical Review A: General Physics
\def\prb{\reff@jnl{Phys.Rev.B}}         % Physical Review B: Solid State
\def\prc{\reff@jnl{Phys.Rev.C}}         % Physical Review C
\def\prd{\reff@jnl{Phys.Rev.D}}         % Physical Review D
\def\prl{\reff@jnl{Phys.Rev.Lett}}      % Physical Review Letters
\def\pasp{\reff@jnl{PASP}}              % Publications of the ASP
\def\pasj{\reff@jnl{PASJ}}              % Publications of the ASJ
\def\qjras{\reff@jnl{QJRAS}}            % Quarterly Journal of the RAS
\def\skytel{\reff@jnl{S\&T}}            % Sky and Telescope
\def\solphys{\reff@jnl{Solar~Phys.}}    % Solar Physics
\def\sovast{\reff@jnl{Soviet~Ast.}}     % Soviet Astronomy
\def\ssr{\reff@jnl{Space~Sci.Rev.}}     % Space Science Reviews
\def\zap{\reff@jnl{ZAp}}                % Zeitschrift fuer Astrophysik
\def\nat{\reff@jnl{Nature}}             % Nature 
\newcommand{\be}{\begin{equation}}
\newcommand{\ee}{\end{equation}}
\newcommand{\bea}{\begin{eqnarray}}
\newcommand{\eea}{\end{eqnarray}}
\newcommand{\bi}{\begin{itemize}}
\newcommand{\ei}{\end{itemize}}
\def\G{G\mu/c^2}

\def\Om{\Omega_m}
\def\Os{\Omega_s}
\def\Ol{\Omega_{\Lambda}}
\def\zd{z_{\rm d}}
\def\zs{z_{\rm s}}
\def\Dd{D_{\rm d}}
\def\Ds{D_{\rm s}}
\def\Dds{D_{\rm ds}}
\def\PI{Paper~I\xspace}
\def\Pcross{P_{\rm cross}}
\def\Pdetect{P_{\rm detect}}
\def\Pstr{P_{\rm string}}
\def\Nstr{N_{\rm string}}
\def\Pbg{P_{\rm bg}}
\def\Prad{P_{\rm rad}}
\def\tres{\theta_{\rm{res}}}
\def\mlim{m_{\rm lim}}
\def\zmin{z_{\rm min}}
\def\zmax{z_{\rm max}}
\def\sex{{\scshape SExtractor}\xspace}

\title%
[Cosmic string lenses]%
{Direct Observation of Cosmic Strings\\ 
via their Strong Gravitational Lensing Effect: \\
II.~Results from the HST/ACS Image Archive}
\label{firstpage}

\author%
[Morganson et al.]%
{Eric~Morganson$^{1}$\thanks{E-mail:ericeric@slac.stanford.edu}, 
Phil~Marshall$^{1,2}$,%\thanks{E-mail:pjm@physics.ucsb.edu} 
Tommaso~Treu$^{2}$,
\newauthor Tim~Schrabback$^{3,4}$,
Roger~D.~Blandford$^{1}$\\
$^{1}$KIPAC, P.O. Box 20450, MS29, Stanford, CA 94309, USA\\
$^{2}$Physics department, University of California, Santa Barbara, CA 93106, USA\\
$^{3}$Argelander-Institut f\"ur Astronomie, Universit\"at Bonn, Auf dem H\"ugel 71, 53121 Bonn, Germany\\ 
$^{4}$Leiden Observatory, Leiden University, Niels Bohrweg 2, 2333 CA Leiden, The Netherlands\\
}

\date{Accepted ---; received ---; in original form \today}

\pagerange{\pageref{firstpage}--\pageref{lastpage}}

\pubyear{2007}

\begin{document}

\maketitle
%-------------------------------------------------------------------------------

\begin{abstract}

We have searched 4.5 square degrees of archival HST/ACS images for 
cosmic strings, identifying close pairs of
similar, faint
galaxies and selecting groups whose alignment is consistent with gravitational
lensing by a long, straight string. 
We find no evidence for cosmic
strings in five large-area HST treasury surveys 
(covering a total of 2.22 square degrees), 
or in any of 346 multi-filter guest observer images (1.18 square degrees). 
Assuming that simulations accurately predict the
number of cosmic strings in the universe, this non-detection
allows us to place upper limits on the unitless Universal 
cosmic string tension of $\G < 2.3\times10^{-6}$, and
cosmic string density of $\Os < 2.1 \times 10^{-5}$ at the 95\% confidence
level (marginalising over the other parameter in each case). 
We find four dubious cosmic string candidates in 318 single filter
guest observer images (1.08 square degrees), 
which we are unable to conclusively eliminate with existing data.
The confirmation of any one of these candidates as cosmic strings 
would imply $\G \approx 10^{-6}$ and $\Os \approx 10^{-5}$. 
However, we estimate
that there is at least a 92\% chance that these string candidates
are random alignments of galaxies. If we assume
that these candidates are indeed false detections, 
our final limits on $\G$ and $\Os$ fall to $6.5\times10^{-7}$ and
$7.3 \times 10^{-6}$. Due to the 
extensive sky coverage of the HST/ACS image archive,
the above
limits are universal.
They are quite sensitive to
the number of fields being searched, and could be further reduced 
% to $\G
% \approx 3 \times 10^-7$ and $\Os \approx 5 \times 10^-6$ 
by more than a factor of two using forthcoming HST data.

\end{abstract}

\begin{keywords}
gravitational lensing --- surveys --- cosmology: observations
\end{keywords}

%-------------------------------------------------------------------------------

\section{Introduction}
\label{sect:intro}

Cosmic strings, which are topological defects formed in the very early universe and
persisting to the present epoch, are a quite generic prediction of many
modern cosmological models \citep{Kib76,Vi+S94,H+K95}. As microscopic
objects, their internal structure can be characterised by a single number, the string 
tension ~$\mu c^2$ which we express in unitless form ~$\G$. Numerical 
simulations of this phenomenon have converged on a 
picture where $\sim10$ horizon-scale strings are consistently predicted to lie 
within the observable universe \citep[see][for a review]{Pol07}. This suggests 
the possibility of a direct detection. 

Currently, the 
best limits on the properties of a putative network of long strings come
indirectly from studies of the CMB power spectrum and pulsar timings. 
The CMB power spectrum is now sufficiently
well-modeled so as to permit only a very small fraction of the power in
temperature fluctuations at the last scattering surface to be due to
strings of any type \citep{PWW04}, and \citet{Pog++06} sets 95\% condfidence limits
on of $\G < 2.7 \times  10^{-7}$. Pulsar timing
experiments like those described in \citet{KTR94,D+V05} give tighter limits, 
but make model-dependent assumptions on the relative densities of string loops and the long 
strings we search for here. Additionally, pulsar 
timing limits rely on string oscillations to produce gravitational radiation in observable 
frequency ranges \citep{A+K89,MWK07}. But some cosmic 
string models decay via Goldstone bosons or other field couplings \citep{Pol04}.

Direct searches for cosmic strings provide an important complement to those 
statistical studies. Direct searches in the CMB rely on the movement of strings, 
which introduce a very small apparent differential between light from other side of
the string \citep{Ki2004}:
\be
\frac{\Delta z}{1+z} = \frac{v_t \gamma}{c} 8\pi\G  \sin{i}\ \frac{\Dds}{\Ds}
\ee

Where $\sin{i}$ is the projection of the string along the line of sight, $\Ds$ is the angular
distance from the observer to the CMB, and $\Dds$ is the angular distance from the string to the CMB. This redshift is potentially observable using the CMB as a backlight. The
instrumentation required to make these observations at high resolution is still 
in its infancy
\citep[][]{Zwa++08,Muc++07,Ste++09,Fow++07,Bou07}. Previous attempts at lower 
resolution have not resulted in
any detections, and produced upper limits on the string tension of 
$3.7 \times 10^{-6}$ \citep{Saz++08,L+W05, J+S07}.
This induced redshift is of order $10^{-10}$, too small 
to be detected in the optical spectroscopic galaxy surveys, 
where it would be dominated by galaxy motions.

Cosmic strings are also directly observable in optical
imaging surveys via their gravitational lensing 
effect~\citep[e.g.][]{Vil84,S+K89}. Cosmic strings produce a unique lensing signal: 
two identical images of a source separated by $\theta$:
\be\label{eq:dtheta}
  \theta = 8\pi\G  \sin{i}\ \frac{\Dds}{\Ds}, 
\ee
Where $\Ds$ is the distance between the observer and the source and $\Dds$ is the 
distance between the string and the source.

In our first paper (\citeauthor{Gas++08}~\citeyear{Gas++08}, \PI), 
we showed how high resolution optical
imaging surveys are capable of producing {\it direct} 
limits on the cosmic string tension
competitive with the {\it indirect}
detection from the CMB power spectrum analysis. For the tensions allowed by 
the CMB power spectrum analysis, this separation is $\lesssim 1$'', and the 
cross-section for lensing is similarly small.
This realisation immediately drives us towards the need to use the most
numerous distant  objects---faint galaxies---as backlights, and to
focus on surveys with high angular resolution. 

Early optical cosmic string searches focused on large image separations and
bright galaxies, but no candidate string lenses have survived inspection at
higher resolution \citep{AHP06,Saz++07}. Ground-based optical surveys are hampered
by limited resolution and cannot probe $\G < 10^{-6}$ \citep{DL1997}.
Conversely, the recent searches by
\citet{Ch++08} using the Hubble Space Telescope (HST) GOODS survey images 
and \citet{Tim++09} using HST COSMOS
survey images did have the angular
resolution to show that no string with $\G > 3.0 \times 10^{-7}$ crossed their survey fields,
but did not have the sky coverage needed to be able to place any universal limit 
on the string tension.
%

%In \PI we showed that the survey area required in order to be able to set 
%a given
%string tension limit is closely related to the angular resolution and depth of the
%survey,  a result that follows from  the low predicted number density of
%strings in the universe.  For the case of HST, our best current source of high

We need both large survey area and high resolution to set universal limits on 
string tension. For the case of HST, our best current source of high
resolution imaging survey data, the solid angle needed to give a universal
limit on the string tension is a few square degrees, provided 
this area is cut up and 
distributed evenly across the sky. As discussed in \PI, a contiguous survey 
of comparable area would provide much weaker limits due to the 
clustering of cosmic string events around the roughly $~10$ long strings in the
sky.

In this work we search the HST/ACS archive, whose images
match all three of the angular resolution, area and pointing 
distribution criteria,
and so produce the strongest direct gravitational
lensing limit on the cosmic string tension and density.

This paper is organised as follows:
in Section~\ref{sect:theory}, we briefly 
review the relevant basic theory of cosmic strings. 
We then discuss how searching for string lensing events can be used to set new limits 
(Section~\ref{sect:limits}).
Regarding our actual search for string lenses,
we explain our use of simulated string lensing images in
Section~\ref{sect:sims}, and in Section~\ref{sect:method} we describe
our string-finding technique. After a brief description of the archival
dataset in Section~\ref{sect:data}, we present our results in
Section~\ref{sect:results} and the
corresponding limits on the string tension and density in
Section~\ref{sect:obslimits}. Finally,
we discuss the implications of this work in Section~\ref{sect:concl}.

%-------------------------------------------------------------------------------

\section{String Lensing Theory}
\label{sect:theory}

When searching for cosmic strings via their gravitational lensing effect, 
we only need to
understand a few basic facts about them. We must know what type of
multiple image system a cosmic string produces, 
and we must understand what string evolution
tells us about the number of strings in the local ($\zd \lesssim 2$) universe.

Cosmic strings, whether topological defects or stretched fundamental strings, can
produce two identical images separated by $\theta$ in Eq. \ref{eq:dtheta} so long 
as they do not ``cut'' the source they are lensing and so long as they are straight on 
scales comparable to the image separation. By ``cutting'' a source, we mean that the 
redshift-dependent strip of sky which the string copies only includes a part of a source,
and the source in incompletely copied. We study this effect via simulation in Section~\ref{sect:sims:pairs}. We discuss our assumption of straight strings below.

% EFM I don't think the next paragraph is worthwhile in this shortened version
%Cosmic strings that extend over large angular scales may give rise to a great
%many image splitting events. In the next section we revisit the discussion
%started in \PI of the size and nature of the expected string lens population.

%where $\Dds$ and $\Ds$ are the usual angular diameter distances between,
%respectively, the lens and the
%source, and the observer and the source (\PI).
% \bea\label{r}
% \Dd(\zd)=\frac{c}{H_0(1+\zd)}
%          \int_0^{\zd}\frac{dz}{\sqrt{\Om(1+z)^3+\Ol}}
%          \nonumber\\%=\frac{I_l}{H_0(1+\zd)}
% \Dds(\zd;\zs)=\frac{c}{H_0(1+\zs)}
%               \int_{\zd}^{\zs}\frac{dz}{\sqrt{\Om(1+z)^3+\Ol}},
%               %=\frac{I_{ls}}{H_0(1+z_i)},
% \eea

% Structure on smaller scales changes the effective lensing image
% separation, but 
% is otherwise not detectable by our method \citep{Ki2004}.
% The transverse
% velocity of a string will introduce a redshift offset between the two images
% of $\Delta z/z \approx (v_t/c) \G$ \citep{Ki2004}, but the tensions of interest 
% ($10^{-6} -- 10^{-7}$) are sufficiently small that this will not give a measurable colour
% difference. 
% - - - - - - - - - - - - - - - - - - - - - - - - - - - - - - - - - - - - - -

\subsection{String network topology: straight strings and loops}
\label{sect:theory:topology}

We can only observe a string by its gravitational lensing effect
if it crosses one of our search fields; we then expect it to be more readily
detectable if it is straight 
on the scale of the field. 
We consult the theory and simulations of cosmic strings to determine whether these
are reasonable possibilities. Horizon scale strings are a generic prediction of 
string network
simulations \citep{Pol04a}, but we must consider the
total angular length of observable string 
and, and to what extent small scale structure (kinks and curvature) will 
affect the image pair alignment.
We also explore the possibility of detecting string loops.

Simulations \citep{A&S90} tend to agree that during the matter dominated epoch and beyond, 
the long string density is:
\begin{equation}\label{eq:rhos}
\rho_s = \Psi \G \rho_m; \Psi = 60 \pm 15,
\end{equation}
where $\Psi$ represents the average length of long, straight string within a cosmological volume,
the concentration of string. These simulations assume that strings ``intercommute'' with 
probability of 1. Lowering this probability raises the simulated value of $\Psi$, but 
we assume unity intercommutation probability, because it is theoretically favoured.

We can relate $\Psi$ to the number of cosmic strings within some redshift 
limit.  Substituting in $\rho_{\rm m}=\Om\frac{3H_0^2}{8\pi G}$ 
for the co-moving matter density, we have:
\begin{equation}
\frac{\rho_s}{\mu} = \Psi \Om\frac{3H_0^2}{8\pi c^2}
\end{equation}
We divide by $2 \Dd$ (where $\Dd$ is the angular diameter distance to the string)
and obtain the concentration of string in angular units,
accounting for projection effects:
\begin{equation}
\frac{d<\theta_s>}{dV} = \Psi \Om\frac{3H_0^2}{16\pi c^2 \Dd}
\end{equation}
Multiplying by the cosmological volume element and integrating over redshift
yields the total expected angular length of string on the sky, in radians,
$<\theta_s>$:
\bea\label{eq:totalstring}
<\theta_s> &=& \Psi \int_0^{z_{\rm{max}}} \frac{3\Om H_0^2}{16\pi c^2 D(z)}\frac{dV}{dz}dz\\
<\theta_s> &=& \Psi \int_0^{z_{\rm{max}}} \frac{3\Om H_0 D(z)(1+z)^2}{4 c \sqrt{\Om(1+z)^3+\Ol}}dz\\
<\theta_s> &=& 0.35  \Psi,\ \rm{for}\ z_{\rm{max}} = 2 \nonumber
\eea

$\theta_s\approx \pi$ corresponds to a 
single straight string stretching across the visible sky. Setting $\Psi = 60$
gives us 21 radians of string within 
redshift 2. This in turn corresponds to roughly 7 straight strings with redshift~$<2$.
Even survey fields as large as a few square degrees would most likely not be
crossed by a string, and would therefore have no chance of seeing a long string
regardless of survey depth or resolution (see Section~\ref{sect:theory:pcross}
for details). A survey of the same area comprising a large number of smaller,
well-distributed fields stands a much better chance of ``hitting'' a string.

Even if a string crosses one of our search fields, detecting it may be 
nontrivial. Cosmic strings that are straight on scales comparable to the 
field size should be readily identifiable:
groups of
aligned galaxy pairs make good candidates for being multiple image systems
generated by the string.
Fields of $\sim 1$~arcmin would then result in the requirement that, for it to
be detectable, a cosmic string be straight on Mpc scales. 

What can we infer from the string network simulations about the likely
straightness of horizon-scale strings? Suppose that cosmic strings follow random
walks.
The total length of a string within the horizon
of a string undergoing a random walk is roughly $d_h^2 / L$ where $d_h$ is the
horizon scale and $L$ is the characteristic length of the random walk.
Assuming a string tension $\G > 10^{-8}$, the random walk length, and similarly
the radius of curvature of the string, must be of order the $d_h$ or $\Psi$
would quickly exceed~60 and $\rho_s$ would be large enough to affect the CMB
spectrum (see Section~\ref{sect:limits}). So $L$ must be large and this random
walk behaviour only affects the Mpc scale perturbatively. The simulations do 
show that a string's self-interactions and its interactions with
other strings can induce smaller scale ``cusps'' which propagate along the
string. However, at least some models suggest that these cusps will leave
strings that are essentially straight on Mpc scales \citep{Roc08}.

String evolution networks also produce loops. Loops smaller than a few Mpc would
not be detectable as  simple alignments of image pairs. \citet{MWK07} provides a
framework for detecting string loops with proposed high resolution radio
surveys. Many models and simulations set the scale of cosmic loops
\citep{C+A1992} to be of order the horizon scale, and so it is possible that
many cosmic string loops are straight on the Mpc scale, but we do not include
them when marginalising over string concentration in our calculations.

% - - - - - - - - - - - - - - - - - - - - - - - - - - - - - - - - - - - - - -

\subsection{String lens observability}
\label{sect:theory:obs}

In this section we summarise the various observable features of cosmic string
lenses, and thus motivate our particular search strategy.

We might hope to detect a cosmic string using a
single, obvious string lensing event, where the images are demonstrably
translated copies of each other \citep[see][for such an analysis]{AHP06}. 
It is only possible to determine if a
single pair is the result of string lensing if the pair separation is several
times larger than the PSF, and if the sources are bright and well-resolved. The
large separation requirement would directly limit our ability to probe the small
(subarcsecond)
string tensions of interest.
The bright source requirement, and the fact that there are
exponentially fewer bright sources than dim sources, would limit our ability
to probe small $\Psi$. 

The key observable feature of a cosmic string is therefore that it
produces {\it many} pairs of images, as it cuts across many background sources
\citep{H+V03,O+T05}. 
We could look for local
overdensities of well-matched (in magnitude, ellipticity and orientation) 
pairs, as is 
expected to occur when a kinked or coiled string lies in the field. This is
the approach taken by \citet{Ch++08}, and only works when the number of string 
pairs 
($\propto \Psi \G$) is large enough that string lensing can be detected
against the very high
background of pairs described by the small-scale correlation function.
% More 
% quantitatively, \citet{Ch++08} can only limit the cosmological $\Os 
% < 0.02$, while even exceedingly conservative CMB limits would have 
% $\Os < \Psi \G \Om < 10^{-4}$.

However, on arcsecond and
sub-arcsecond scales the background 
sources are highly correlated \citep{M+B2008}. 
The increase in the abundance of apparent 
galaxy pairs due to the presence of a cosmic string is, as we shall see, quite
small. Only for the expected straight strings from the previous section
do we expect a measurable phenomenon: the
appearance of groups of apparent galaxy pairs {\it aligned with each other}. 

The similarity of the members of each aligned
pair should still be a useful indication
of string lensing.
At image separations of $0.3''$, we are probing down to the scale of the source
size itself, and so many multiple image systems 
will contain incompletely copied sources.
If these images were well-resolved, we could detect a sharp edge, but small,
incompletely copied sources look unremarkable.  Pairs with incomplete images
can, in rare cases,  have significant differences in their magnitudes and
ellipticities. In Section~\ref{sect:sims:pairs}, we quantify these differences.

\section{Estimating the expected number of string detection}
\label{sect:limits}

In order to set limits on string parameters, we calculate the expected number
of string detections in a given dataset, $\Nstr(\G,\Psi)$. 

Each (assumed rectangular) field is defined by three parameters, two
spatial extents $\theta_1$ and $\theta_2$, and its limiting magnitude,~$m$. 
A given string
is defined by its tension, $\G$, its inclination along the line of sight, $i$, and
its redshift, $\zd$. The intersection between the string and the field is defined
by $\theta_c$, the projected angular length of string that crosses the field.

To calculate $\Nstr$, we must first calculate the probability that a randomly
oriented string crosses a field with crossing length $\theta_c$,
$\Pcross(\theta_1, \theta_2, \Psi, \sin{i}, \zd, \theta_c)$. We then calculate the
probability that a string which crosses our field with length $\theta_c$ is
detected, $\Pdetect(m, \G, \sin{i}, \zd, \theta_c)$. Knowing these two
quantities, we can integrate over all possible strings to obtain $\Nstr$ for
a randomly oriented field:
\bea
\Nstr &=& \int d \theta_c \int d \zd
  \int_0^{\pi/2} \sin{i}\ d i,\ 
  \Pcross \Pdetect\\
   &=& \int d \theta_c \int d \zd
  \int_0^1 \frac{\sin{i}\ d \sin{i}}{\sqrt{1-\sin^2{i}}}.
  \Pcross \Pdetect\nonumber
\eea
In Section~\ref{sect:theory:survey} below, we discuss how we transform our expression
for the probability of a detection in a single field to our 95\% detection limits
for a multi-field survey.

% - - - - - - - - - - - - - - - - - - - - - - - - - - - - - - - - - - - - - -

\subsection{The Probability of a String Crossing a Field}
\label{sect:theory:pcross}

We define $\Pcross \equiv \Pcross(\theta_1, \theta_2, \Psi, \sin i, \zd, \theta_c)$ 
as the probability that a string at redshift $\zd$ will cross our 
$\theta_1 \times \theta_2$ field with length $\theta_c$ at inclination angle,~$i$, 
to the line of sight. We assume that our fields are much smaller than a 
radian and that our strings are straight on scales longer than a few arcminutes
(the size of a single field). We break down the probability of a string crossing 
our field with overlap length $\theta_c$ into two terms: the probability of the 
field crossing a randomly oriented line that is 1 radian long and the projected 
length of string at redshift~$\zd$ (in radians):
\be
\Pcross = \Prad(\theta_1,\theta_2,\theta_c) \cdot
          \frac{\rho_s \sin{i}}{\mu D}\frac{D^2 dD}{dz}
\ee
$\Prad$ is a quantity derived solely from two dimensional spherical geometry
without any cosmological input. It is the probability of a randomly positioned
$\theta_1 \times \theta_2$ rectangle being crossed by a randomly oriented $1$~rad
arc with an overlap length of $\theta_c$. The angular length 1 radian is arbitrary
and chosen for mathematical convenience. We provide its (somewhat cumbersome)
geometrical derivation in the appendix.

The remaining terms represent the expected projected angular length of string on
the sky at every redshift and the cosmological volume element.  If we plug in our
model value of $\rho_s$ (equation~\ref{eq:rhos}) 
and the $\Lambda_{\rm{cdm}}$ volume element, we obtain:
\be
\Pcross = \Prad(\theta_1,\theta_2,\theta_c) 
        \Psi\frac{3\Om H_0}{2 c}\frac{\sin{i} D(z)(1+z)^2}{\sqrt{\Om(1+z)^3+\Ol}}
\ee

% - - - - - - - - - - - - - - - - - - - - - - - - - - - - - - - - - - - - - - --

\subsection{The Probability of a String Crossing being observable}

Assuming that a string crosses a field with overlap $\theta_c$, we define the
probability of  that string producing a detectable lensing signature as $\Pdetect
\equiv \Pdetect(m, \G\sin{i}, \theta_c, \zd)$. To obtain this function, we must
calculate the expected number  events for our string and field. This is just a
($\zs$-dependent) lensed area times a ($zs$-dependent) source density, integrated
over $\zs$.

The lensing cross sectional area~$\sigma$ of a string is its length, $\theta_c$, times the
string lensing width in equation~\ref{eq:dtheta}:
\be
\sigma \equiv \sigma(\zd,\zs, \G, \sin i, \theta_c) = 
     \theta_c 8\pi\G  \sin{i}\ \frac{\Dds}{\Ds}
\ee
We model the source density, $n(\zs,m)$, using the redshift distribution
distribution from \citet{Lea++07} and a fit to the magnitude distribution of
COSMOS:
\bea\label{eq:zdist}
n(\zs,m) &=& n_0(m) \frac{2 z^2}{3 z_0^2} e^{\left(\frac{z}{z_0}\right)^{-1.5}}\\
n_0(m) &=& e^{0.67(m-34.9)}\rm{arcsec}^{-2}\nonumber\\
z_0 &=& 0.13 m - 2.3\nonumber
\eea
This equation assumes magnitudes $m$ observed in the F814W filter. For fields
imaged in a different filter, we measure the source density and assume the 
F814W limiting magnitude that would yield an identical density.

We multiply the cross sectional area by the (redshift-dependent) density of sources and integrate over
source redshift to obtain the expected number of lensing events
\be
<N_{events}> = \theta_c \int_{\zmin}^\infty d\zs 
     \int_0^{\mlim}dm\ d\zs  \sigma n(\zs,m)
\ee
Here, $\zmin$ is the minimum source redshift that will produce a
resolvable lensing event:
\be
\frac{\Ds(\zmin)}{\Dds(\zd,\zmin)} = 8\pi \frac{G\mu \sin{i}}{c^2 \tres}
\ee
$\tres$ is the ``effective resolution,'' or the 
angular separation at which actual faint pairs 
in our survey can realistically be deblended. \citet{M+B2008} found 
$\tres \approx 0.3$'', and we use this value when deriving detection limits.

Most of our events will involve faint sources that are fairly weakly clustered 
on arcminute scales, so we can assume a Poisson event rate. 
We require three aligned pairs of similar galaxy images to claim a string
detection, so we calculate the probability of making three detections or more:
\begin{equation}
\Pdetect = 1-e^{-<N_{\rm{events}}>}
           \left(1+<N_{\rm{events}}>+\frac{<N_{\rm{events}}>^2}{2}\right)
\end{equation}

When probing a potential string detection, we incorporate the information
we get from the detection into the above analysis. The expected number of 
lensing events becomes
\be
<N_{events}> = \theta_c \int_{\zmin}^{\zmax} d\zs 
     \int_0^{\mlim}dm\ d\zs  \sigma n(\zs,m)
\ee
Where, $\zmin$ and $\zmax$ are the observed source redshifts that will produce 
the minimum and maximum observed lensing separations:
\bea
\frac{\Ds(\zmin)}{\Dds(\zd,\zmin)} &=& 8\pi \frac{G\mu \sin{i}}{c^2 \theta_{\rm{min}}}\\
\frac{\Ds(\zmax)}{\Dds(\zd,\zmax)} &=& 8\pi \frac{G\mu \sin{i}}{c^2 \theta_{\rm{max}}}\nonumber
\eea
In addition, we define $P_{\rm{observation}}$ to include the observed number of detected pairs:
\begin{equation}
P_{\rm{observation}} = e^{-<N_{\rm{events}}>}\frac{<N_{\rm{events}}>^{N_{\rm{observed}}}}{N_{\rm{observed}}!}
\end{equation}

% - - - - - - - - - - - - - - - - - - - - - - - - - - - - - - - - - - - - - - --

\subsection{Expected number of string detections and string limits in a multifield survey} 
\label{sect:theory:survey}

In this paper we aim to set limits on $\G$ using multifield surveys, and
as an intermediate step, we must calculate $N_{\rm{string\ survey}}(\G,\Psi)$, the 
expected number of string detections across our complete dataset.
$\Nstr(\G,\Psi)$ must be calculated numerically for every field at each tension $\G$, 
but it is linearly 
proportional to $\Psi$. For each field, we calculate $\Nstr(\G,\Psi=1)$, and then
compute  
$\Nstr(\G,\Psi) = \Psi \Nstr(\G,\Psi=1)$.

The expected number of string detections of multiple fields will add linearly so long as
the fields are randomly distributed, so we have:
\bea
N_{\rm{string\ survey}}(\G,\Psi) &=& \sum_{\rm{fields}\ i} N_{\rm{string\ i}}(\G,\Psi)\\
&=& \Psi N_{\rm{string\ survey}}(\G,\Psi = 1)\nonumber
\eea
To see what $\G$ and $\Psi$ give us 95\% chance of detecting a string, 
we assume a Poisson distribution for the number
of detections. A random variable taken from a Poisson distribution with a mean of 
$\log{20} = 3.00$ only has a 5\% chance of being 0. So for a given $\G$, we set:
\bea
N^{95\%}_{\rm{string\ survey}}(\Psi^{95\%}) &=& \log{20}\\
\longrightarrow \Psi^{95\%} &=& \frac{\log{20}}{N_{\rm{string\ survey}}(\Psi = 1)}
\eea

To compute proper Bayesian confidence limits, we must must assume prior 
distributions for $\G$  
and $\Psi$. We assume $\G$ is log-uniformly distributed between $10^{-8}$ and $10^{-5}$, 
and we take a logarithmic version the $\Psi$ distribution from Eq. \ref{eq:rhos}:

\bea
P(L) &=& \frac{1}{\sqrt{2\pi}\sigma} e^{-\frac{(L-L_0)^2}{2\sigma^2}}\label{eq:pofl}\\
L &=& \log(\Psi)\nonumber\\
L_0 &=& \log(60)\nonumber\\
\sigma &=& \frac{\log(2)}{2}\nonumber
\eea

The probability of a null detection is $e^{-\Nstr(\G,\Psi)}$ and the probability of 
observing a detection in one field and no detections in all other fields is 
$e^{-\Nstr(\G,\Psi)} P_{\rm{observation}}(\G,\Psi)$. We use the above priors, these 
formulae and a standard Bayesian formulation to produce confidence limits in
Section~\ref{sect:obslimits}.

%-------------------------------------------------------------------------------

\section{Simulated Images}
\label{sect:sims}

As we discussed in \PI and Section~\ref{sect:theory} above,
to detect a cosmic string lensing event one must
be able to detect close pairs of faint objects. Developing and testing search
algorithms, and then characterising their respective selection functions,
requires sets of simulated images in which we know the location of the 
sources and the strings. These mock images are based on those used in the 
faint source correlation function measurement by \citet{M+B2008}. These
simulations reproduce the magnitude, area and ellipticity distributions of the
faintest galaxies in HST/ACS GOODS images; we adapt them
for cosmological settings with cosmic strings by including 
magnitude-dependent clustering, stochastic redshift assignment and a 
string lensing effect. We provide a very brief outline of the 
image simulation procedure we use in this section, and refer the reader 
to the paper by \citet{M+B2008} for more details.

% - - - - - - - - - - - - - - - - - - - - - - - - - - - - - - - - - - - - - -

\subsection{Image Production}
\label{sect:sims:method}

Our standard simulated images are square, consisting of  $8192 \times 8192$
pixels. Each pixel is a $0.03$'' square, so our mock frames are $246$'' on a
side.  In each mock image we place a single straight string (as described in 
Section~\ref{sect:sims:method}), with known orientation. We use these images to
test our string detection algorithms and to measure the detection
probability.  We also produce a set of narrow $10$''$\times 246$'' images,
each containing a single string, in order to produce a high density of
multiply-imaged faint galaxies and allow us to quantify the string lensing
effect on the galaxy images themselves  (see Section~\ref{sect:sims:pairs}).

We choose source redshifts and magnitudes using the distributions in 
equation~\ref{eq:zdist}. Each 
source is assigned an independent redshift, based solely on its magnitude, 
and printed into the image with the nearest discrete 
$\zs =$ (0.1, 0.3, 0.6, 1, 1.5, 2, 2.5, 3, 3.5, 4, 5, 6). 
We have one string per image. The strings are located at a constant redshift
(the analog of the thin lens approximation). 
Each $\zs$ image
is lensed by copying and shifting half of the image by the 
amount given equation~\ref{eq:dtheta}.
All simulated strings are absolutely straight and vertical. 
We add the images at each
$\zs$ to produce an image with all sources. 

Finally, we convolve the lensed and combined images with a point spread function
(PSF) determined by an average of many model PSFs calculated using the ``Tiny
Tim'' software package. 

%We then add a convolved Gaussian noise background, the amplitude of which we vary to 
% produce various desired limiting magnitudes.
%
We then add a noise background which has been convolved with a separate PSF that mimics the image 
combination process. We vary the noise amplitude to produce images of various desired limiting magnitudes.

%%%%%%%%%%%%%%%%%%%%%%%%%%%%%%%%%%%
\begin{figure}
\begin{minipage}{\linewidth}
\centering\includegraphics[width=0.6\linewidth]{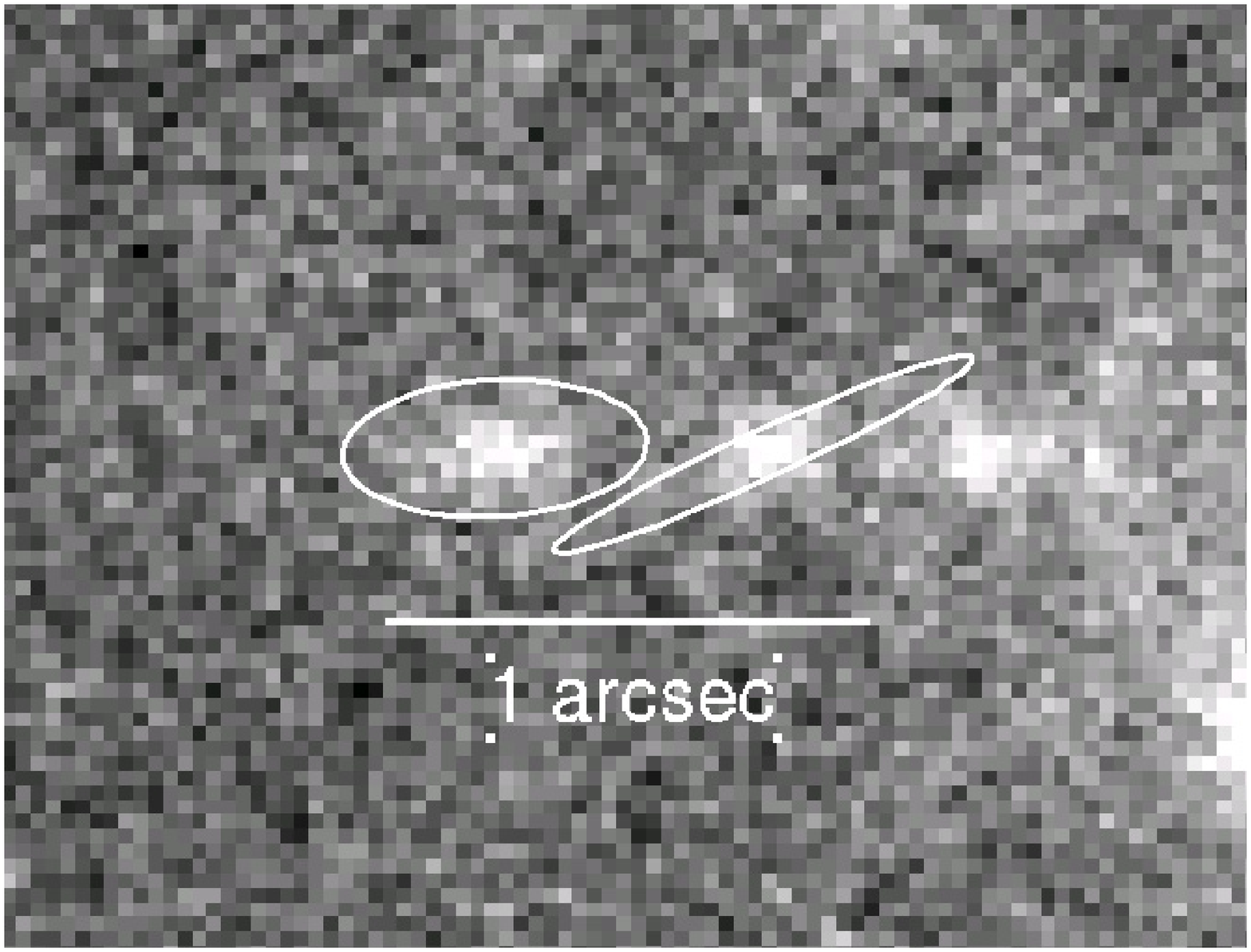}\medskip

\centering\includegraphics[width=0.6\linewidth]{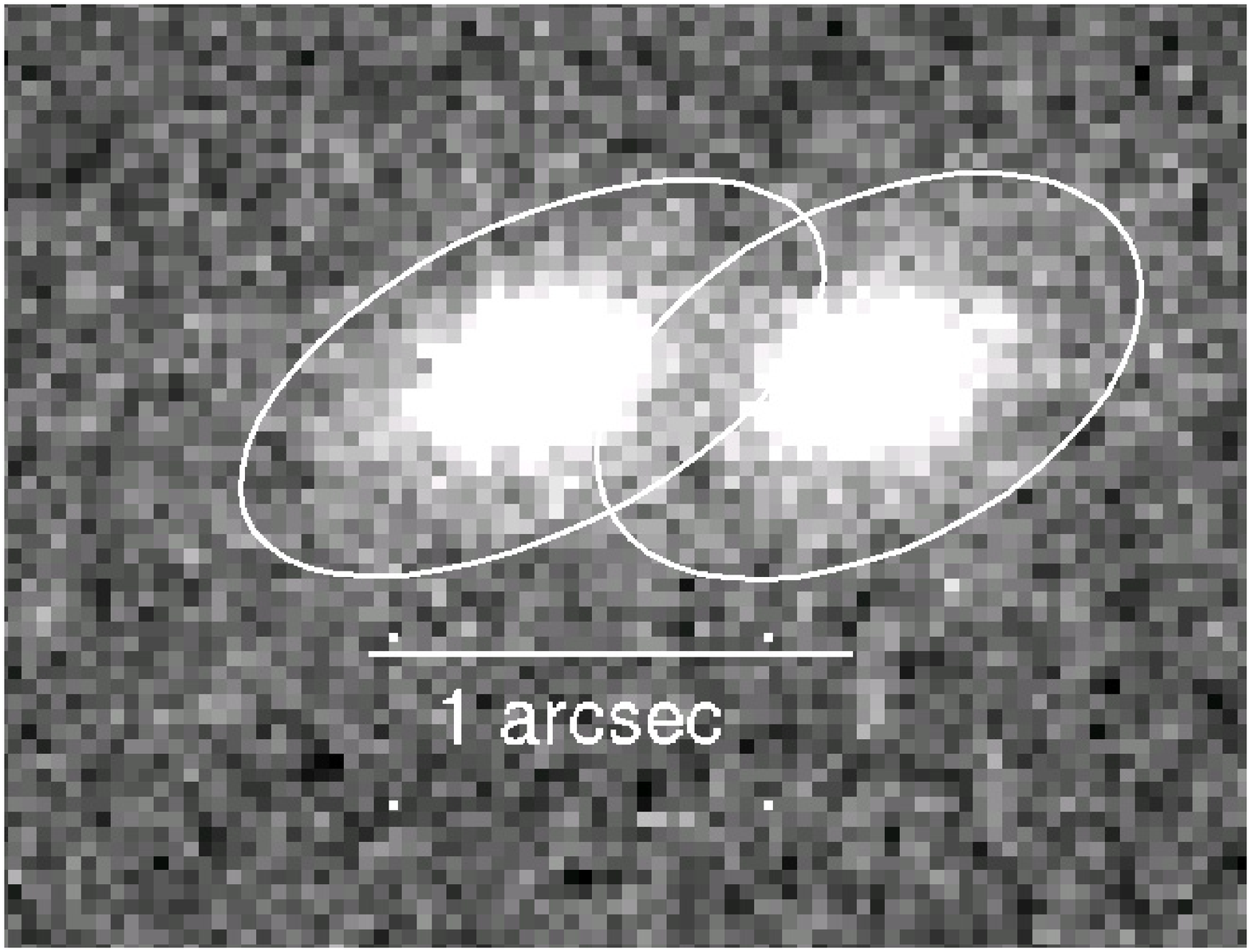}\medskip

\centering\includegraphics[width=0.6\linewidth]{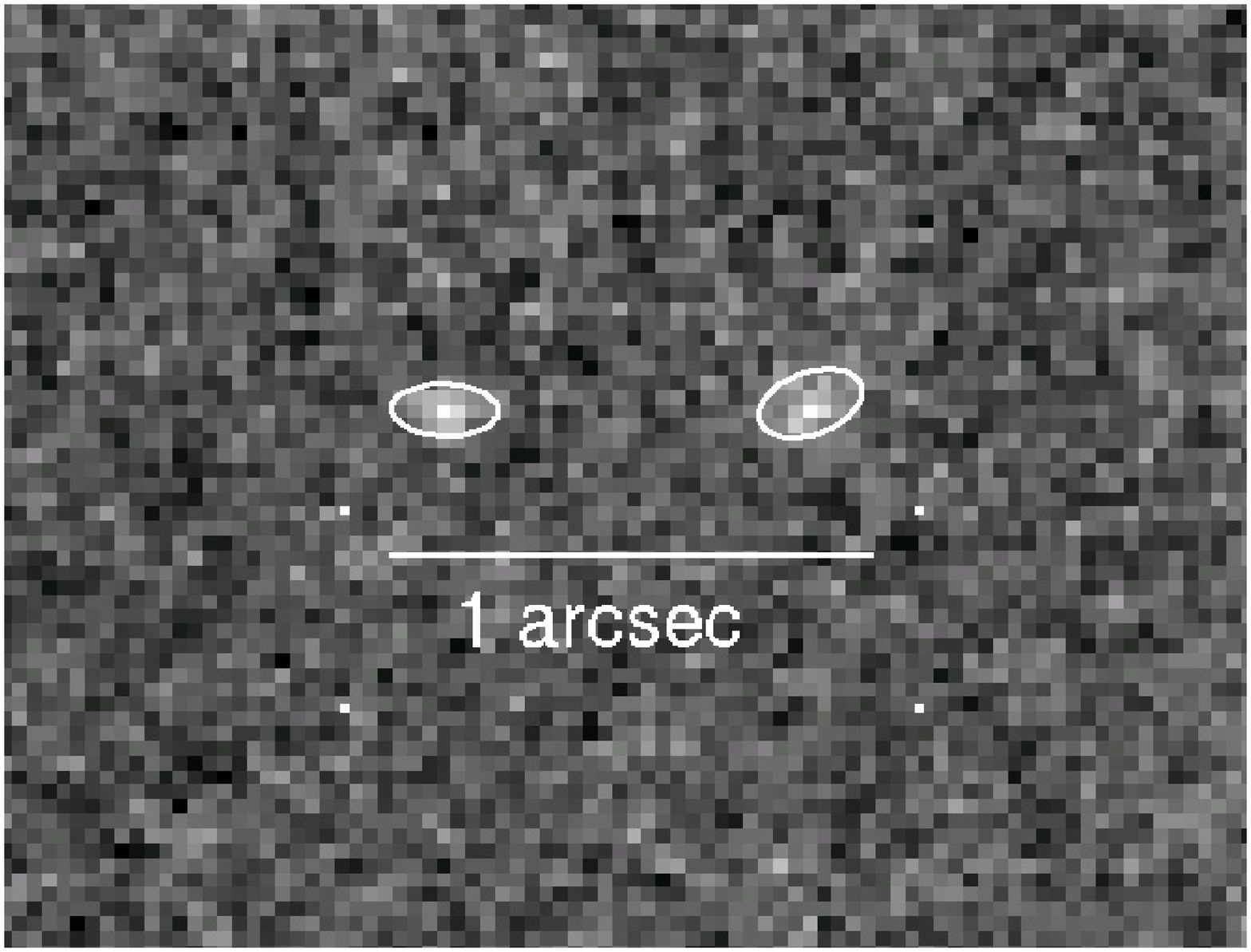}\medskip

\centering\includegraphics[width=0.6\linewidth]{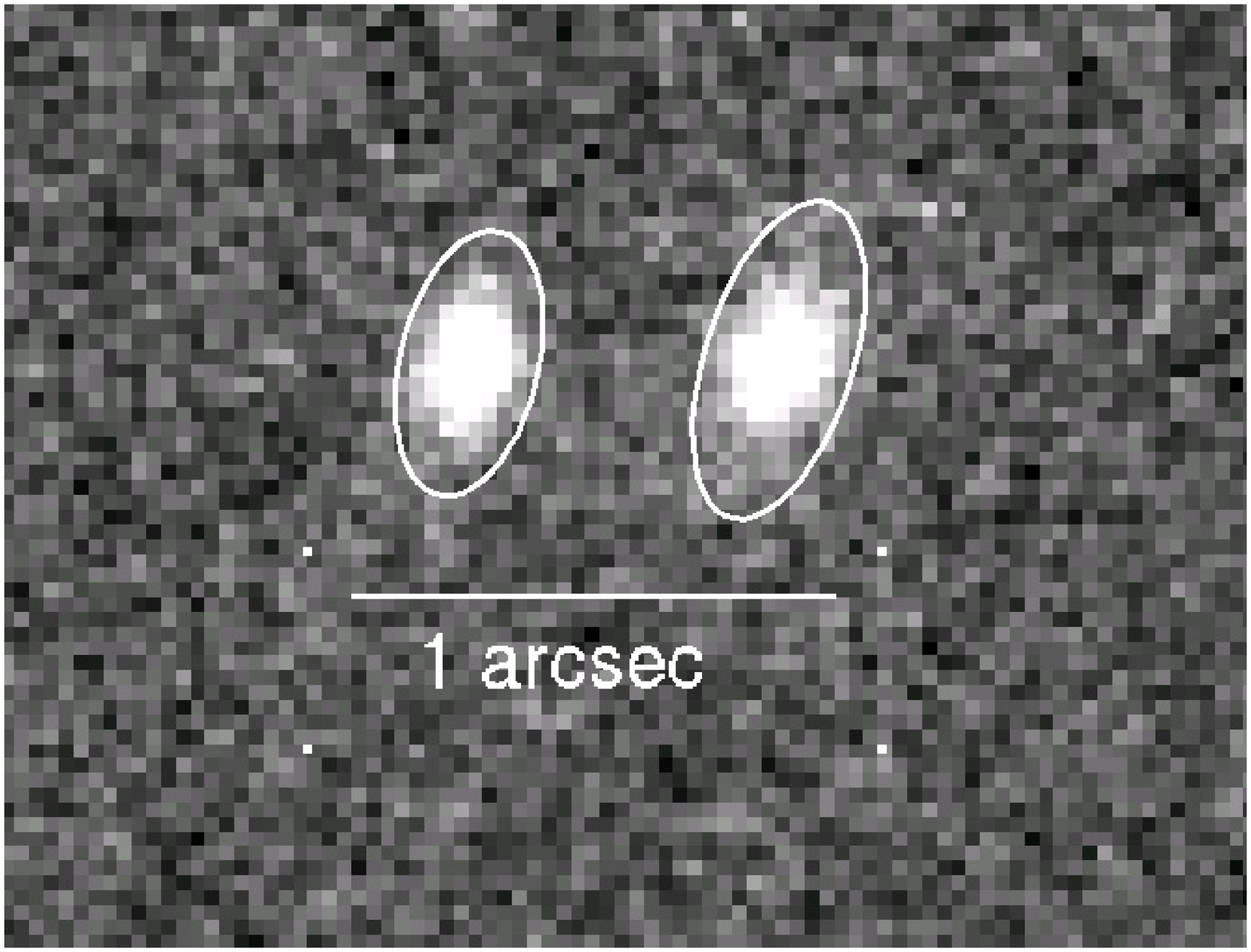}\medskip

\centering\includegraphics[width=0.6\linewidth]{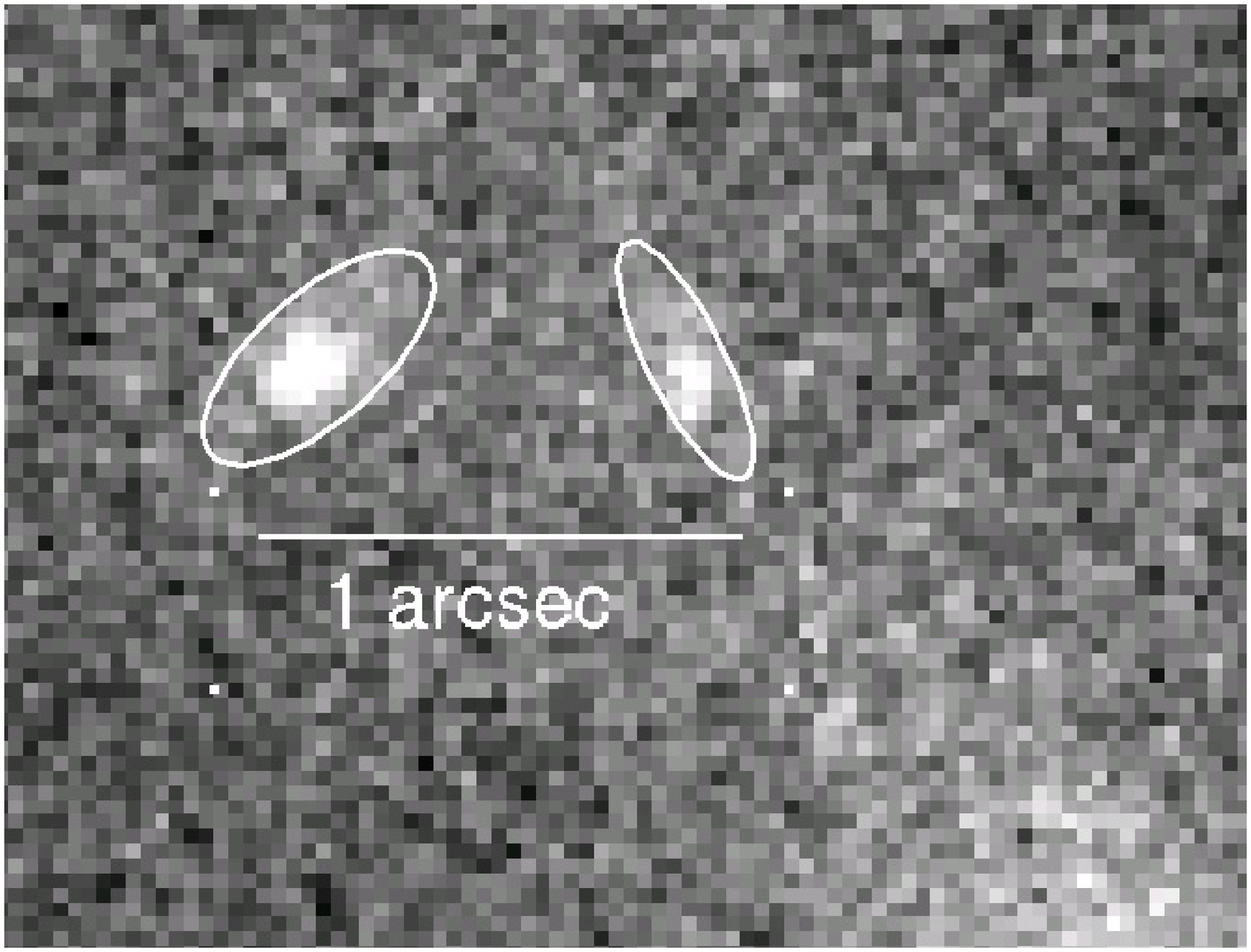}
\end{minipage}
\caption{Example simulated string-lens image pairs 
caused by a vertical string with
$8\pi \G = 1$'' at redshift 0.5. The ellipses show the 
ellipticity and orientation of each object, as measured with \sex.}
\label{fig:pairs}
\end{figure}
%%%%%%%%%%%%%%%%%%%%%%%%%%%%%%%%%%%

In Fig.~\ref{fig:pairs}, we show 5 examples of typical string-lensed image
pairs
from one of our simulated fields. Overlaid are ellipses representing the
orientation and ellipticity measured for each source using the \sex program
\citep{B+A1996}. We see that some pairs will have well-measured and well-matched magnitude,
size, ellipticity and orientation, and that these should be closely-matched
between conjugate images in the multiple-image system. However, due to the
prevalence of faint, poorly-measured sources, and to strings splitting
sources, most pairs are not as readily recognizable as string lensed pairs. We
quantify this below.

% - - - - - - - - - - - - - - - - - - - - - - - - - - - - - - - - - - - - - -

\subsection{Properties of simulated string-lensed objects}\label{sect:sims:pairs}

One important application of our simulations is to compare the statistical
properties of string lens multiple-images with those of unlensed, background
galaxy pairs. 
An idealised string lensing event would produce two identical galaxy
images, but incomplete copying of sources leads to non-ideal
string-lensing. Still, as we show in this section, string-lensed pairs 
tend to be more similar than random pairs.

We compare catalogs of string-lensed image pairs and unlensed close galaxy pairs 
generated from a large number of simulated images.
These images are the long, narrow strips mentioned in Section~\ref{sect:sims}. 
We made a set of thin ($10$''$\times 246$'') simulated 
images with a string running along the long axis. For each combination of
lens redshift $\zd = $ (0.1, 0.5, 1, 1.5, 2, 3), 
$8\pi\G =$(0.3'', 0.4'', 0.5'', 1'', 2'')
and the F606W-band AB limiting magnitude $V_{\rm{lim}} = $ (25, 26, 27, 28), 
we made 100 images, for a total of 120000. 

In each image, we identify the string-lensed
pairs as those whose centre lies 
within $8\pi \frac{G \mu}{2 c^2}$ of the string, and whose separation 
vector is of length $\theta < 8\pi \G$ 
and within 0.05 radians of being perpendicular
to the straight, simulated string. For the brightest magnitude limit,
$V_{\rm{lim}} = 25$, this latter condition is relaxed to 0.1 radians, because 
the bright, large sources have larger uncertainty in their positions. 
In all we generate 30000 simulated string-lensed image pairs. 

Those sources whose 
centre is further than $8\pi \frac{G \mu}{2 c^2}$ from the string and whose 
separation is $\theta < 8\pi \frac{ G\mu}{2 c^2}$ are chosen as sample 
background pairs, regardless of orientation. We find 500000 such background 
pairs.

We divide our catalogs of string pairs and background pairs along three
different axes. First, we split the data into different limiting magnitude
bins (25, 26, 27, 28). The fainter our limiting magnitude, the smaller
the sources we tend to detect. We then further divided the pairs 
into two classes of separation
of the string pair, ``close'' and ``far,'' with $\theta = 0.5$'' (a
characteristic source size) being the
cutoff separation. Pairs separated by
distances much larger than this threshold are well-resolved and deblended, and
so should be more similar than close pairs. Finally, we divide
our sources into two classes: ``dim'' sources that are within one magnitude 
of the limit, and ``bright'' sources that are more than one magnitude above 
the limiting
magnitude. Bright sources tend to have more distinct morphology 
which allows us to distinguish lensed pairs from random pairs more easily.

We quantify the similarity between sources in pairs using three parameters. 
The first two are the measured
magnitude difference between the sources in a pair,
\begin{equation}
\Delta V = |V_1-V_2|,
\end{equation}
and the pseudo-vector ellipticity dot product, 
\begin{equation}
E \equiv  \vec{\varepsilon_1}\cdot \vec{\varepsilon_2} =  \varepsilon_1 \varepsilon_2 \cos(2(\phi_1-\phi_2)),
\end{equation}
where $\varepsilon_i$ is the \sex ellipticity 
$1-R_{\rm minor}/R_{\rm major}$ of the $i^{\rm th}$ source, 
and $\phi_i$ is its orientation. 

We also use $\Delta \phi$, the angle between the pair separation vector 
and a line
perpendicular to the string, to determine how well-aligned a string pair
is with its string candidate. Note that two well-resolved, well-separated and 
well-measured string-lensed conjugate
images will have $\Delta \phi = 0$, $E = \varepsilon_1 \varepsilon_2$, and $\Delta V = 0$.

%%%%%%%%%%%%%%%%%%%%%%%%%%%%%%%%%%%%%%%%%%
\begin{figure*}
\begin{minipage}[t]{0.33\linewidth}
\centering\includegraphics[width=\linewidth]{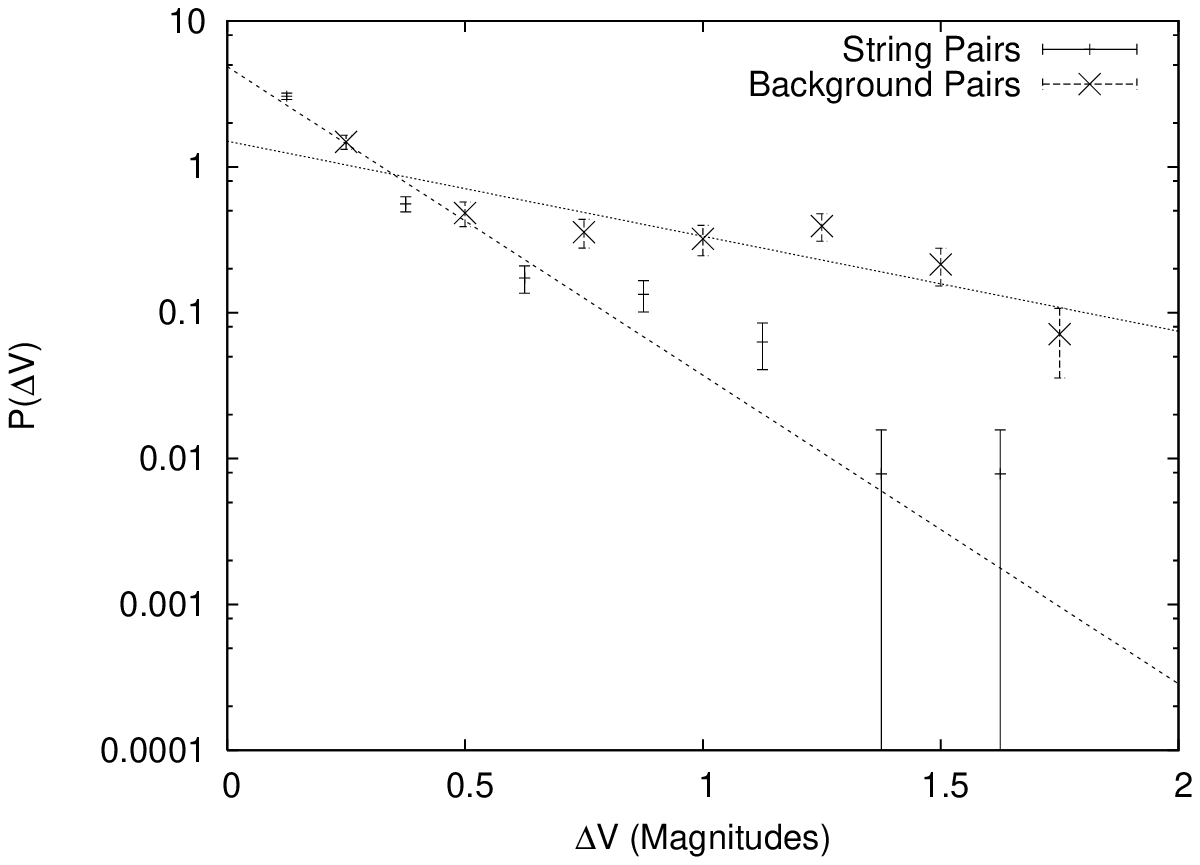}
\end{minipage}
\begin{minipage}[t]{0.33\linewidth}
\centering\includegraphics[width=\linewidth]{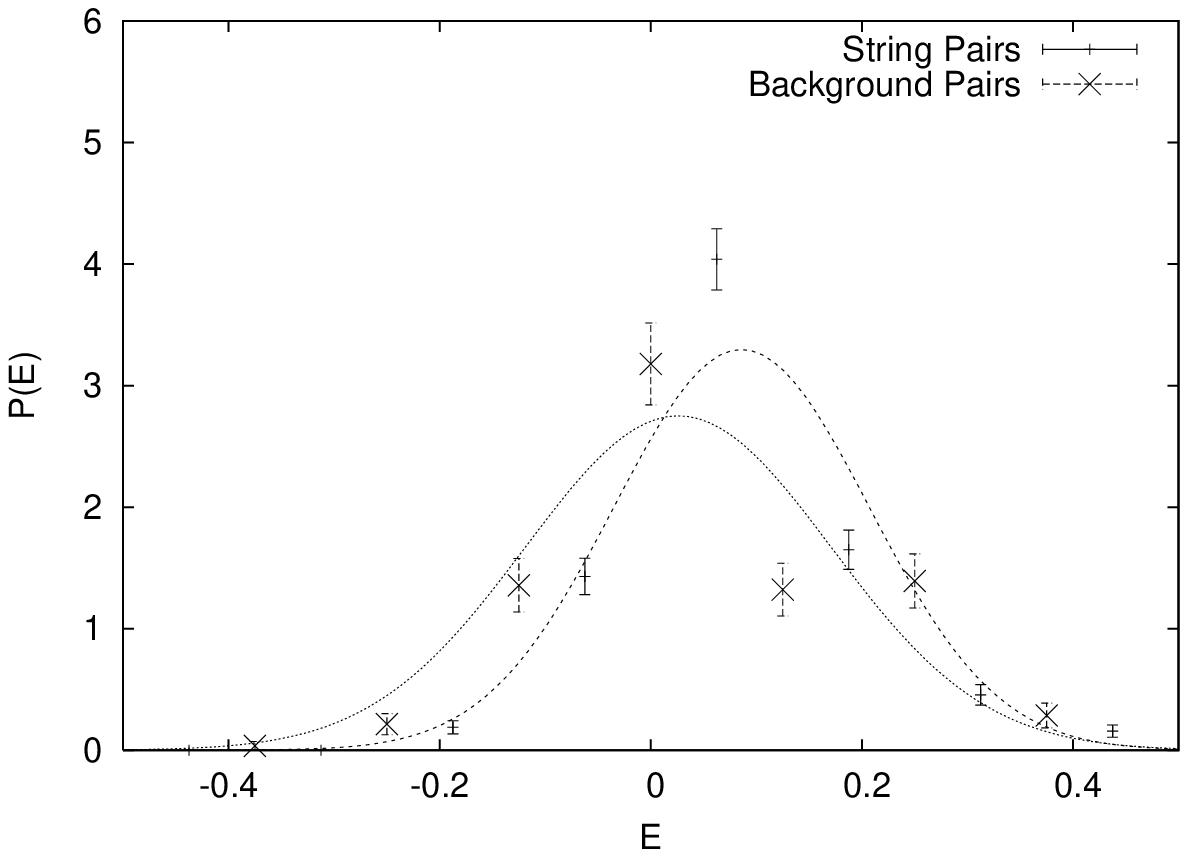}
\end{minipage}
\begin{minipage}[t]{0.33\linewidth}
\centering\includegraphics[width=\linewidth]{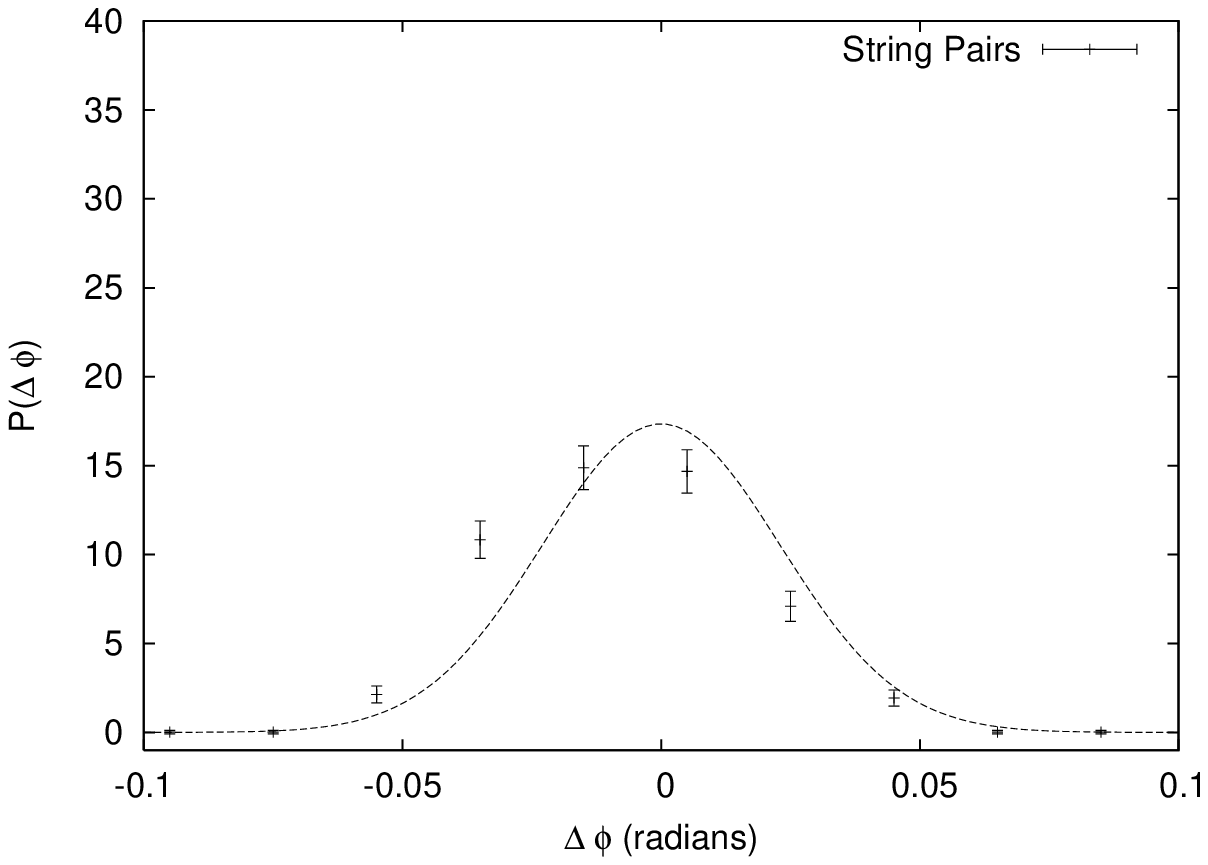}
\end{minipage}
\begin{minipage}[t]{0.33\linewidth}
\centering\includegraphics[width=\linewidth]{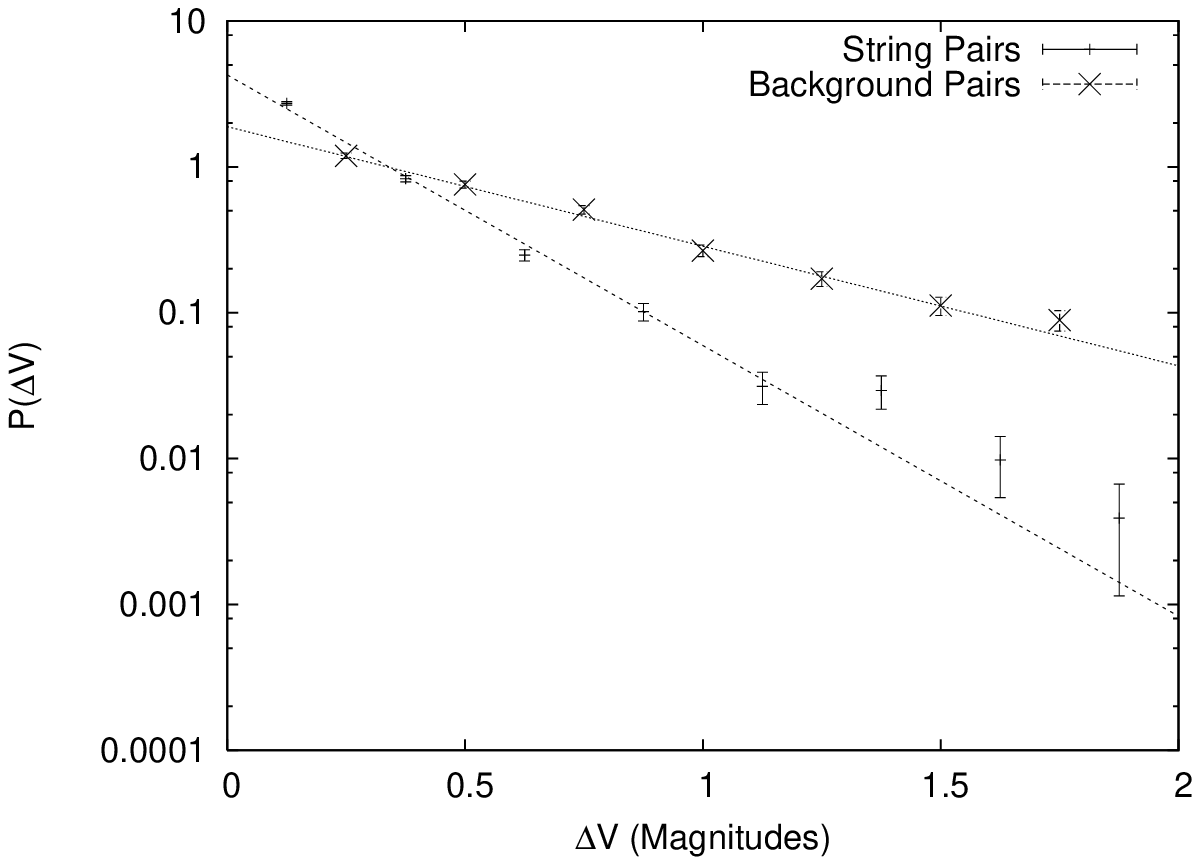}
\end{minipage}
\begin{minipage}[t]{0.33\linewidth}
\centering\includegraphics[width=\linewidth]{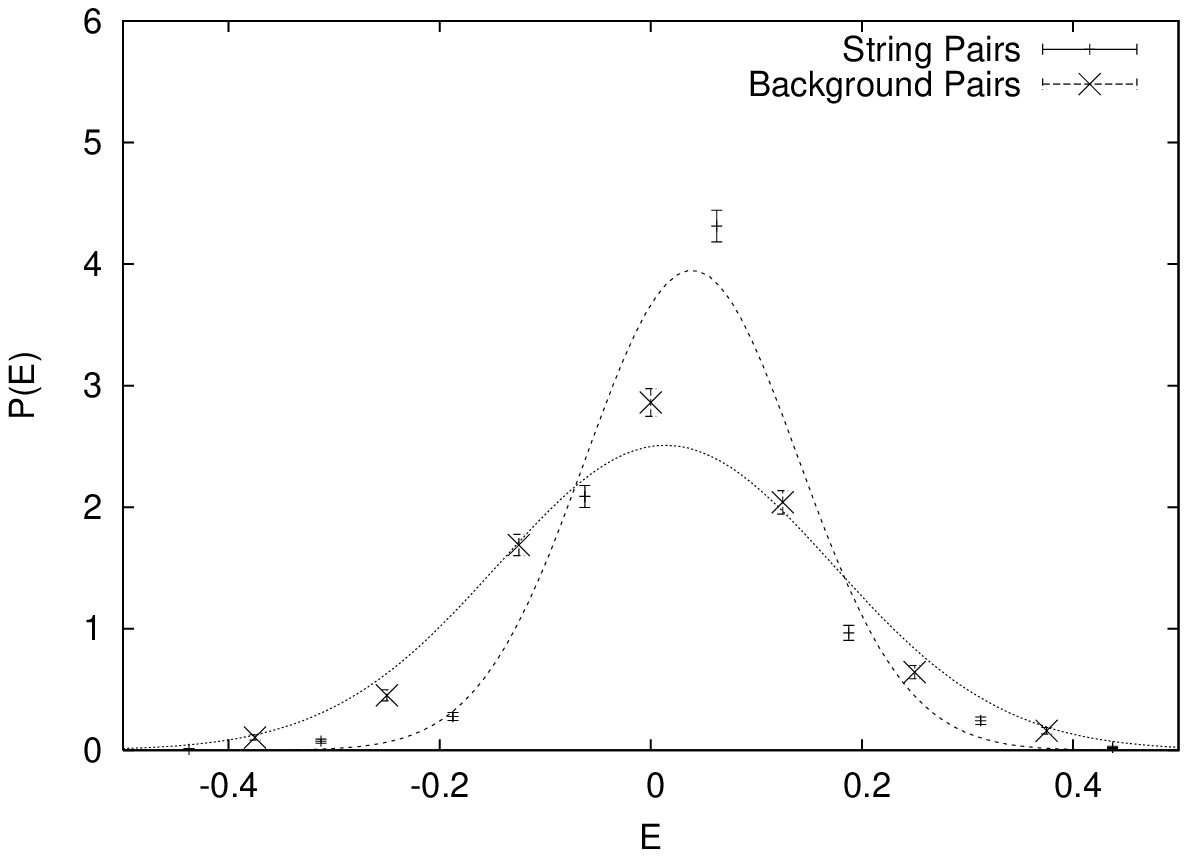}
\end{minipage}
\begin{minipage}[t]{0.33\linewidth}
\centering\includegraphics[width=\linewidth]{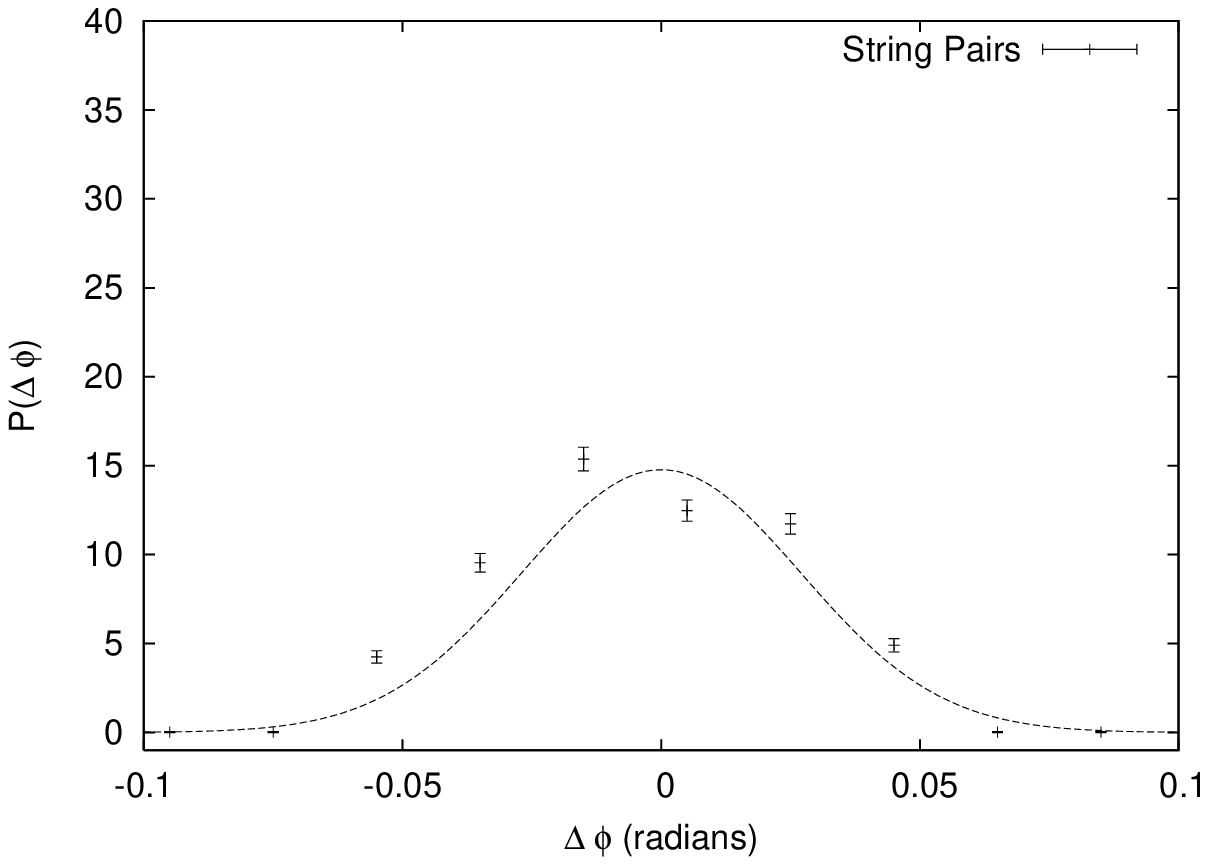}
\end{minipage}
\begin{minipage}[t]{0.33\linewidth}
\centering\includegraphics[width=\linewidth]{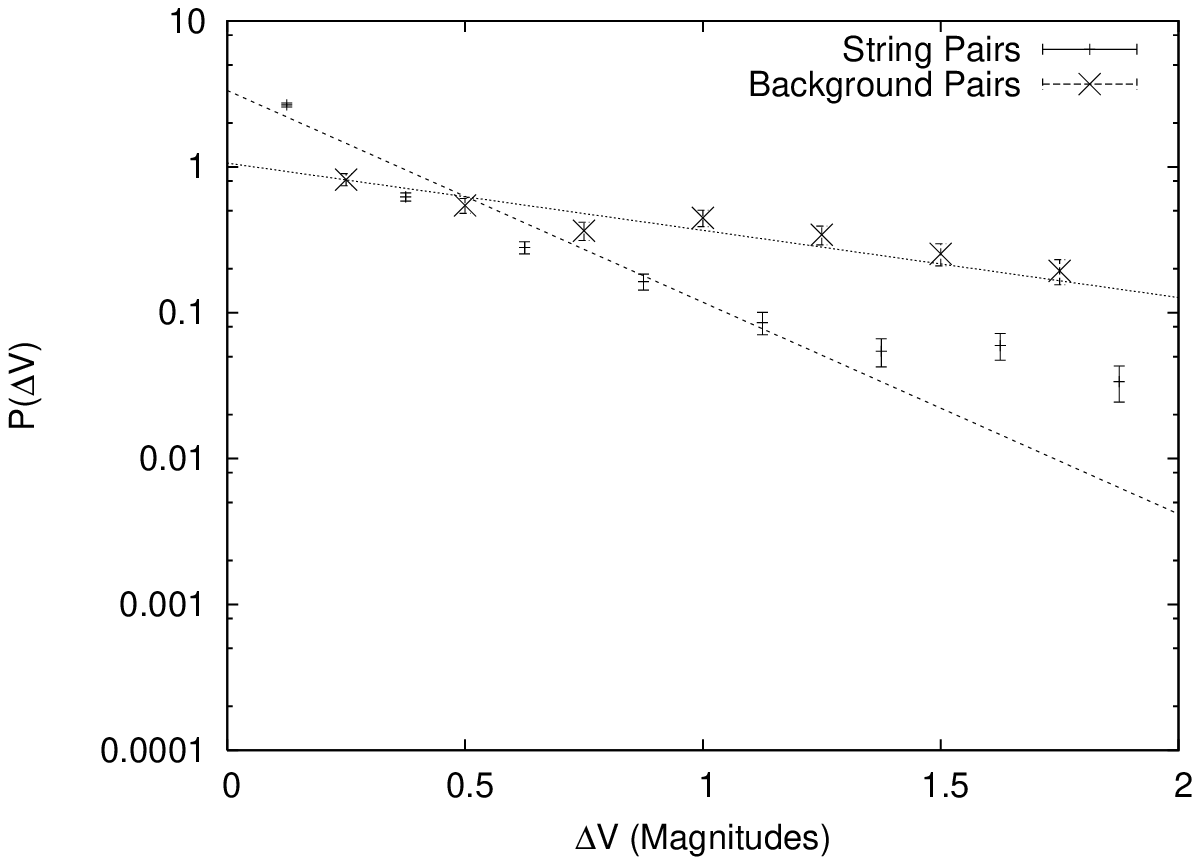}
\end{minipage}
\begin{minipage}[t]{0.33\linewidth}
\centering\includegraphics[width=\linewidth]{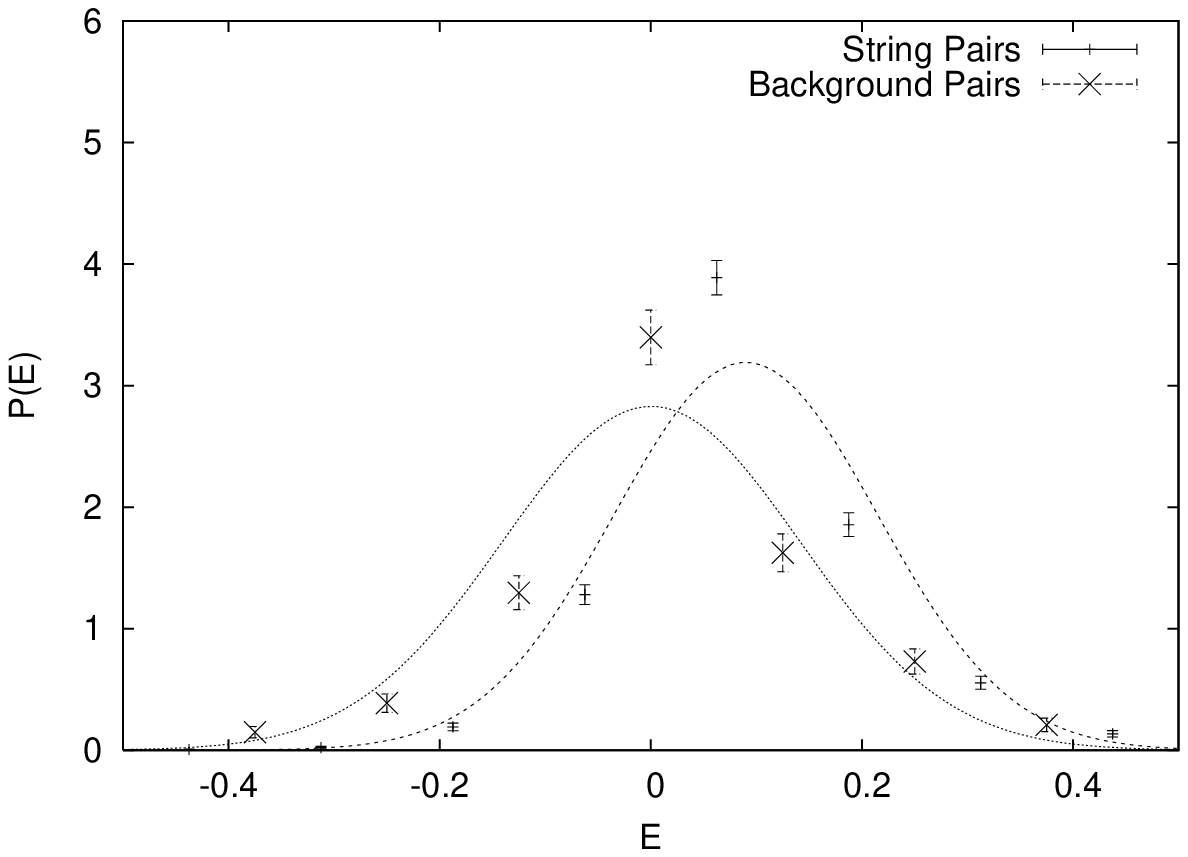}
\end{minipage}
\begin{minipage}[t]{0.33\linewidth}
\centering\includegraphics[width=\linewidth]{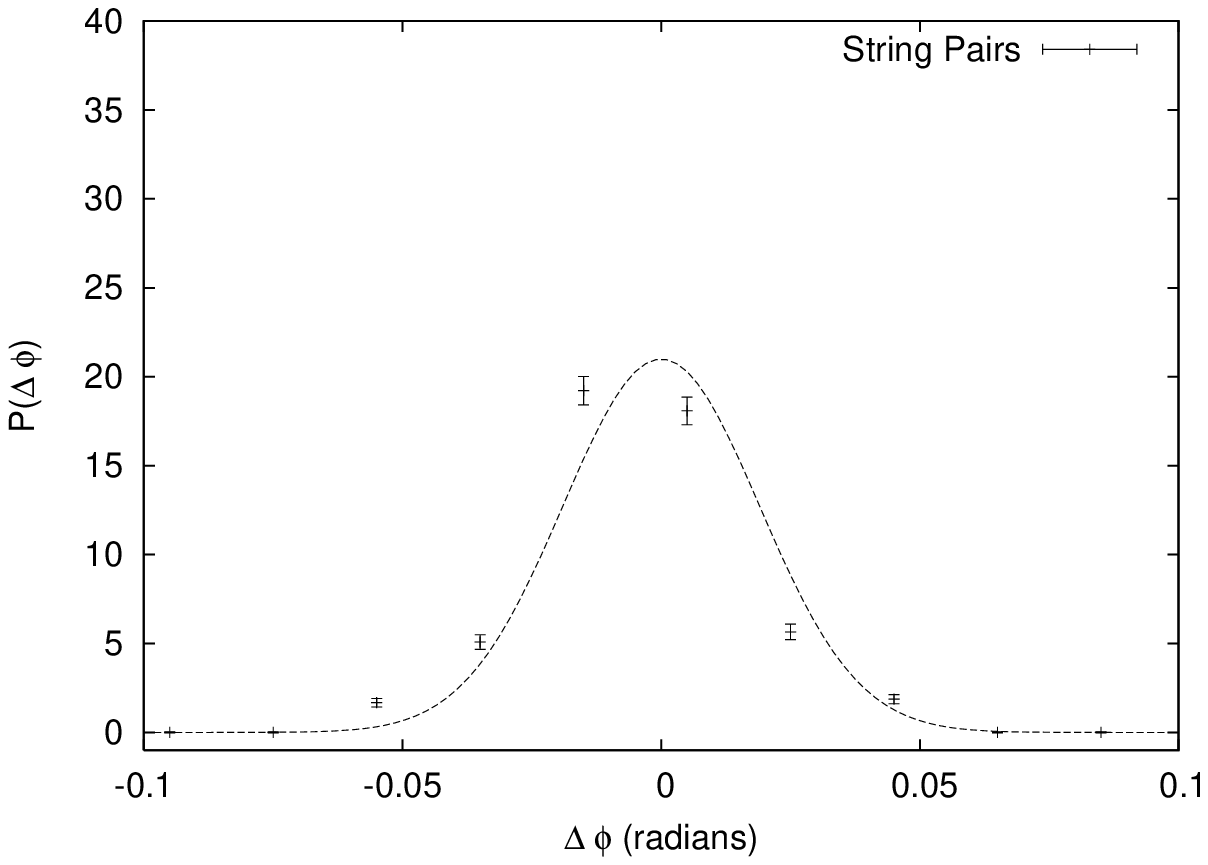}
\end{minipage}
\begin{minipage}[t]{0.33\linewidth}
\centering\includegraphics[width=\linewidth]{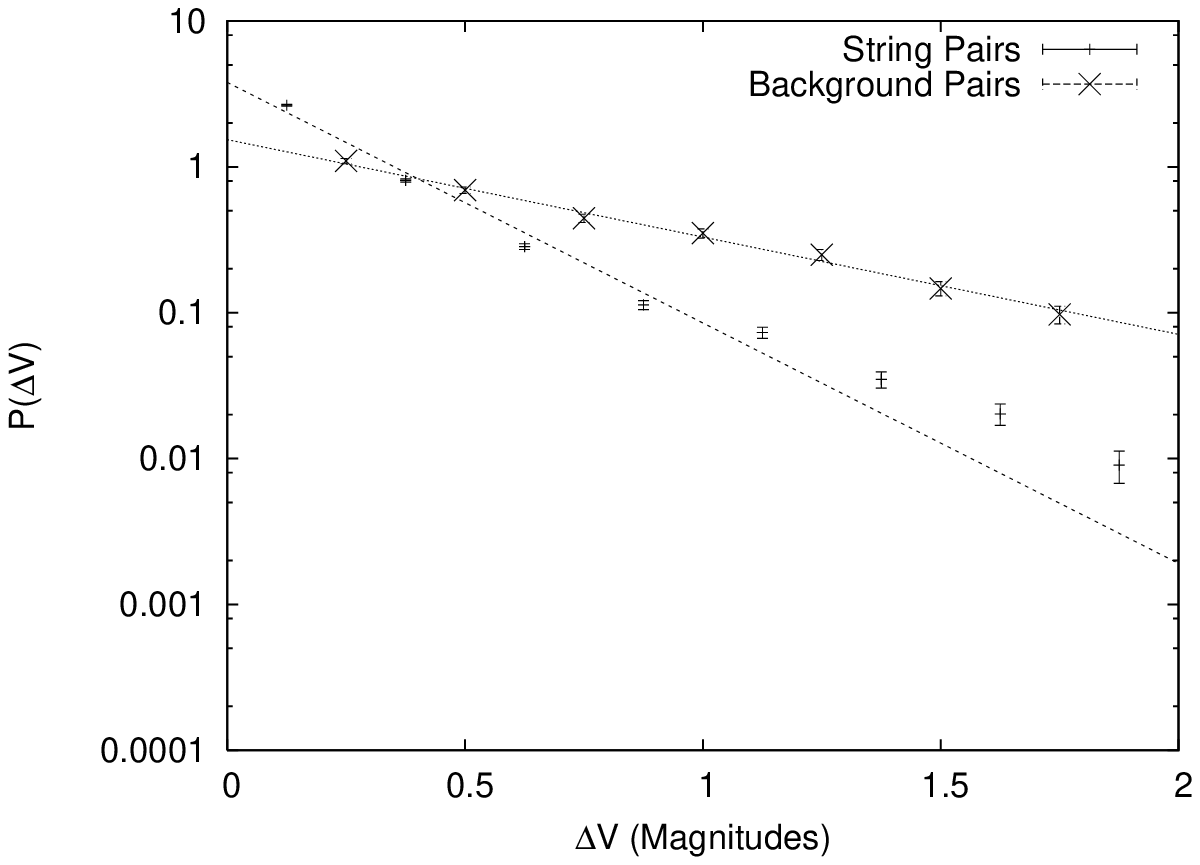}
\end{minipage}
\begin{minipage}[t]{0.33\linewidth}
\centering\includegraphics[width=\linewidth]{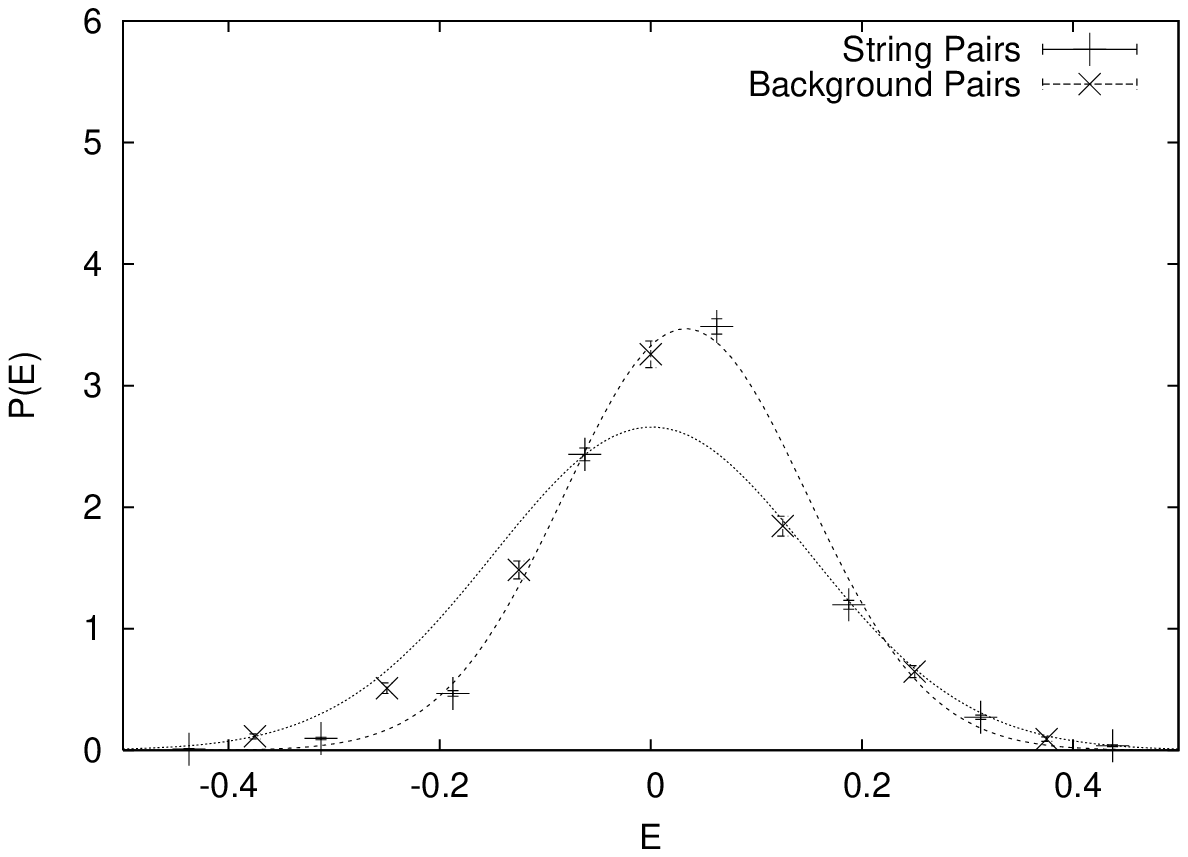}
\end{minipage}
\begin{minipage}[t]{0.33\linewidth}
\centering\includegraphics[width=\linewidth]{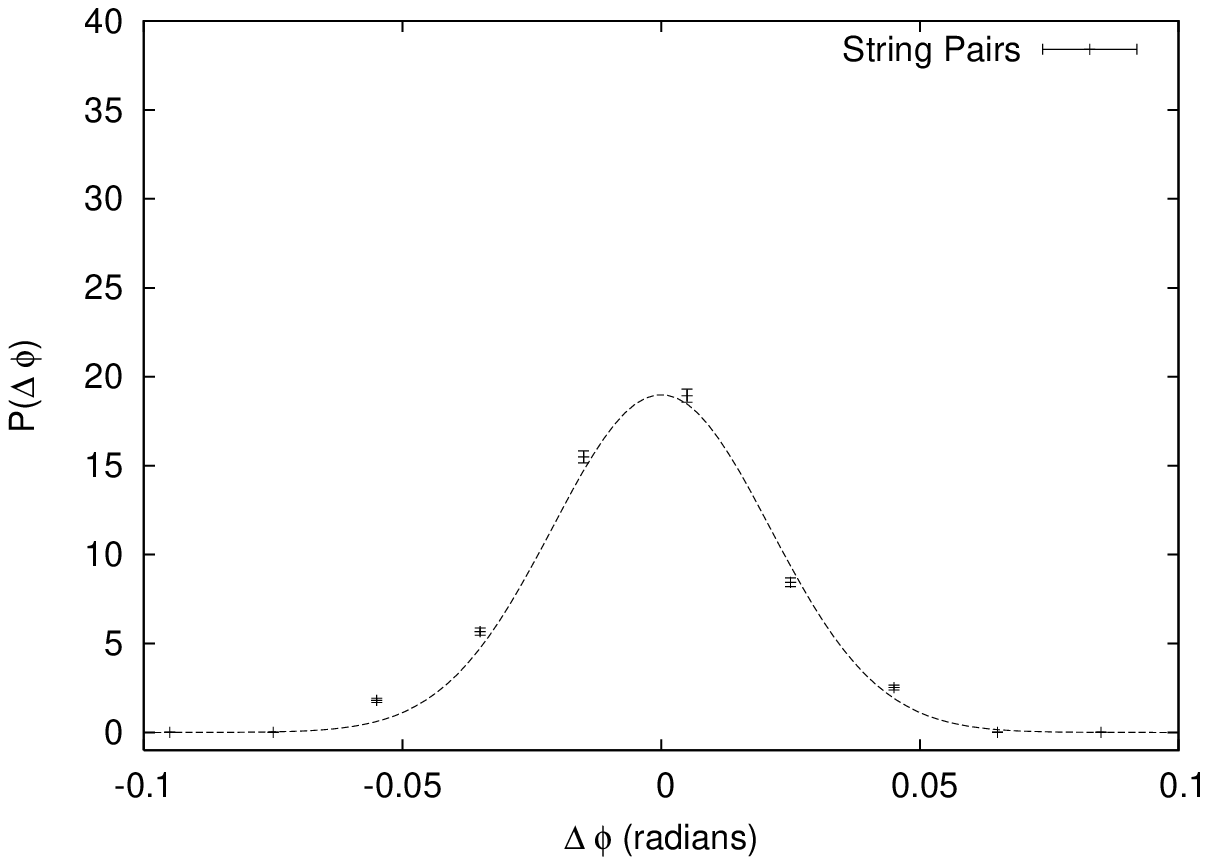}
\end{minipage}
\caption{In each row, we plot the distributions of 
$\Delta V$, $E$ and $\Delta \phi$ for pairs of faint objects detected in
limiting magnitude $V_{\rm lim} = 27$ simulated images. 
The rows correspond to the different
classes of pairs defined in the text.
Top row: close, bright pairs; second row: close, faint pairs; third row:
far,
bright pairs; bottom row: far, faint pairs.}
\label{fig:compare}
\end{figure*}
%%%%%%%%%%%%%%%%%%%%%%%%%%%%%%%%%%%%%%%%%%

The distributions of these three parameters are shown in
Fig.~\ref{fig:compare}.
We fit these distributions with the following functions:
\begin{eqnarray}
P(\Delta V) &=& \frac{1}{\Delta V_0}e^{-\frac{\Delta V}{\Delta V_0}},\\
P(E) &=& \frac{1}{\sqrt{2\pi}\sigma}e^{-\frac{1}{2}\left(\frac{(E-E_0)}{\sigma}\right)^2},\\
\Pstr(\Delta \phi) &=& C+\frac{1}{\sqrt{2\pi}\Delta \phi_0}e^{-\frac{1}{2}\left(\frac{\Delta \phi}{\Delta \phi_0}\right)^2},\label{eq:Pstring}\\
\Pbg(\Delta \phi)&=& \frac{1}{2\Delta \phi_{\rm{max}}}.
\end{eqnarray}
We measure separate $V_0$, $E_0$ and $\sigma$ for string pairs and
background pairs; 
$\Delta \phi_{\rm{max}}$ is the maximum $\Delta \phi$ we accept (0.05 or 0.1
rad, depending on $V_{\rm lim}$). The resulting model distributions are
overlaid on Fig.~\ref{fig:compare}, and the fit parameters tabulated in 
Table~\ref{tab:compare}.

In this table we can see that sources in string pairs have lower
average $\Delta V$ and higher $E$ than random pairs. We also see that $\Delta
\phi$ for string pairs is strongly clustered within $|\Delta \Phi| < 0.1$.
Random pairs have a uniform random $\Delta \phi$ and their distribution is
not shown. 
In Fig.~\ref{fig:compare} we
see that our model distributions 
give good fits in all cases. We use these models in the computation of
string detection probability in 
Section~\ref{sect:method:assess1}.

%%%%%%%%%%%%%%%%%%%%%%%%%%%%%%%%%%%%%%%%%%
\begin{table*}
\begin{tabular}{cc|ccccccc} 
Limiting Magnitude & Subsample & $\Delta V_{0, \rm{str}}$ & $\Delta V_{0, \rm{bg}}$ & $E_{0,\rm{str}}$ & $E_{0,\rm{bg}}$ & $\sigma_{\rm{str}}$ & $\sigma_{\rm{bg}}$ & $\Delta \phi_0$\\
\hline\hline
25 & Close, Bright & 0.350 & 0.604 & 0.086 & 0.006 & 0.095 & 0.109 & 0.046 \\
   & Close, Faint  & 0.355 & 0.511 & 0.010 & 0.010 & 0.126 & 0.152 & 0.052 \\
   & Far, Bright   & 0.327 & 0.997 & 0.057 & 0.002 & 0.154 & 0.157 & 0.036 \\
   & Far, Faint    & 0.382 & 0.569 & 0.016 & 0.000 & 0.134 & 0.157 & 0.044 \\
\hline
26 & Close, Bright & 0.215 & 0.707 & 0.081 & 0.019 & 0.131 & 0.126 & 0.026 \\
   & Close, Faint  & 0.284 & 0.534 & 0.041 & 0.013 & 0.111 & 0.153 & 0.028 \\
   & Far, Bright   & 0.285 & 0.907 & 0.069 & 0.004 & 0.121 & 0.129 & 0.021 \\
   & Far, Faint    & 0.324 & 0.633 & 0.020 & 0.001 & 0.125 & 0.153 & 0.023 \\
\hline
27 & Close, Bright & 0.205 & 0.666 & 0.086 & 0.026 & 0.122 & 0.145 & 0.023 \\
   & Close, Faint  & 0.234 & 0.530 & 0.039 & 0.014 & 0.101 & 0.159 & 0.027 \\
   & Far, Bright   & 0.299 & 0.942 & 0.090 & 0.000 & 0.125 & 0.141 & 0.019 \\
   & Far, Faint    & 0.263 & 0.651 & 0.033 & 0.001 & 0.115 & 0.150 & 0.021 \\
\hline
28 & Close, Bright & 0.200 & 0.684 & 0.100 & 0.023 & 0.116 & 0.138 & 0.022 \\
   & Close, Faint  & 0.233 & 0.545 & 0.039 & 0.015 & 0.093 & 0.158 & 0.026 \\
   & Far, Bright   & 0.259 & 0.991 & 0.113 & 0.009 & 0.138 & 0.144 & 0.018 \\
   & Far, Faint    & 0.249 & 0.668 & 0.038 & 0.004 & 0.109 & 0.147 & 0.020 \\

\hline
\end{tabular}
\caption{Fitted parameters for model distributions of the properties of
string pairs (str) and background pairs (bg).
The close/far dividing separation is $0.5$,'' 
and the bright/faint division is at one magnitude brighter
than the limiting magnitude.}\label{tab:compare}
\end{table*}
%%%%%%%%%%%%%%%%%%%%%%%%%%%%%%%%%%%%%%%%%%

Incidentally, the term~$C$ in equation~\ref{eq:Pstring} represents the
background pairs which are uniformly distributed in $0 < \Delta \phi < 2\pi$. We
use it to find that there is no more than $5\%$ background pair contamination in
any of the string pair subsamples.

%-------------------------------------------------------------------------------

\section{Lens-finding Methodology}
\label{sect:method}

As explained in Section~\ref{sect:theory},
we look for strings that are straight on scales 
at least as large as $4$' (corresponding to $\gtrsim 1.5$~Mpc at $\zd=0.5$)
by correlating close pairs of faint sources along
straight lines across small-field images. We give a brief
summary of the search methodology, before describing it in more detail.

% - - - - - - - - - - - - - - - - - - - - - - - - - - - - - - - - - - - - - - --

\subsection{Summary}
\label{sect:method:summary}

For each field, we first create a catalog optimised to contain as many faint
sources as possible, deblending close pairs aggressively. From  this we
produce a catalog of close ($\theta < 5$'') pairs of objects.  
The line which bisects each pair is then parametrised 
by its position and orientation in
the image --- this line represents
a putative cosmic
string. The change of variables between image pair position and orientation to
bisector line parameters is a modified ``Hough transform:'' pairs straddling the same
straight line will appear very close together in the Hough space \citep{Ball96}. 
We therefore identify clusters of points in the Hough space as indicators 
of possible strings. We then assign a score to each string candidates by 
evaluating the similarity of constituent pairs and their alignment with 
the possible string. We  visually inspect high-scoring candidate strings, 
remove non-physical image pairs and retest them against a higher threshold.
In the next two subsections we explain this process in more detail.

% - - - - - - - - - - - - - - - - - - - - - - - - - - - - - - - - - - - - - - --

\subsection{Image pair detection}
\label{sect:method:detection}

We aim to find many sources and not to exclude pairs. We achieve both these 
goals by emulating the catalog production techniques in
\citet{M+B2008} which are in turn influenced by \citet{Be++2003}. Roughly
speaking, we used the \sex program \citep{B+A1996} to produce catalogs
of faint sources. After we make an initial source catalog, we find all pairs
separated by less than $\theta = 5$'' and make a catalog of pairs. 
We do not probe 
above 5'' because this corresponds to a minimum 
$\G = 1 \times 10^{-6}$ and a typical 
$\G > 5 \times 10^{-6}$ (including projection and redshift effects) which is not an 
interesting  area of parameter space, because it is ruled out with high
significance by \citet{J+S07}.

Our source extraction parameters are summarised in Table~\ref{tab:sex}. We make
three changes to the setup used by \citet{M+B2008}. To make our faint 
detections very sensitive, we lowered our DETECT\_THRESH from $1.7$ to $1.6$ 
for WEIGHT type weight images. However, we primarily use RMS weight images
(which were not available for the GOODS data that our catalog techniques
were initially designed for), and the two types are 
normalised differently. Setting the DETECT\_THRESH for RMS weight images to
$0.895$ produces roughly identical catalogs as those produced above with weight
type WEIGHT. We also changed DEBLEND\_MINCONT from $0.03$ to $0.0003$ to promote
more aggressive deblending. 

These changes do not produce many false counts from noise peaks in our catalogs.
Running \sex on ``negative'' images where the pixels have been multiplied
by $-1$ produces  $0.0002$ false sources for every source detected in real
images. However, complex sources  are often deblended into several sources. At
this stage, the catalogs are inclusive as possible;  false pairs will be rejected later
in the lens-finding process. 

%%%%%%%%%%%%%%%%%%%%%%%%%
\begin{table}
\begin{center}
\begin{tabular}{cc}
%                 & Data Images \\
\hline
WEIGHT\_TYPE            & MAP\_RMS \\
Filtering FWHM (Pixels) & 1.5 \\
DETECT\_THRESH          & 0.895 \\
DETECT\_MINAREA         & 10 \\
DEBLEND\_NTHRESH        & 32 \\
DEBLEND\_MINCONT        & 0.0003 \\
\hline
\end{tabular}
\end{center}
\caption{The \sex parameters we use to find faint sources and close
pairs. }\label{tab:sex}
\end{table}
%%%%%%%%%%%%%%%%%%%%%%%%%

From our initial faint source catalog, we produce a catalog of close pairs.  
In this catalog, we derive various pair parameters including the centre of 
each pair, $(x_0, y_0)$, and the separation vector $(\Delta x, \Delta y)$. 
We also compute each pair's $\Delta V$, $E$ and $\Delta \Phi$ parameters 
(see Section~\ref{sect:sims:pairs}) and interpolate between limiting magnitudes 
to obtain the statistical properties of string pairs and background pairs in 
Table~\ref{tab:compare}. 

String-lensed image separations can be $0 < \theta < 8 \pi \G \sin i$, so we never 
set a minimum $\theta$ when looking for string pairs. However, the number of 
background pairs increases with the separation limit, so we search for strings with 
different maximum $\theta$ to reduce background pairs when searching for
small separation strings. We produce 15 pair subcatalogs for each field with 
$\theta < (0.4, 0.6, 0.8, 1.0, 1.2, 1.4, 1.6, 1.8, 2.0, 2.5, 3.0, 3.5, 4.0, 4.5, 5.0)$ arcsec. All the work in 
Section~\ref{sect:method:hough} and beyond is performed on each pair subcatalog 
separately.

% - - - - - - - - - - - - - - - - - - - - - - - - - - - - - - - - - - - - - - --

\subsection{The Modified Hough Transform to String Parameter Space}
\label{sect:method:hough}

We cannot make a detailed analysis of every possible string orientation in every 
field and must focus on straight lines with many string lens candidates along them. 
An efficient way to find significant numbers of aligned pairs is to 
perform a suitable modified Hough transform on all detected pairs, and then
accumulate a Hough space image (histogram) and search this image for
significant peaks. These peaks represent 
multiple image pairs which could have all been lensed by the same straight 
string.

\begin{figure}
\centering\includegraphics[width=\linewidth]{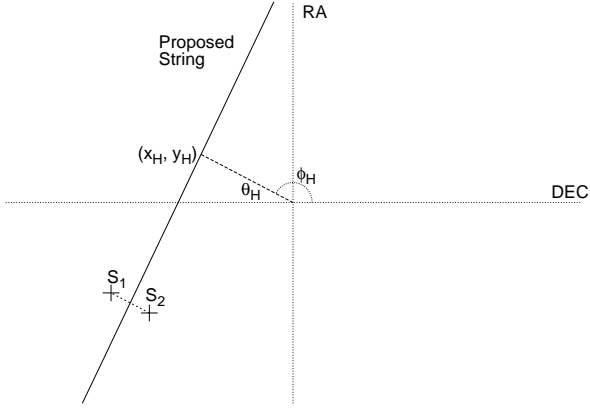} 
\caption{We schematically show the transformation from a close pair (in physical
space) to a potential string (in Hough space). Our pair is represented by the two 
points $S_1$ and $S_2$. The most likely string is the perpendicular bisector of 
this pair. Regardless of the location of the pair along this string, this string 
can be defined by the coordinates of the point ($x_H$, $y_H$) where it is closest 
to the origin (the centre of the data image). 
When we convert this impact parameter to
polar coordinates ($\theta_H$,$\phi_H$) we can produce a Hough space where the
points from random pairs are evenly distributed.}
\label{fig:houghdiagram}
\end{figure}

To convert a single image pair to its corresponding putative lensing string, 
we find the straight line which bisects the separation vector between 
the objects in the pair. For a pair at mean position $(x_0, y_0)$ and 
separation vector $(\Delta x, \Delta y)$, this line is defined as:
\begin{eqnarray}
y& =& -\frac{\Delta x}{\Delta y}(x-x_0)+y_0\\
 &=& -\frac{x_H}{y_H}(x-x_H)+y_H\nonumber\\
(x_H, y_H) &=& (\Delta x, \Delta y)\frac{\Delta x\ x_0 + \Delta y\ y_0}{\Delta
x ^2 +\Delta y^2}
\end{eqnarray}
where $(x_H, y_H)$ are the coordinates of the point where the bisector line
(string) comes closest to the origin (the centre of the field). 
We convert the impact parameter vector $(x_H, y_H)$ to
polar coordinates $(\theta_H, \phi_H)$, because a set of randomly oriented and
distributed pairs will produce a uniform random field in these coordinates. 
$\theta_H$ is then the minimum distance (in arcsec) from the bisector line to
the origin, while $\phi_H$ is the polar angle to the point of closest approach 
between the string candidate and the origin. $\phi_H$ is
measured anticlockwise from the horizontal image axis.
We
illustrate 
this transformation graphically in Fig.~\ref{fig:houghdiagram}. 
 This  transform is equivalent to a standard Hough transform of point sources, 
except
 that each pair, in addition to having a position in image
 space, also has an implied orientation angle. When transformed into Hough space, each of 
these pairs becomes a point source representing a single line which goes through a source point rather
than an extended Hough source representing every line that goes through that source.

%%%%%%%%%%%%%%%%%%%%%%%%%%
%%%%%%%%%%%%%%%%%%%%%%%%%%

In this new mathematical space, each image 
pair corresponds to a single source. 
We would like to weight these sources by their likelihood of being a
string pair rather than a background pair.
To this end,
we assign each source a non-unit intensity~$I$ given by:
\bea
I &=& 1 + \gamma \log\left(\frac{\Pstr(\Delta V, E)}{\Pbg(\Delta V, E)}\right)\\
I &=& 1 + \gamma \log\left(\frac{\Delta V_{0\ \rm{bg}} \sigma_{\rm{bg}} \Delta \phi_{0\ \rm{bg}}}{\Delta V_{0\ \rm{str}} \sigma_{\rm{str}} \Delta \phi_{0\ \rm{str}}}\right)+\frac{\Delta V}{\Delta V_{0\ \rm{bg}}}\\
 && -\frac{\Delta V}{\Delta V_{0\ \rm{str}}}+\frac{(E-\mu_{\rm{bg}})^2}{2\sigma_{\rm{bg}}^2}-\frac{(E-\mu_{\rm{str}})^2}{2\sigma_{\rm{str}}^2}\nonumber
\eea
(Note that we approximate  $P(\Delta V, E) = P(\Delta V) P(E)$, since 
$\Delta V$ and E are uncorrelated for all but the brightest, most completely 
copied pairs). By construction, all
pairs bisected by the same line will appear at the same point in Hough space.
If $\Pstr = \Pbg$ for every pair, each pair would have intensity of 1 and the
total Hough space flux of a feature (string)
would be the number of pairs along a string.
However, pairs of similar sources where $\Pstr > \Pbg$ are weighted more
significantly than other  pairs. The parameter $\gamma = 0.55$, and we discuss
its purpose and optimisation in Section~\ref{sect:method:assess1}.

\begin{figure}
\centering\includegraphics[width=\linewidth]{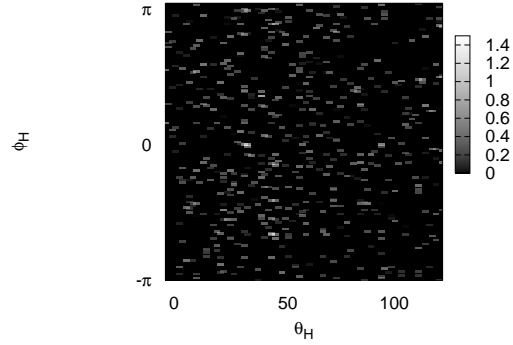} 
\caption{Example Hough space image for one of our simulated single 
string-lensed
fields. The image extends from 
$-4$''$ > \theta_H < 120$'' along the horizontal axis,
 and $-\pi < \phi_H < 3.3$  along the vertical axis. 
 The slight asymmetries in boundaries allow us to account for various edge 
 effects. The simulated string in this field is at
 $(\phi,d) = (25'',0.0)$, and can be seen as a 
 high-significance object in the the
 Hough space image (halfway down and left of centre).}
\label{fig:hough}
\end{figure}

After accumulating individual points in Hough space, and assigning them
intensities, we convolve the resulting $(\theta_H,
\phi_H)$ space image by a $2'' \times 4^\circ$ kernel so that nearby points blend
together to make multi-pair Hough sources. Four degrees is the typical $\Delta 
\phi_H$ between the pair  separation vector and the actual direction of the string. 
$2$'' is the typical error in $\theta_H$ caused by projecting the string back 
towards the origin incorrectly due to $\Delta \phi_H$. The resulting convolved 
images contain bright sources corresponding to lines in the real space images
which have many pairs along them; an example is shown in Fig.~\ref{fig:hough}. 
We extract bright sources from these images using \sex with DETECT\_MINAREA = 7 
and DETECT\_THRESH = 0.16. These settings are optimised to find objects with total 
Hough space flux of 2 units (indicating 2 or more pairs of well-matched galaxy 
images lying along a line in real space). Every peak we detect in this Hough space 
becomes a string candidate.

%%%%%%%%%%%%%%%%%
%%%%%%%%%%%%%%%%%

% - - - - - - - - - - - - - - - - - - - - - - - - - - - - - - - - - - - - - - --

\subsection{Candidate Assessment: Scoring Potential Strings and Score Thresholds}
\label{sect:method:assess1}

Having eliminated the vast majority of possible string orientations 
and identifying likely string candidates as 
significant peaks in the $(\theta_H, \phi_H)$ Hough space image, 
we examine each candidate more carefully using information
from each galaxy image 
pair to assign it a score based on its relative likelihood of
it being a string lensed pair and a random pair. 
Our first step
is to find all pairs which could possibly be associated with our string
candidate. We cast a wide net finding all pairs with $\phi_H$ and $\theta_H$
within $0.1$ and $10$'' of our ideal strings.
This allows even very
incompletely copied pairs to be included for consideration as string lensing
candidate pairs.

We then maximise a string ``score'' 
defined as:
\bea
S &=& \sum_{i} S_{i}\\
S_{i} &=& 1 + \gamma \log\left(\frac{\Pstr(\Delta V_i, E_i, \Delta \phi_i, y_i)}{\Pbg(\Delta V_i, E_i, \Delta \phi, y_i)}\right)\\
S_{i} &=& 1 + \gamma \log\left(\frac{\Delta V_{0\ \rm{bg}} \sigma_{\rm{bg}} \phi_{\rm{max}\ \rm{bg}} y_{\rm{max}\ \rm{bg}}}{\pi \Delta V_{0\ \rm{str}} \sigma_{\rm{str}} \Delta \phi_{0\ \rm{str}} \Delta y_0}\right)\label{eq:Si}\\
&&+\frac{\Delta V_i}{\Delta V_{0\ \rm{bg}}}-\frac{\Delta V_i}{\Delta V_{0\ \rm{str}}}+\frac{(E_i-\mu_{\rm{bg}})^2}{2\sigma_{\rm{bg}}^2}\nonumber\\
&&-\frac{(E_i-\mu_{\rm{str}})^2}{2\sigma_{\rm{str}}^2}-\gamma\log(2\phi_{\rm{max}})-\frac{\Delta \phi_i^2}{2\Delta \phi_0^2}-\frac{y_i^2}{2 y_0^2}\nonumber
\eea

Except for $y_i$ and $y_0$, the terms in the above equation are discussed in
Section~\ref{sect:sims:pairs} and \ref{sect:method:hough} and represent statistical 
distributions of string pairs and background pairs. $y_i$ is the
distance of the centre of the pair from the string. To simplify the calculation
of $y_i$, we rotate our coordinates by $\phi_0 = \sum_i \phi_i$ so that the
string angle $\phi = \phi_0 + \Delta \phi$ and each pair angle $\phi_i = \phi_0
+ \Delta \phi_i$. All the $\Delta \phi$'s are small enough to permit linear
approximations. After this rotation, $x_i$ is the distance of the pair along the 
string, $\theta$ is the $y$-intercept of the string and $y_i$ is the of the centre 
of each pair from the potential string.

In parameterising our string, we only maximise $S$ with respect to $\Delta \phi$ 
and $\theta$. All but the last two terms of Eq. \ref{eq:Si} are constant, and our
score is maximised by solving:

\bea
\left[\begin{array}{cc}\sum_{i}\left(\frac{x_i^2}{y_{0i}^2}+\frac{1}{\Delta\phi_{0 i}^2}\right)&\sum_i \frac{x_i}{y_{0i}^2}\\ \sum_i \frac{x_i}{y_{0i}^2}&\sum_i \frac{1}{y_{0i}^2}\end{array}\right]
\left[\begin{array}{c}\Delta \phi\\\theta\end{array}\right]&=&\label{eq:linefit}\\
\left[\begin{array}{c}\sum_i\left(\frac{x_i y_i}{y_{0i}^2}+\frac{\Delta \phi_i}{\Delta\phi_{0 i}^2}\right)\\\sum_i \frac{y_i}{y_{0i}^2}\end{array}\right]&&\nonumber 
\eea

At this point, the pair with the most negative $S_i$ --- corresponding to a much 
higher likelihood of being a background pair than a lensed pair --- is rejected,
and $\phi$ and $\theta$ are recalculated until all pairs have  $S_i > 0$. 
Only strings with at least three positively scored pairs are considered possible
strings. This three pair requirement is included in our analysis in 
Section~\ref{sect:limits}.

Two terms in this algorithm, $y_0$ and $\gamma$ are not taken directly from
comparisons of simulated string pairs and background pairs. In defining $y_0$
we aim to penalize strings which are farther from the proposed string than could
be allowed by the pair separation, its small uncertainty in position or a slight
curvature of the string. 
The centroid of legitimate string pairs can be up to
$0.5\theta$ (where $\theta$ is the pair separation) away from the string. The
angular size 
of a typical faint galaxy is roughly $0.2$'' and this leads to a
similarly sized uncertainty in position when a galaxy image is cut by
a string. The radius of curvature of a horizon-scale cosmic string should be of
order 1~Gpc \ref{sect:theory:topology}. 

Across a typical $4' \approx 4$ Mpc GO field, a string with a 1 Gpc radius of curvature 
will bend 2 kpc or an observed
$0.3$'' from being a straight line. We do not want to penalize strings for being
less than $0.5$'' or half their pair separation $\theta$ from the string. We
therefore set $y_0 = \rm{max}(0.5''+.5\theta,\theta)$.

The parameter~$\gamma$ determines
the relative weighting between the existence of a pair
along a string, and the degree to which the pair resembles an ideal string pair.
We derive the optimal value of $\gamma$ from our
simulations. Our goal is to produce the highest score differential between the
string pair and the highest scored background pair. We vary $\gamma$ in
increments of $0.05$ and run our algorithm on 15 simulations with limiting
magnitude 27, a string at $z = 0.5$ and at string tensions of $8\pi\G = (0.5'',
1'', 2'')$ (five each). $\gamma = 0.55$ produces the highest average
$S_{\rm{string}}-\max(S_{\rm{background}})$.

\begin{figure}
\centering\includegraphics[width=\linewidth]{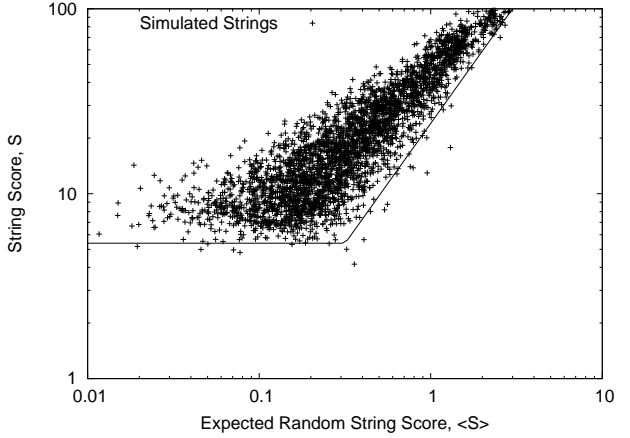} 
\caption{The observed string score versus the expected string score of a randomly oriented line for simulated strings. The threshold line is below 99.5\% of observed strings.}
\label{fig:scorethresh}
\end{figure}

Having defined a metric, S, to evaluate string candidates, we use simulated
string scores to set a threshold score for
all potential strings. This score will be dependent on the limiting magnitude,
maximum pair separation and field size (although in practice, the field size is
fairly constant). 
We combine these factors into $S_0$, the
expected score for a randomly oriented string in our field:
\be
S_0 = \frac{A_{\rm{H\ PSF}}}{A_{\rm{H}}}\sum_{\rm{pairs}\ i} S_i = \frac{1}{1000}\sum_{\rm{pairs}\ i} S_i
\ee
Where the term $\frac{1}{1000}$ accounts for the number of resolvable strings
(the number of PSF areas that fit in our Hough space).
and we are  summing over all the pairs separated by some  $\theta <
\theta_{\rm{max}}$ which has fifteen values for each field as described in 
Section~\ref{sect:method:detection}. We plot the observed scores of all 
simulated strings
with three or more pairs versus the expected score in 
Fig.~\ref{fig:scorethresh} and determine an empirical threshold which 
accepts 99.5\% of simulated strings:
\be
S_{\rm{threshold}}(S_0) = \rm{max}(5.4,24 S_0^{1.3})\label{eq:threshhold}
\ee
%%%%%%%%%%%%%%%%%%%%%%%%%%%
%%%%%%%%%%%%%%%%%%%%%%%%%%%

Scores will be modified slightly in Section~\ref{sect:method:assess2} after string candidates have been inspected by eye and bad pairs removed, so at
this point, we pass all pairs with:
\be
S_i > 0.9 S_{\rm{threshold}}(S_0)\label{eq:threshhold2}
\ee
We learned by trial that random alignments of two pairs are too common to
consider as strings, so we also require that each potential string have three or
more constituent pairs (three pairs detections are assumed when we use our 
results to set limits). 

% - - - - - - - - - - - - - - - - - - - - - - - - - - - - - - - - - - - - - - - 

\subsection{Candidate Assessment: Initial Human Inspection}
\label{sect:method:assess2}

\begin{figure*}
\begin{minipage}[t]{0.49\linewidth}
\centering\includegraphics[width=\linewidth]{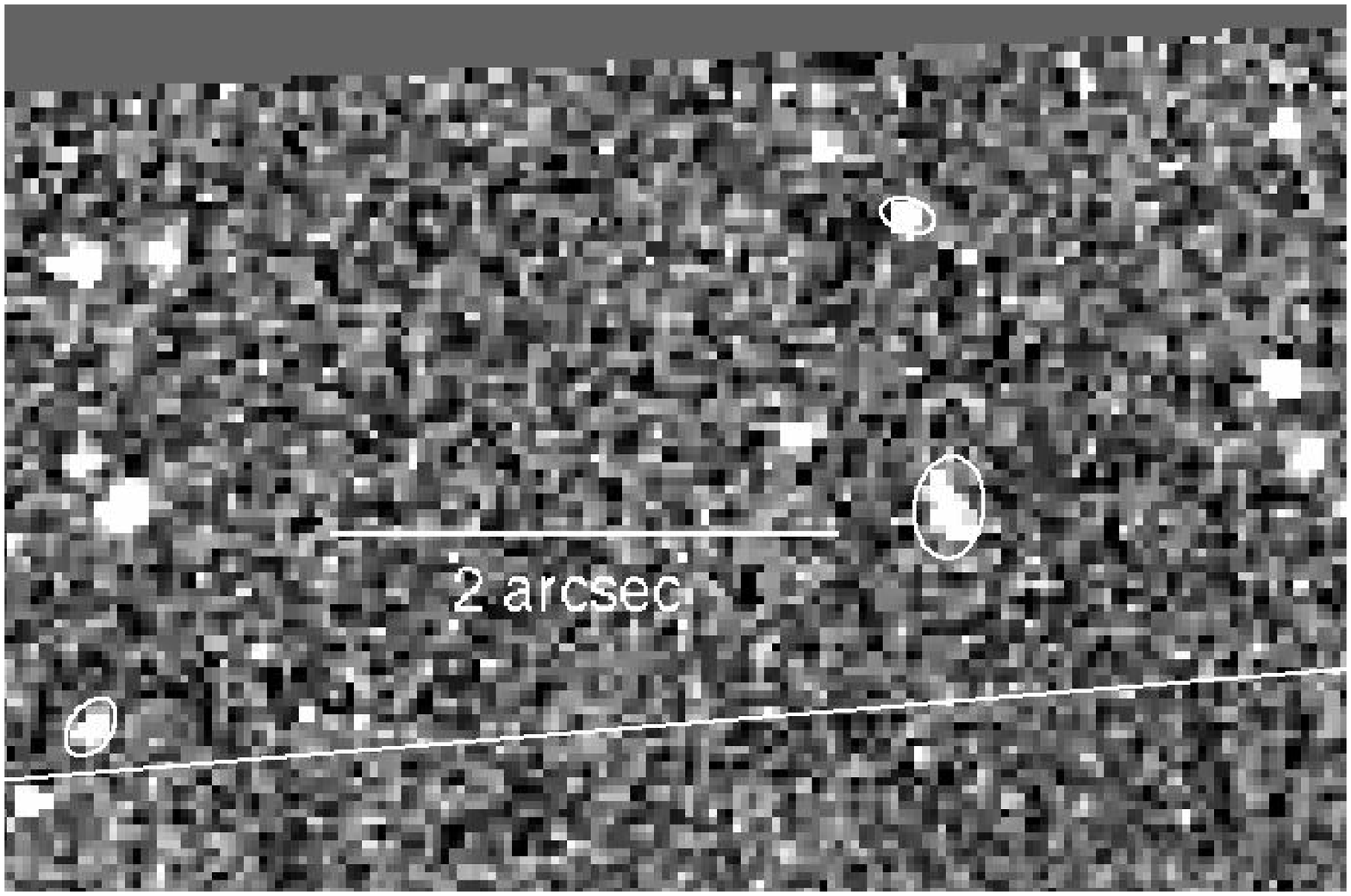}
\end{minipage}
\medskip
\begin{minipage}[t]{0.49\linewidth}
\centering\includegraphics[width=\linewidth]{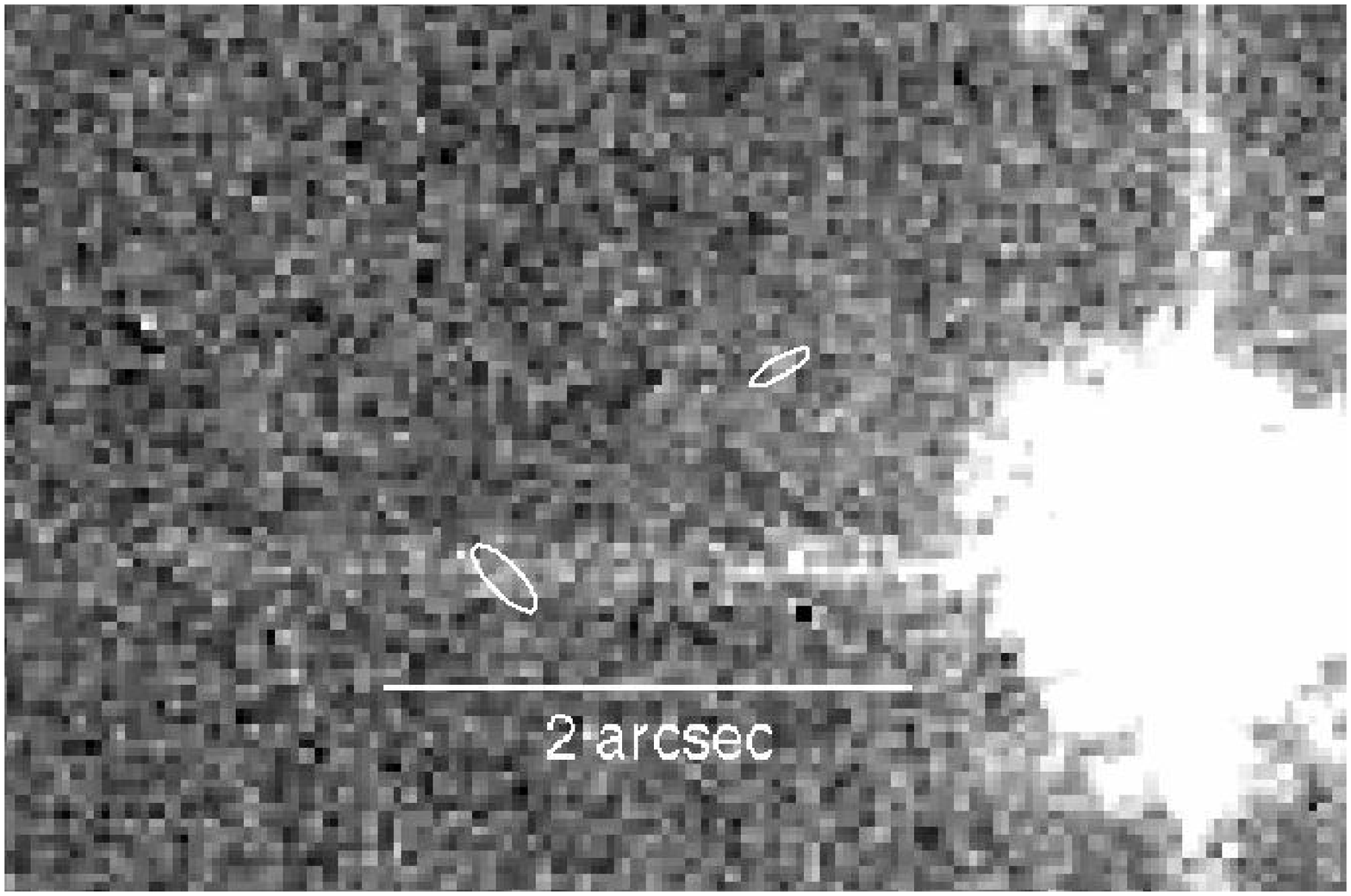}
\end{minipage}
\medskip
\begin{minipage}[t]{0.49\linewidth}
\centering\includegraphics[width=\linewidth]{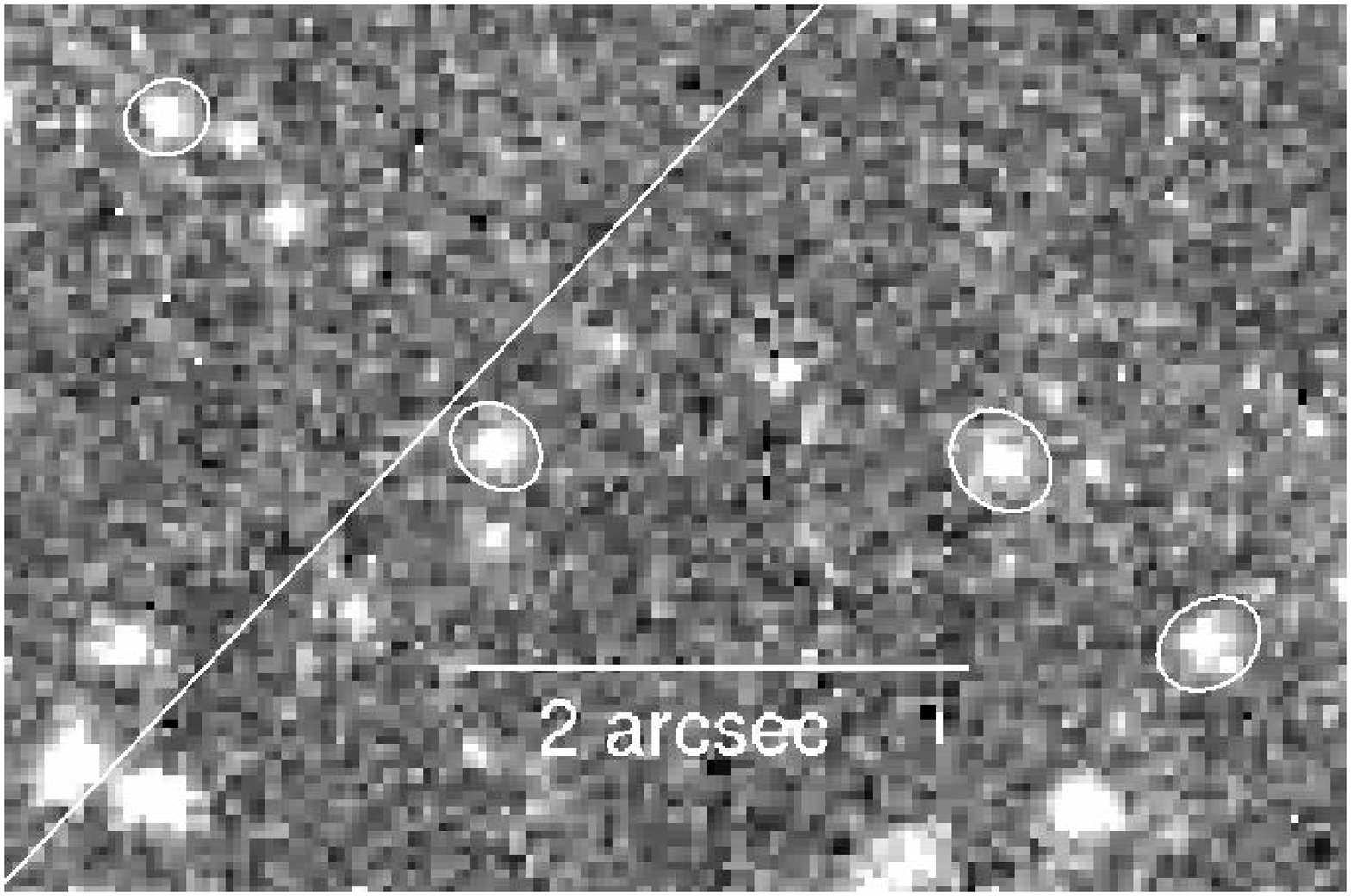}
\end{minipage}
\medskip
\begin{minipage}[t]{0.49\linewidth}
\centering\includegraphics[width=\linewidth]{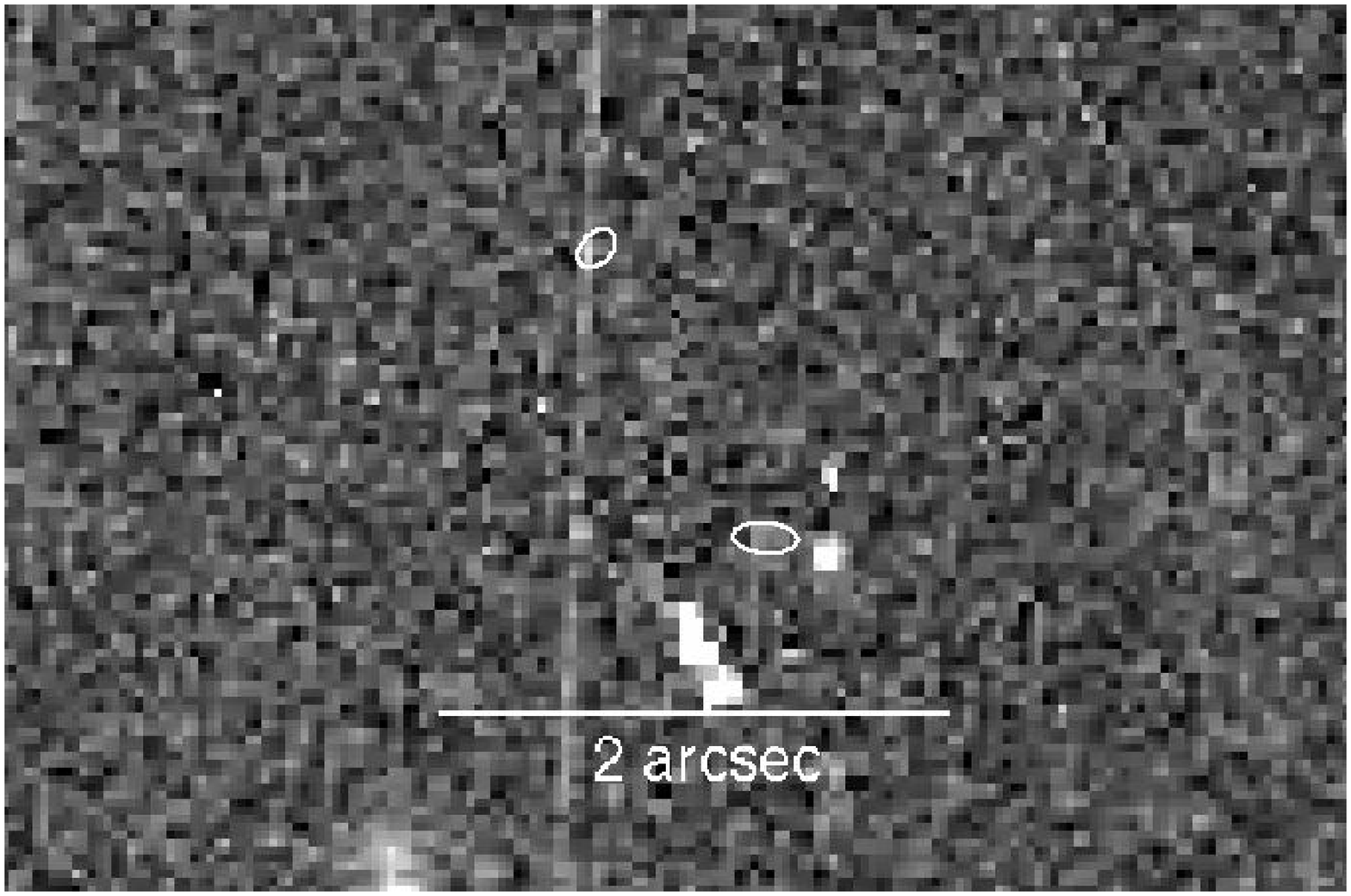}
\end{minipage}
\medskip
\caption{The four basic types of false pairs rejected
during human inspection: cosmic rays not removed by our automated
algorithm (upper
left), a diffraction spike (upper right), a foreground cluster (lower left)
and a line of bad pixels due to poor field processing (lower right).}
\label{fig:badpairs}
\end{figure*}

String candidates which exceed the threshold in equation~\ref{eq:threshhold2}
are passed along to more detailed statistical analysis, but first the
constituent pairs of each string are evaluated by eye to exclude false pairs
from future consideration. The false pairs we exclude come in four types:
residual cosmic rays, diffraction spikes, large foreground objects 
(technically
not false pairs, but generally not cosmologically distant) and misprocessed
data. We show examples of each in Fig.~\ref{fig:badpairs}. A common problem 
with
these false pairs is that they can produce an overdensities along
straight lines, thus 
mimicking the cosmic string signal.
Strings with false pairs are not automatically eliminated at this
stage unless their total number of valid pairs goes below three.

%%%%%%%%%%%%%%%%%%%%%%%%%%%%%%%%%%%
%%%%%%%%%%%%%%%%%%%%%%%%%%%%%%%%%%%

% - - - - - - - - - - - - - - - - - - - - - - - - - - - - - - - - - - - - - - 

\subsection{Candidate Assessment: Detailed String Analysis}
\label{sect:method:assess3}

Having eliminated bad pairs from the potential stings, we can execute two more
rigorous cuts knowing that our results will not be significantly influenced by
false pairs. The first cut is to require that the string go between each pair
(or nearly so) while allowing for some string curvature. The second cut is to
require that the score per pair is similar to that observed in simulation.

To fit a curved string to a set of pairs, we first fit a straight line using
Eq.~\ref{eq:linefit}. We then rotate the pairs by $-\Delta\phi$ so 
that they
are perpendicular to this fitted line. We define a Cartesian system in which $x$ is along the linear approximation
of a string candidate and $y$ is perpendicular to that line. We fit the pair positions with:
\begin{equation}
y_i = \theta + a x_i^2
\label{eq:quadform}
\end{equation}

We intentionally do not fit the curve with three degrees of freedom at once.
The spatial terms $x_i$ are centered so that $x_i = 0$ is in the middle of the
string. Fitting the quadratic term in equation~\ref{eq:quadform} allows for
symmetric curvature across the field, as we would expect for a string with a
very large radius of curvature. A full three-term quadratic fit has the
undesirable property of allowing the fit
to curve out sharply at either end.

As noted in Section~\ref{sect:theory:topology}, the curvature of the string is
believed to be small. We quantify this by assigning the following Gaussian
prior on $a$:
\be
P(a) \propto e^{-5\times10^8 a^2}
\ee
The most probable curved string given the data is then given by:
\bea
\left[\begin{array}{cc}5\times10^8+\sum_{i}\left(\frac{x_i^4}{y_{0i}^2}+\frac{4 x_i^2}{\Delta\phi_{0 i}^2}\right)&\sum_i \frac{x_i^2}{y_{0i}^2}\\ \sum_i \frac{x_i^2}{y_{0i}^2}&\sum_i \frac{1}{y_{0i}^2}\end{array}\right]
\left[\begin{array}{c}a\\\theta\end{array}\right]&=&\label{eq:curvefit}\\
\left[\begin{array}{c}\sum_i\left(\frac{x_i^2 y_i}{y_{0i}^2}+\frac{2 x_i \Delta \phi_i}{\Delta\phi_{0 i}^2}\right)\\\sum_i \frac{y_i}{y_{0i}^2}\end{array}\right]&&\nonumber 
\eea
Once this fit is obtained, we examine all pairs in the field 
that this curve does 
{\it not} pass through and whose closest source is more than $0.2$'' 
away from the curve. The pair whose closest source is farthest from the curve
is excluded. The fit is recalculated, first as a line and then as a curve
until all pairs are either split by the curve or within $0.2$'' of being split
by the curve. This more strict filtering eliminates many high separation
string candidates, but does not affect a significant number of simulated straight
strings.

\begin{figure}
\centering\includegraphics[width=\linewidth]{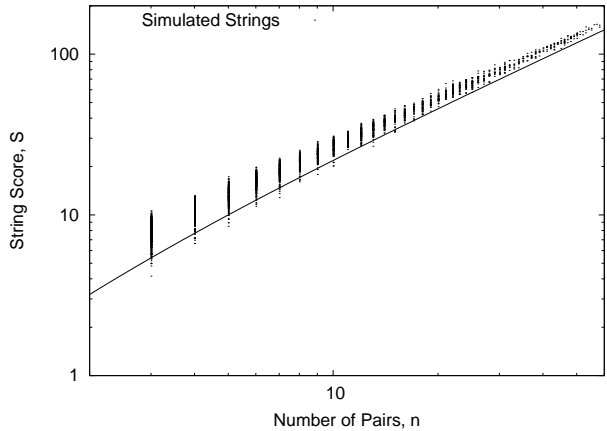} 
\caption{The observed string score versus the number of pairs. The
threshold line is below 99.5\% or observed strings.}
\label{fig:scoren}
\end{figure}

Our final method of string candidate rejection is to assert 
a cutoff on string score as
a function of the number of pairs associated with the string,~$S(n)$. 
In Fig.~\ref{fig:scoren}, we plot
$S(n)$ for our simulated strings and find that
\be
S_{\rm{Threshold}}(n) = \frac{2.4 n^2}{1+n}
\ee
is a threshold that accepts more than 99.5\% of strings with 3 or more
observed pairs.

%%%%%%%%%%%%%%%%%%%%%%%%%%
%%%%%%%%%%%%%%%%%%%%%%%%%%

String candidates that survive these two cuts  are passed along to a final
round of field-dependent inspection in which neighboring fields in surveys and
multi-band images are used (where possible) to eliminate string candidates.
These methods require more human involvement and are not generally applicable,
so we present them separately in Section~\ref{sect:results}. Having outlined
the end-to-end string candidate generation procedure, we next describe the
imaging dataset we use in our string search.

%-------------------------------------------------------------------------------

\section{Survey data: the HST/ACS archive}
\label{sect:data}

We now describe the input data for our cosmic string search.  As previously
discussed, we use HST 
optical imaging data for its high resolution and
background source density. For ease and homogeneity of processing we focus on
images taken with the Advanced Camera for Surveys(ACS). The
total sky area imaged with ACS is approximately 12 square degrees. However, we
require the images to be sufficiently deep and high galactic latitude  to be
able to detect significant numbers of faint galaxies at redshift 1--2,
and for those images to be largely 
free from confusing cosmic rays. We selected 4.5
square degrees of ACS imaging data that met these criteria (although see
Section~\ref{sect:data:cr} below for further discussion of cosmic ray
contamination); 
about half of this
data comes from the large programs GEMS \citep{Rix++04}, GOODS \citep{Gia++04},
AEGIS \citep{Dav++07} and COSMOS \citep{Sco++07}.  The remainder are data
from  extragalactic Guest Observer (GO) observations. For reasons explained
in Section~\ref{sect:results} we divide the GO images into GO-M (fields with imaging
in multiple filters) and GO-S (single filter fields). Approximately 0.75
square degrees of the GO-M imaging had more than 2000 seconds exposure time, 
and was made public before November
2005: this subset was searched for galaxy-scale strong gravitational lenses by
Marshall et al.\ (in preparation) as part of the HST Archive
Galaxy-scale Gravitational Lens Search (HAGGLeS).  Here, we use a larger
set that includes more recently-observed fields, fields that were only
observed in one filter and fields with a little as 1500 seconds exposure time. 
We retain the need for 3 exposures per field for
effective cosmic ray rejection. We summarise the HST/ACS imaging data used in
Table~\ref{tab:data}.

\begin{table}
\begin{center}
\begin{tabular}{ccccc}
\footnotesize
Survey      & No.\ of fields & Area        & Mean depth \\
            &                & (deg$^2$)   & (AB mag)   \\
\hline\hline
GO-S        & 318$^a$        & 1.08        & 25.8       \\
GO-M        & 346$^a$        & 1.18        & 26.3       \\
COSMOS      & 575            & 1.84$^b$    & 26.5       \\
AEGIS       & 63             & 0.20        & 25.8       \\
GEMS        & 63$^c$         & 0.18        & 25.3       \\
GOODS       & 35             & 0.10        & 28.6       \\
\hline
Overall:    & 1400           & {\bf 4.48}  & 26.2                                    &                 \\
\hline
\end{tabular}
\end{center}
\caption{HST/ACS archive data used. See text for our source detection parameters, and the
definition of ``depth.'' 
Notes: $a)$~This is the number of fields remaining after visual inspection for
high cosmic ray or star density (Section~\ref{sect:data:cr}).
$b)$~A pointing overlap area of 4\% was assumed for each of the large surveys, as in 
\citet{Sco++07}.
$c)$~For the purposes of this paper, GEMS does not include GOODS-S area: we
have subtracted the overlap area (which we estimate to be about 0.015~deg$^{-2}$,
\citeauthor{Rix++04}~\citeyear{Rix++04})
between the two surveys from the GEMS area.}
\label{tab:data}
\end{table}

% % Plotting dndzs:
% 
% set surveys = ( GO    COSMOS  AEGIS   GEMS  GOODS  Mean )
% set depths  = ( 26.1  26.5    25.8    25.3  28.6   26.2 )
% 
% foreach i ( `seq $#surveys` )
% echo "0.0\
% 20,$depths[$i]\
% 0.0\
% 0" | CriticalDensity >& /dev/null
% mv criticaldensity_dndzs.txt dndzs_${surveys[$i]}.txt
% mv criticaldensity_dpdzs.txt dpdzs_${surveys[$i]}.txt
% \rm -f criticaldensity*
% wc -l d?dzs_${surveys[$i]}.txt
% end
% % 200 dndzs_GO.txt
% % 200 dndzs_COSMOS.txt
% % 200 dndzs_AEGIS.txt
% % 200 dndzs_GEMS.txt
% % 200 dndzs_GOODS.txt
% % 200 dndzs_Mean.txt
% 
% echo "GO\
% COSMOS\
% AEGIS\
% GEMS\
% GOODS" > dpdzs.legend
% echo "1 4 1\
% 2 3 1\
% 3 3 1\
% 5 3 1\
% 4 3 1" > dpdzs.style
% 
% lineplot.pl -l -xmax 4.9 -ymax 0.9 \
% dpdzs_GO.txt dpdzs_COSMOS.txt \
% dpdzs_AEGIS.txt dpdzs_GEMS.txt dpdzs_GOODS.txt \
% -o dpdzs.ps \
% -xlabel "source redshift z\ds\u" \
% -ylabel "Probability density Pr(z\ds\u)" \
% -legend dpdzs.legend -style dpdzs.style 
% 
% mv dpdzs.ps ../figs/

%%%%%%%%%%%%%%%%%%%%%%%%%%%%%%%%

%%%%%%%%%%%%%%%%%%%%%%%%%%%%%%%%
\begin{figure}
\centering\includegraphics[width=\linewidth]{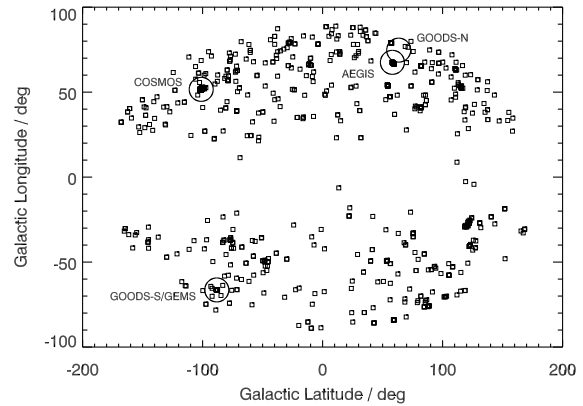}
\caption{HAGGLeS HST/ACS field positions, in galactic coordinates. The
symbols are not to scale.}\label{fig:data}
\end{figure}
%%%%%%%%%%%%%%%%%%%%%%%%%%%%%%%%

The geometry of the GO survey is shown in Fig.~\ref{fig:data}. The avoidance
of the galactic plane area can be seen, but otherwise the HST/ACS archive does
a reasonable job of uniformly sampling the sky. 

% - - - - - - - - - - - - - - - - - - - - - - - - - - - - - - - - - - - - - - --

\subsection{Survey field depths and areas}
\label{sect:data:fields}

We initially search through individual fields from large surveys as though they
were independent entities. When determining the limits implied by null detections  
in each survey, however, we view each contiguous survey as one large area (with GOODS 
North and GOODS South being two areas). This method automatically accounts for
field overlap in surveys, reducing areas by 4\% \citep[as appropriate for the
largest survey, COSMOS,][]{Sco++07}. We could search for strings across full 
surveys as though they were individual fields, but our measurement of $\theta_H$ 
becomes less accurate as fields get larger. In 
Section~\ref{sect:results:inspection}, we discuss how we use the 
contiguous nature of 
large surveys to eliminate string candidates. The GO fields are
predominantly single ACS pointings, although in some cases we mosaiced
overlapping fields together. 

Very few instances of GO field overlap remain. We
account for overlaps between the large program fields by working with a single
catalog of unique sources for each large program area; we then correct the
large program survey areas downwards by 4\% \citep[as appropriate for the
largest survey, COSMOS][]{Sco++07}  when computing string densities and
parameter constraints.

%%%%%%%%%%%%%%%%%%%%%%%%%%%%%%%%
\begin{figure}
\centering\includegraphics[width=0.9\linewidth]{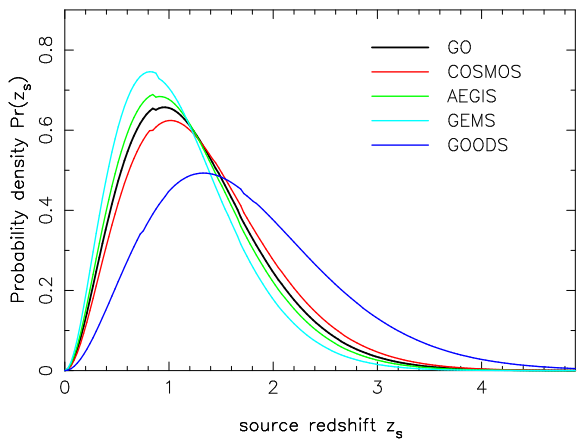}
\caption{Source redshift distributions assumed for the different surveys.
The ``GO'' curve shows the mean redshift distribution: in practice, we used a
different distribution for each field according to its depth.}
\label{fig:dpdzs}
\end{figure}
%%%%%%%%%%%%%%%%%%%%%%%%%%%%%%%%

The ``depth'' of each field is a label allowing the sensitivities to source
detection to be quantified in each field. It is defined as the F814W AB
5-sigma limiting magnitude for an extended source measured within a 0.4" 
radius circle, were it a faint blue galaxy at redshift~1.5. This definition
was adopted to allow comparisons between fields observed in different filters:
specifically we assume the Scd template spectral energy distribution of
\citep{CWW80} for the transformations. The depths we derive match well the
extended source F814W AB limiting magnitudes calculated by the AEGIS project
team  \citep{Dav++07} and the 90\% faint galaxy completeness limit estimated
for their images by the COSMOS team \citep{Lea++07}. 
We chose the F814W filter for consistency with the model redshift distribution
of \citep{Lea++07} that we have adopted. This is plotted for each of the
surveys in Fig.~\ref{fig:dpdzs}.

% - - - - - - - - - - - - - - - - - - - - - - - - - - - - - - - - - - - - - - --

\subsection{Image processing and Cosmic Ray rejection}
\label{sect:data:cr}

In the case of the large programs, we used the high level archive
science product 
images provided by the project teams. The GO images were reprocessed from
the uncalibrated data using the multidrizzle pipeline developed for the
HAGGLeS project (Marshall et al.\ in preparation).  This processing included a
visual inspection step to remove cosmic ray clusters, satellite trails,
scattered light and other confusing artifacts. 

Many of the GO fields were produced from only a
few raw images. While HAGGLeS aims to produce uniformly high-quality data, many
fields suffer from cosmic ray corruption not recorded in the weight image. Since
we are not interested in perfectly accurate catalogs but rather rare, high-level
correlations, we use a simple method to extract the vast majority of these
sources and accept some impurity.

\begin{figure}
\centering\includegraphics[width=\linewidth]{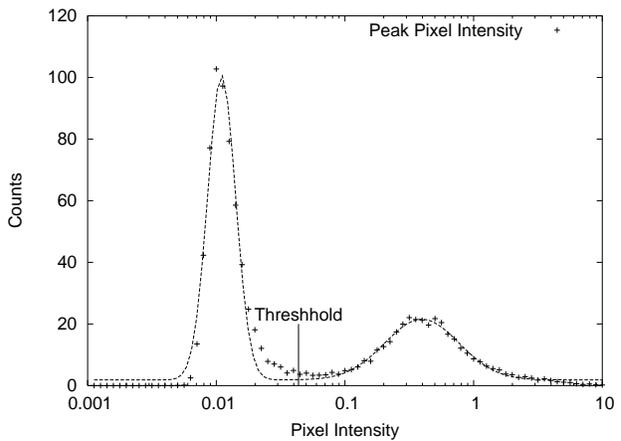} 
\caption{
The distribution of brightest pixel intensity distribution in field
ACSJ002418-020750. We see a characteristic double Gaussian distribution 
with the
brighter of the two peaks being due to cosmic rays.
}
\label{fig:cosmicrays}
\end{figure}

We have many processed images with false sources caused by cosmic rays.
Typically, these detections are around the edges of images where the input raw
images did not overlap, but in some fields, cosmic rays corrupt the image interior as
well. The HAGGLeS weight images usually mark these pixels as having high
variance, but this process is likely to fail when there are fewer input exposures.
Fortunately, cosmic ray events tend to deposit more energy on a pixel than all
but the bright sources. Examining the intensity of the brightest pixel in every
detect source in a typical HAGGLeS GO field produces a double Gaussian
distribution as shown in Fig.~\ref{fig:cosmicrays}.

%%%%%%%%%%%%%%%%%%%%%%%
%%%%%%%%%%%%%%%%%%%%%%%

The higher intensity peak is due to cosmic rays,
while the lower intensity peak is due to astronomical sources. 
We fit the
distribution with two Gaussians centered at $\log I = \mu_1$ and $\mu_2$ and
with widths $\sigma_1$ and $\sigma_2$, and reject sources with $\log I > \mu_0$
where:
\bea
\mu_0 &=& \frac{\mu_1+2\sigma_1+\mu_2-2\sigma_2}{2},\ \rm{for}\ \mu_1 < \mu_2 -2\sigma_1-2\sigma_2\\
\mu_0 &=& \frac{\mu_1+\sigma_1+\mu_2-\sigma_2}{2},\ \rm{for}\nonumber\\ 
&&\mu_2 -2\sigma_1-2\sigma_2 < \mu_1 < \mu_2 -\sigma_1-\sigma_2\nonumber
\eea
We do not filter out bright sources if the peaks cannot be resolved at the
$\sigma_1+\sigma_2$ level. Generally, these fields have very low cosmic rays
counts, but some fields are dominated by cosmic rays and were excluded 
during visual inspection. After we have identified a population of cosmic rays,
we search for significant overdensities around the edges of our image and remove
these edges if an overdensity exists. This filtering is imperfect
since the cosmic ray intensity peak has a long negative tail and the source peak
intensity has a long positive tail, but we estimate roughly 95\% purity in
typical samples. We eliminated 74 of the 738 GO fields as being
unsuitable for string detection due to excessive cosmic rays, stars or foreground 
clusters (which prohibit string detection just like excess cosmic rays) in this way. 

%-------------------------------------------------------------------------------

\section{Results}
\label{sect:results}

The search methods described in Section~\ref{sect:method} are successful at
greatly reducing the number of potential strings in our surveys, but we must
address the small number of remaining candidates in a rigorous way in order to
produce detections or robust upper limits on string tension and concentration.
Our methods sort through millions of possible strings in 1400 fields, leaving
only 50 string candidates which we must study in more depth. These remaining 
candidates show every indication of being the statistical tail of coincident 
alignments. 

In Table~\ref{tab:npairs}, we show how many string candidates existed at each
stage of our analysis. Each of our fifteen Hough images can resolve roughly 1000
strings, so we are in effect examining 15000 potential strings per field. Only
1638 potential strings (1.2 per field) pass initial automated inspection
described in Section~\ref{sect:method:assess1}. These candidates were all
examined by eye and had any obvious bad pairs removed. Only 50 of the remaining
candidates passed the final automated test in 
Section~\ref{sect:method:assess3} with at least 3 pairs and a score per pair
consistent with simulated strings. The 3-pair requirement is particularly
effective at eliminating potential strings in bright
limiting magnitude fields
(AB Mag $<$ 26) and small separation strings, because there
are fewer background pairs in these cases. Potential strings in high limiting magnitude
fields (AB Mag $>$ 27), and those that produce high angular separations, have
higher thresholds which require more chance alignments, and so are also
efficiently excluded. It is the fields with 
intermediate limiting magnitudes (26
$<$ AB Mag $<$ 27) that produce the most automatically-assessed 
candidates, and we have relatively few candidates remaining from the 
GO survey (which has many
shallow and deep fields) and none from GOODS (which is uniformly deep).

%%%%%%%%%%%%%%%%%%%%%%%%%%%%%%%%%%%%%%%%%%
\begin{table*}
\begin{tabular}{c|ccccc} 
Survey & Number of Fields & Resolvable Strings & Passed Assessment 1  & Passed Assessment 2 & Possible Strings \\
\hline
GO-S$^a$ & 318    & 4.6 $\times$ 10$^6$  & 323  & 4  & 4 \\
GO-M$^b$ & 346    & 5.3 $\times$ 10$^6$  & 382  & 8  & 0 \\
COSMOS   & 575    & 8.7 $\times$ 10$^6$  & 631  & 31 & 0 \\
GEMS     & 63     & 1.0 $\times$ 10$^6$  & 135  & 5  & 0 \\
AEGIS    & 63     & 1.0 $\times$ 10$^6$  & 120  & 2  & 0 \\
GOODS    & 35$^c$ & 0.3 $\times$ 10$^6$  & 46   & 0  & 0 \\
\hline
Total    & 1400   & 21.9 $\times$ 10$^6$ & 1637 & 50 & 4 \\
\hline
\end{tabular}
\caption{The number of fields from each survey of HST data, the
approximate number of strings we can resolve in our Hough space, the number of
candidates which pass the assessment described in 
Section~\ref{sect:method:assess1}, the number of candidates which pass the assessment
described in Section~\ref{sect:method:assess3} and the number of candidates
that we cannot reject. Notes: $a)$~The GO-S fields are the GO fields with only
a single color filter of data. $b)$~The GO-M fields are the GO fields with
multiple color filters taken.  $c)$~The number of GOODS fields is the number of
panels produced by the GOODS team and does not represent individual HST
exposures.}\label{tab:npairs}
\end{table*}
%%%%%%%%%%%%%%%%%%%%%%%%%%%%%%%%%%%%%%%%%%

The numbers of pairs in the 50 remaining string candidates suggest that they
are false detections. While we probe separations in the range $0.3$''$< \theta < 5$'', 
we only find candidates with maximum separations of $0.6$'' or greater. If a string
were making $0.6$'' separated image pairs of high redshift sources, we would expect 
it to make roughly:
\be\label{eq:exnpairs}
<n_{\rm{pairs}}> = \frac{0.5 \times 5000\ \theta_{\rm{max}}}{210''} = 7.2
\ee
Where the factor $0.5$ is to account for the fraction of sources which are 
significantly behind the string, $210$'' and 5000 are typical ACS width and
the number of sources in a panel of our shallow surveys, respectively. The
majority of our string candidates exhibit larger separations, indicating that
we would expect even more than 7 pairs. But of our 50 string candidates, 38 
consist of 3 pairs, 11 consist of 4 pairs and 1 consists of 5 pairs. 

The above argument is not quantitative enough to explicitly figure into our analysis.
But we can use other objective methods to show that string candidates in large surveys
and fields imaged in multiple filters do not represent real cosmic strings.

% - - - - - - - - - - - - - - - - - - - - - - - - - - - - - - - - - - - - - -

\subsection{Detailed string candidate inspection}
\label{sect:results:inspection}

The 50 candidate strings described above were selected by geometry and single filter
image pair morphology alone. To test the string lens hypothesis further in each 
case
we use different methods depending on whether the
string candidate in question is in a large survey or an isolated field, and
whether it has been imaged with multiple filters or just one. After all string
candidates are examined, we can eliminate all but four candidates using
objective criteria, and we believe that even these 
final four candidates could be
eliminated with more data as we discuss in 
Section~\ref{sect:results:implications}.

\begin{figure}
\centering\includegraphics[width=\linewidth]{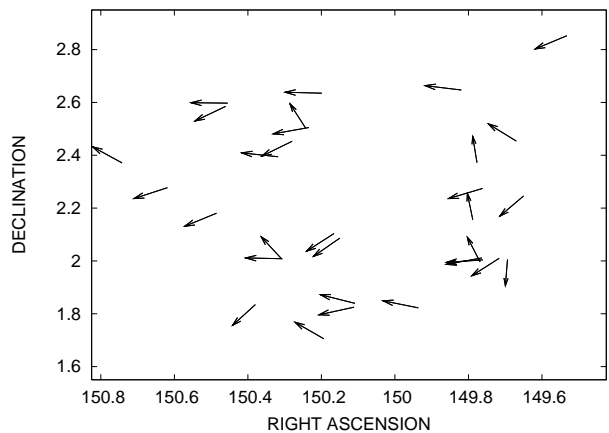}
\caption{The 31 string candidates in COSMOS with the location of the
string at the foot of the vector and the string projection along the line of
the vector. The graphing area is roughly the same as the COSMOS survey. Every
string candidate, when extrapolated, would cut
across a significant fraction of the survey and so should
thus be detected in multiple fields. There is no strong pattern of string
candidates in a line to indicate this. }
\label{fig:COSMOSstrings}
\end{figure}

\begin{figure}
\centering\includegraphics[width=\linewidth]{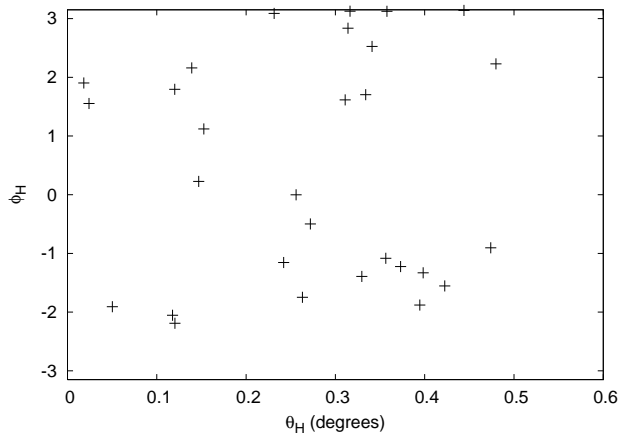}
\caption{The Hough image of potential COSMOS strings. We only find
one near overlap of string candidates. Examination of the candidates in detail
shows that they are parallel lines detected in the same field and separated by
more than an arcminute.}
\label{fig:COSMOSstringshough}
\end{figure}

In large contiguous surveys, our assumption of long, straight
strings requires that any strings detected in a single field be detected as
collinear strings in neighbouring fields. 
We illustrate this idea for COSMOS in Fig.~\ref{fig:COSMOSstrings} 
by plotting the extrapolation of every string candidate into the rest of the
survey.
While there are some coincidentally aligned strings, there is no pattern
of strongly collinear string candidates to indicate a real cosmic string. 
To make this
analysis more precise, we use the techniques in Section
\ref{sect:method:hough} to make a single survey-wide Hough image of every 
string candidate in the survey
(Fig.~\ref{fig:COSMOSstringshough}) and see if there are any significant
alignments. 
Our measurements of string candidate orientation
angle are precise to $0.01$~radians in a field,
but we want to allow for modest curvature so we look for pairs whose $\phi_H$
are within $0.1$ of each other and whose $\theta_H$ are within $0.14^\circ$
(allowing the uncertainty in $\phi_H$ to be projected across the $1.4^\circ$
survey).
Only one pair of string candidates survives this cut. This pair is
composed of two candidates in acs\_100036+0205\_unrot, 
which are actually parallel but
separated by more than an arcminute, rather than being collinear. 
Analogous
Hough space images for GEMS and AEGIS also show no alignment of string
candidates.

%%%%%%%%%%%%%%%%%%%%%%%%%%
%%%%%%%%%%%%%%%%%%%%%%%%%%

%%%%%%%%%%%%%%%%%%%%%%%%%%
%%%%%%%%%%%%%%%%%%%%%%%%%%

This argument breaks down for any string candidates that cut across the corner of 
a large survey, or any that cut across a fairly
thin survey. Fortunately, none of our string candidates cut across single
corner fields. The survey string candidate least likely to be observed in
neighboring fields is the second string detected in the EGS-16-02. The AEGIS
survey is 3 fields wide and 21 fields long, and the projection of this string
cuts across only three fields. This string candidate consists of three
$1.0$'' pairs
and the typical AEGIS field contains 5000 sources, so, analogous to Eq. 
\ref{eq:exnpairs}  it should produce an
average of 11.9 events per field as above. The chance of such a string
producing only three events in one field and less than three events in two
neighboring fields is essentially zero.

\begin{figure}
\begin{minipage}{\linewidth}
\centering\includegraphics[width=0.6\linewidth]{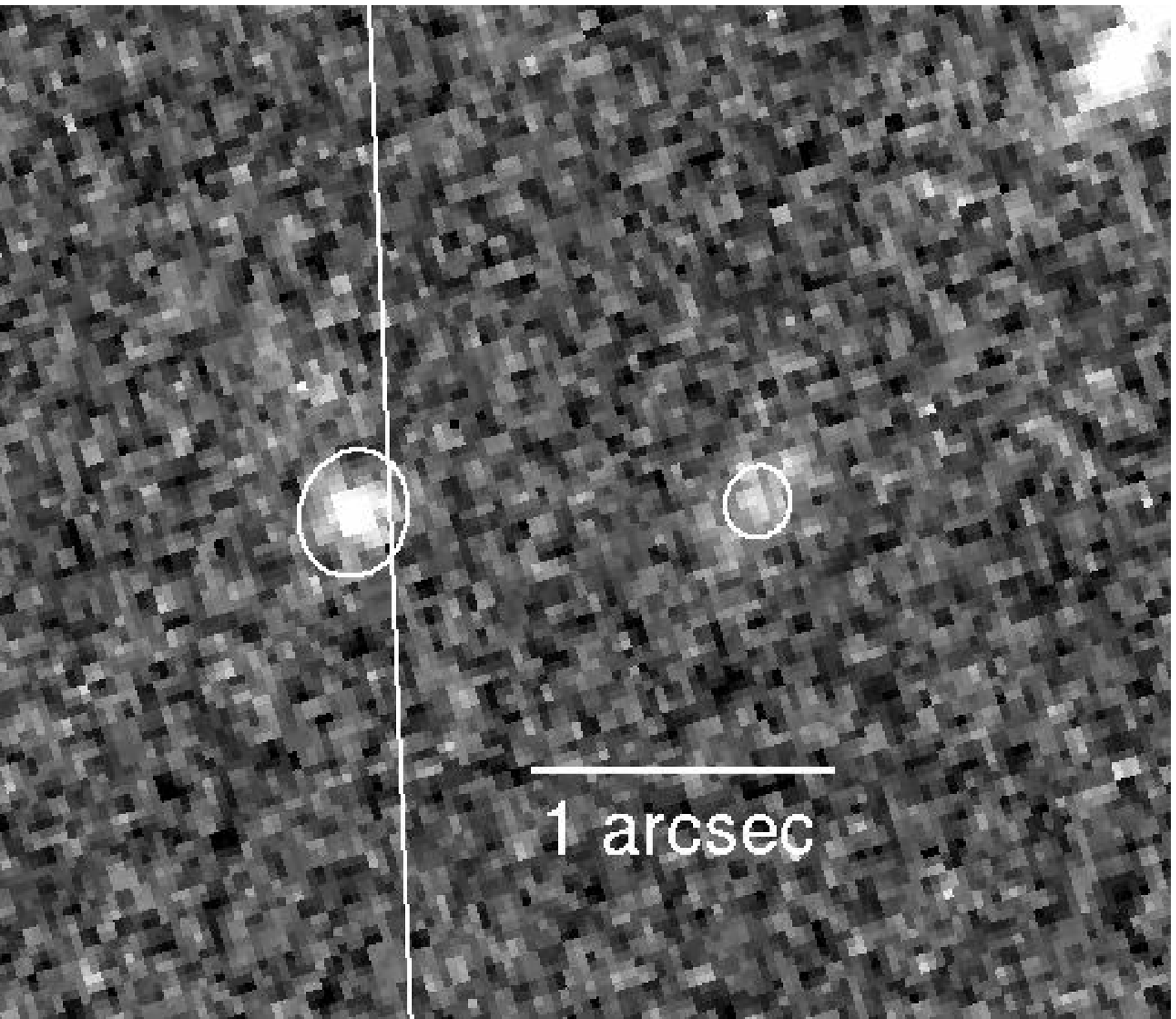}\medskip

\centering\includegraphics[width=0.6\linewidth]{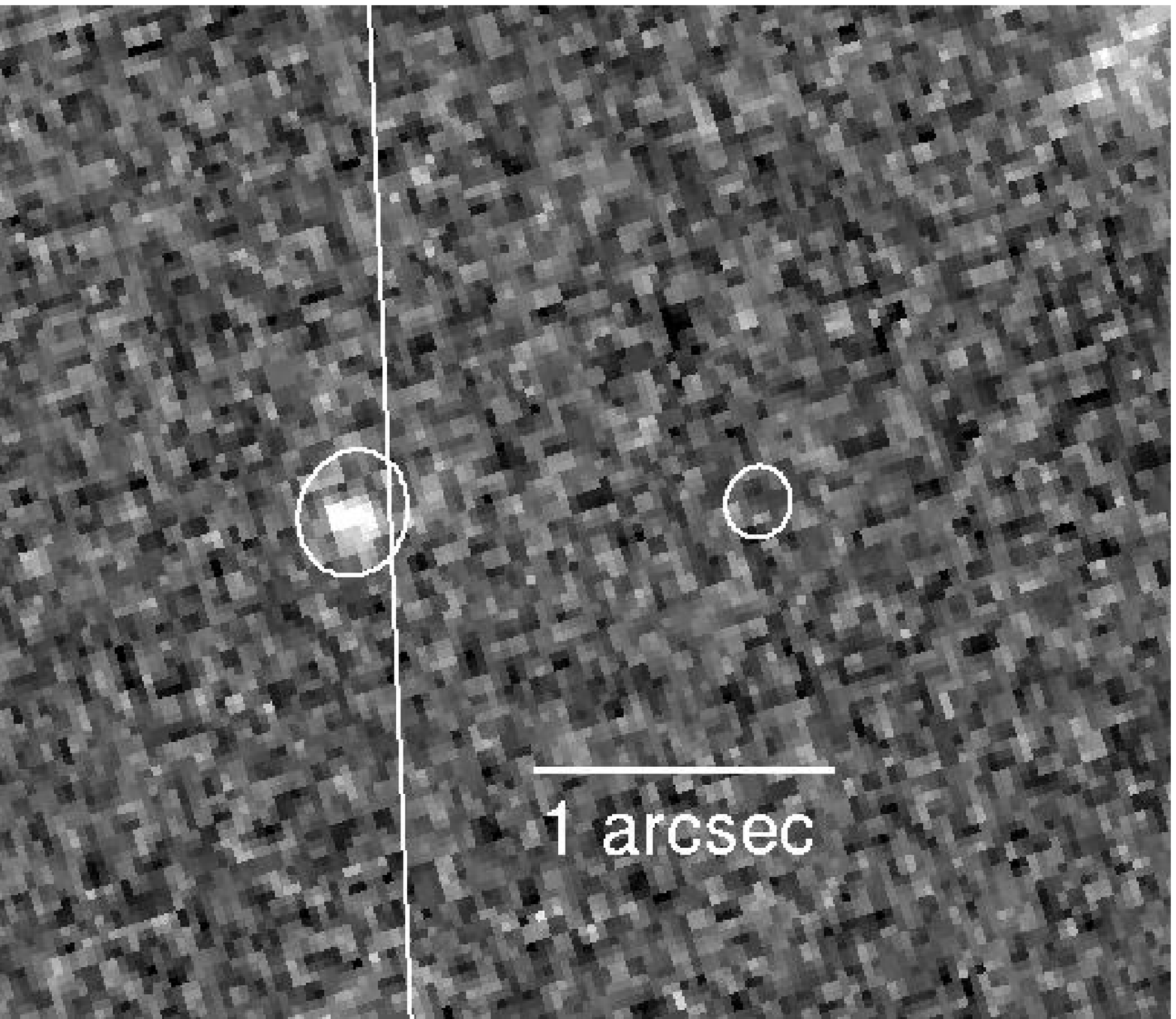}
\end{minipage}
\caption{Example multi-band dropout in the GO field 
ACSJ065819-555630. The F814W-band detection image shows a pair of
two sources (top), but in
the F435W filter (bottom) only one source is apparent.}
\label{fig:colorpair}
\end{figure}

Eliminating string candidates across the GO fields requires a different
approach, because the GO fields are not contiguous. For the GO-M fields, we can 
eliminate candidates by finding pairs for
which one member of the pair has significantly different colour than
the other. String lensing is ideally an achromatic effect, but it is
possible that a galaxy could be incompletely copied so that, perhaps, the blue
star-forming core only appeared in one image, and the two images could have
very different colors. For this reason, we do not automate the process and
search by eye for sources with different colour that did not seem to be cut in
half. We eliminated enough pairs from each potential string to bring the pair
total below three. We show an example of a colour-rejected
pair in Fig.~\ref{fig:colorpair}. This
multi-band analysis could reasonably be automated, and points to the usefulness
of multi-band analysis in future string searches. 

%Ideally, this analysis would
%be included earlier in the string-searching process, but since most of our
%data was single filter, we prioritised methods that were discriminating in the
%single filter case. If an interesting
%candidate survives the monochromatic rejection, imaging at a different
%wavelength is an excellent follow-up tool.

%%%%%%%%%%%%%%%%%%%%%%%%%%
%%%%%%%%%%%%%%%%%%%%%%%%%%

% - - - - - - - - - - - - - - - - - - - - - - - - - - - - - - - - - - - - - - 

\subsection{ The four remaining string candidates}
\label{sect:results:implications}

\begin{figure*}
\begin{flushleft}
\begin{minipage}[t]{0.22\linewidth}
\centering\includegraphics[width=\linewidth]{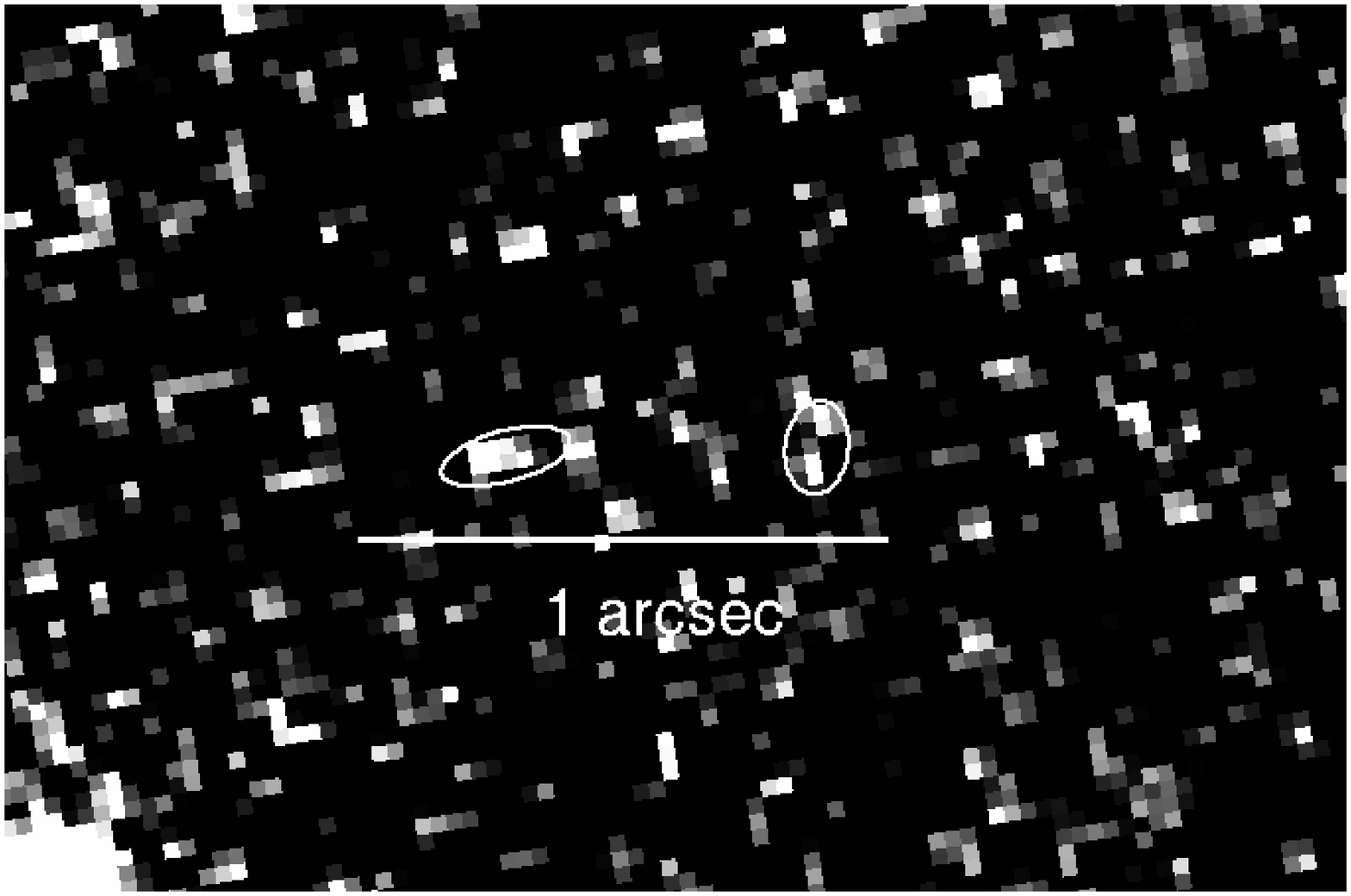}
\end{minipage}
\medskip
\begin{minipage}[t]{0.22\linewidth}
\centering\includegraphics[width=\linewidth]{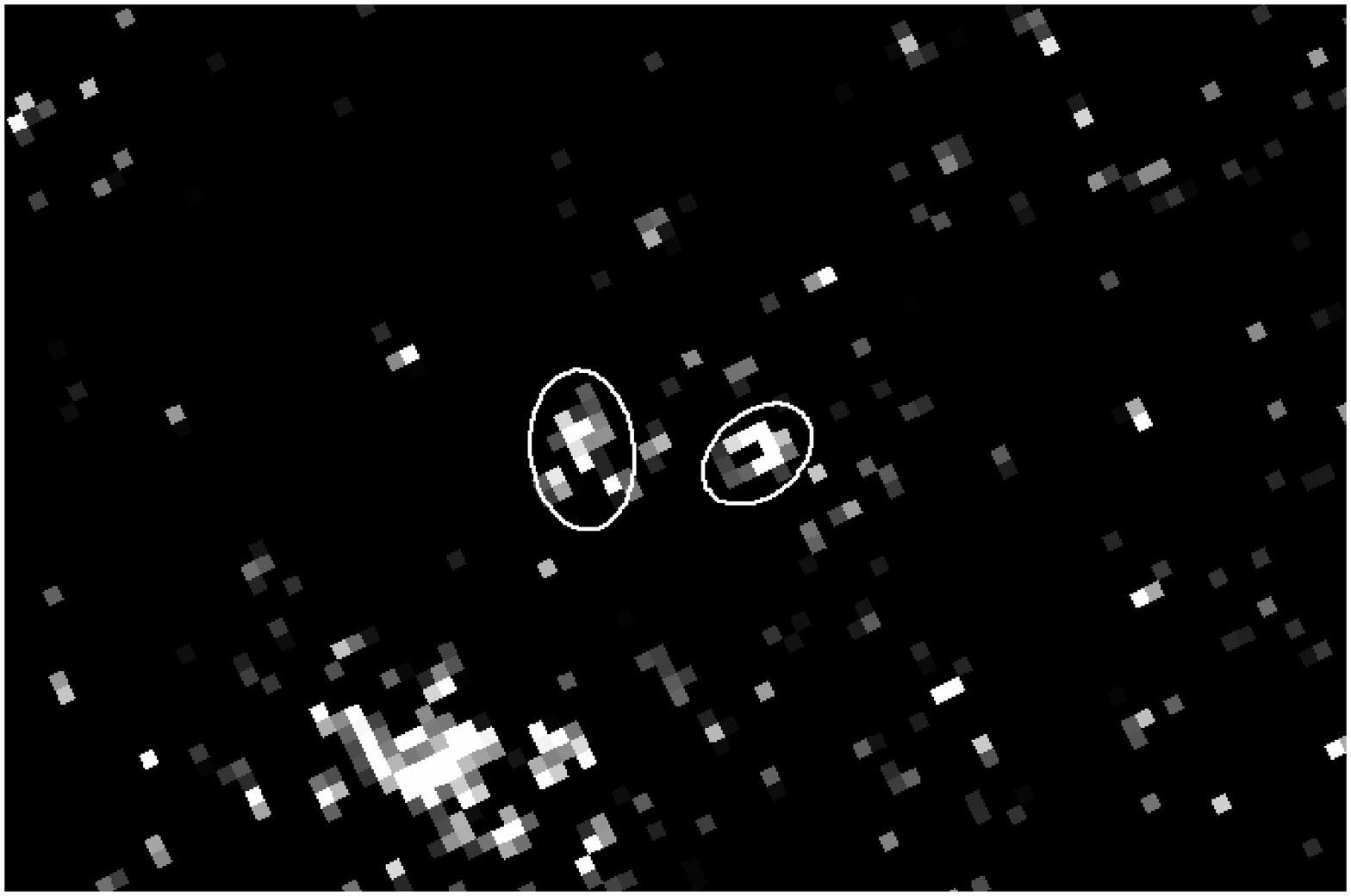}\medskip
\end{minipage}
\medskip
\begin{minipage}[t]{0.22\linewidth}
\centering\includegraphics[width=\linewidth]{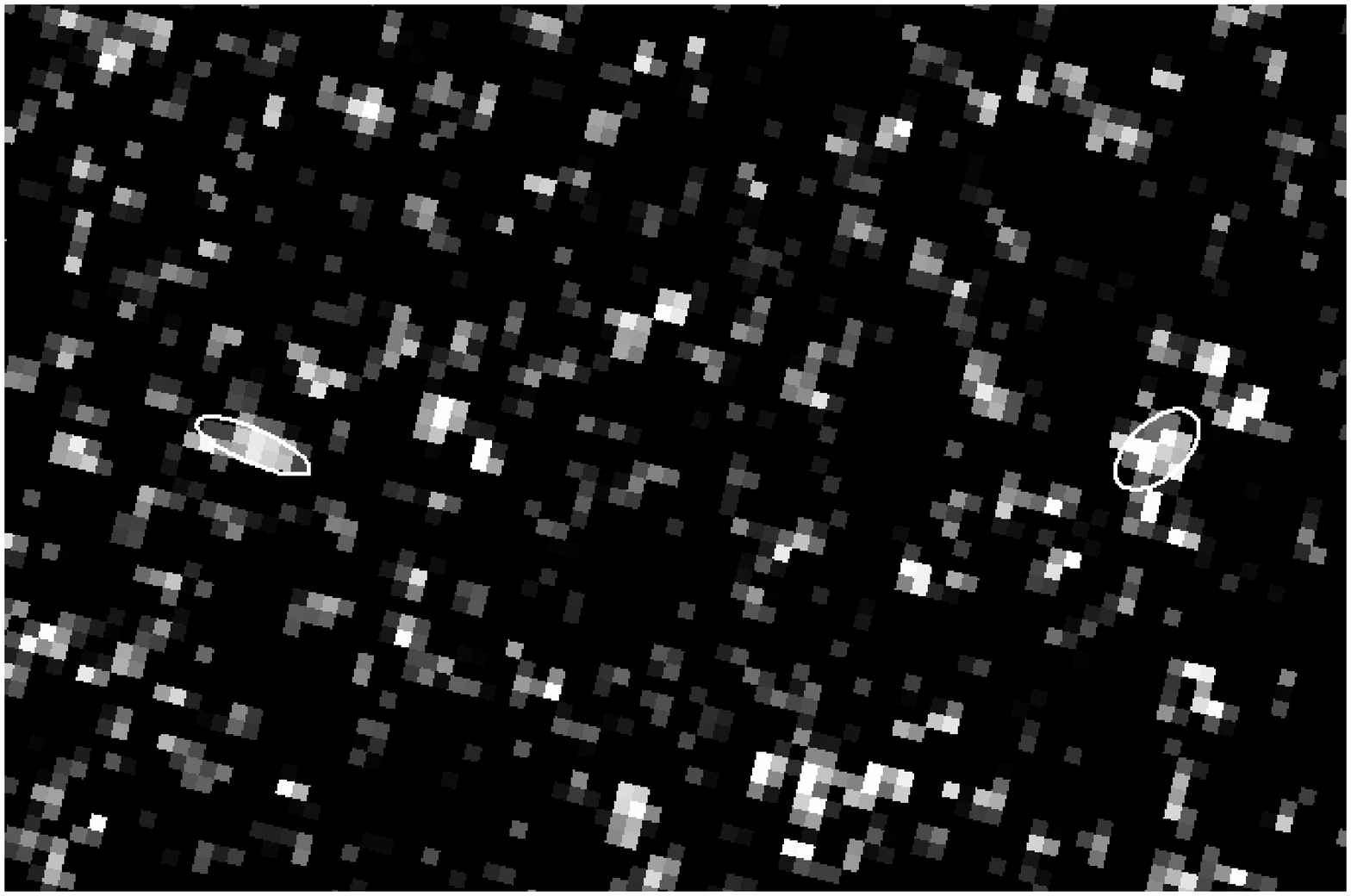}\medskip
\end{minipage}
\medskip
\begin{minipage}[t]{0.295\linewidth}
\centering\includegraphics[width=\linewidth]{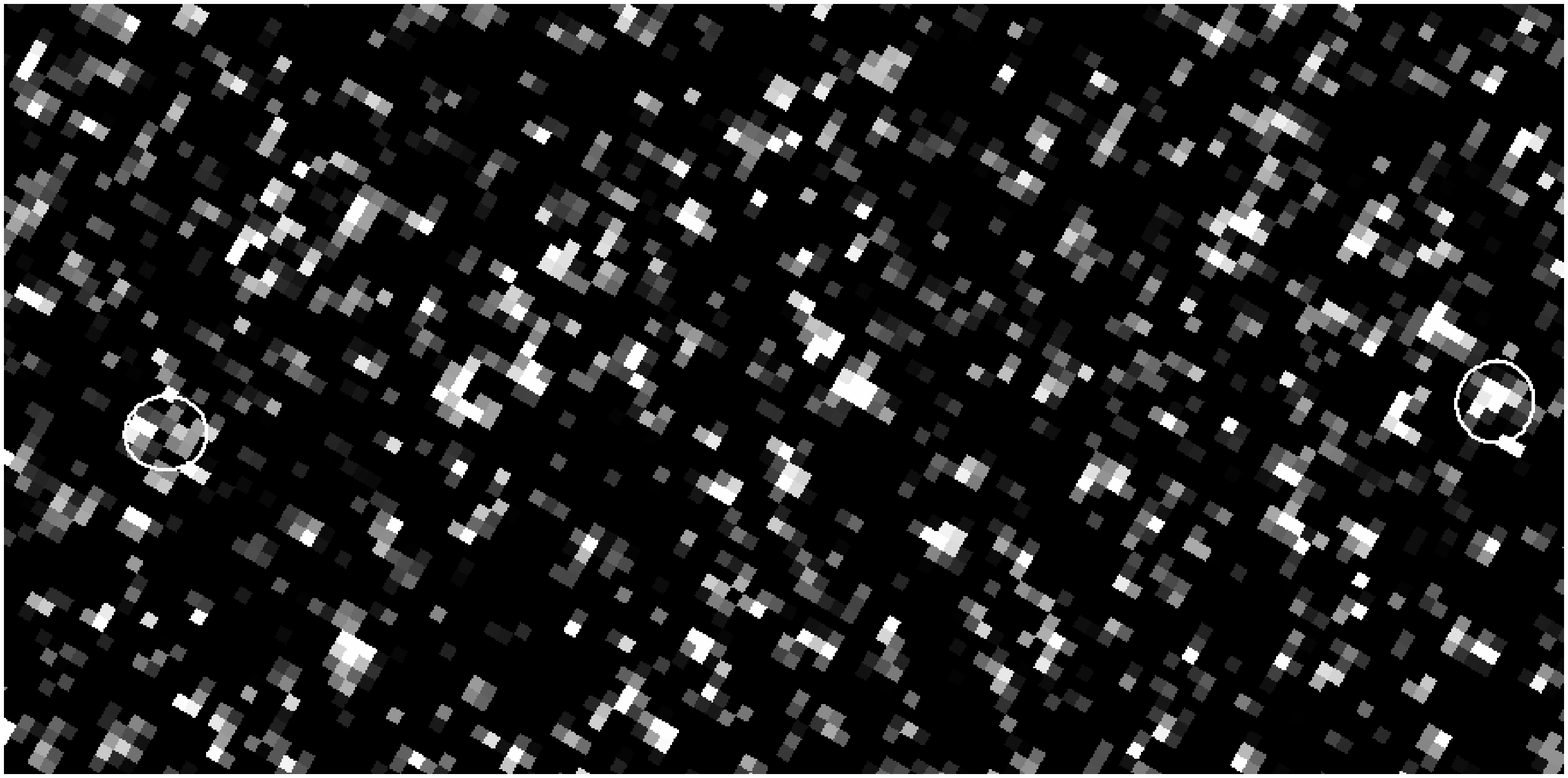}\medskip
\end{minipage}
\medskip
\begin{minipage}[t]{0.22\linewidth}
\centering\includegraphics[width=\linewidth]{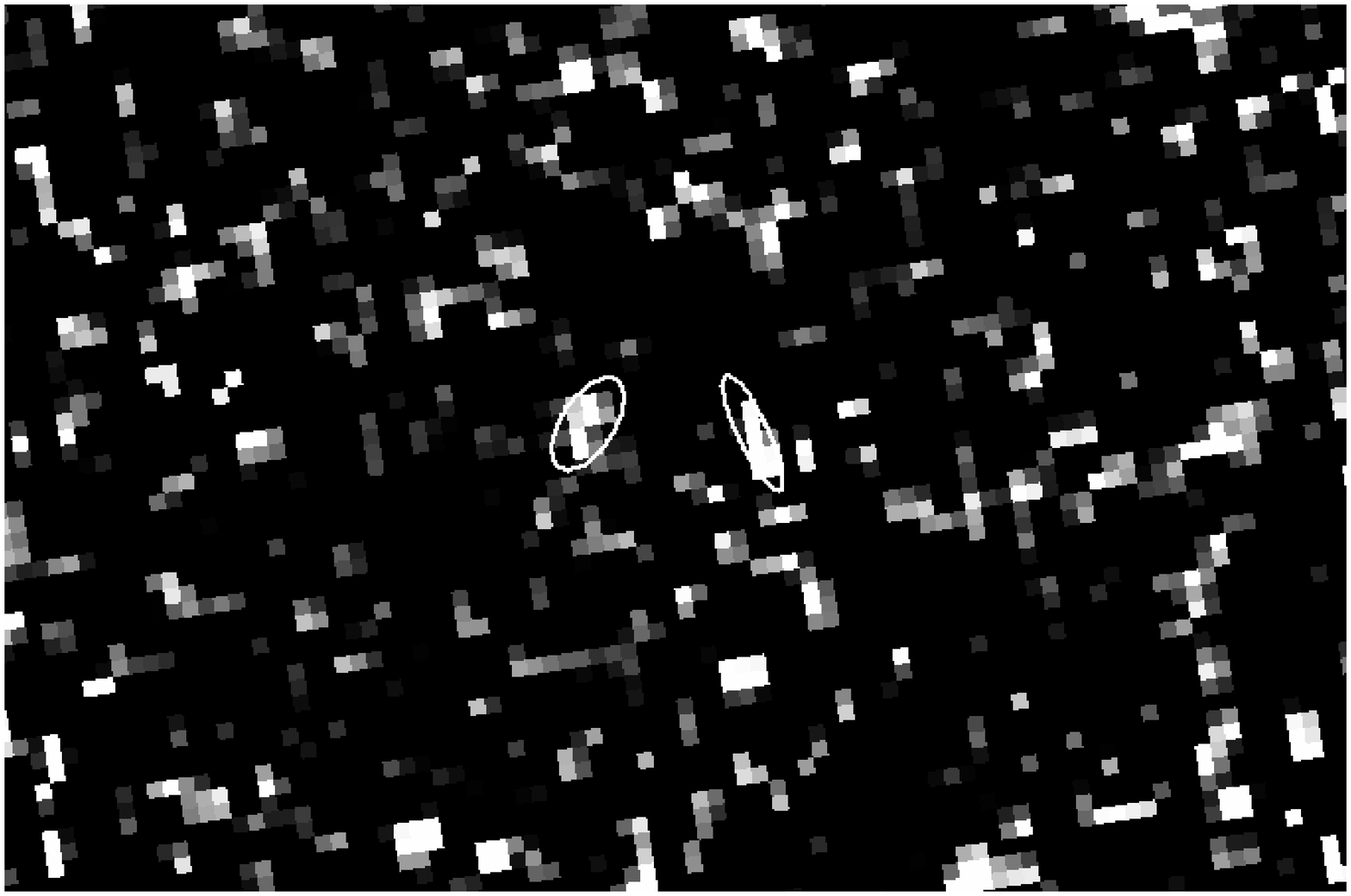}\medskip
\end{minipage}
\medskip
\begin{minipage}[t]{0.22\linewidth}
\centering\includegraphics[width=\linewidth]{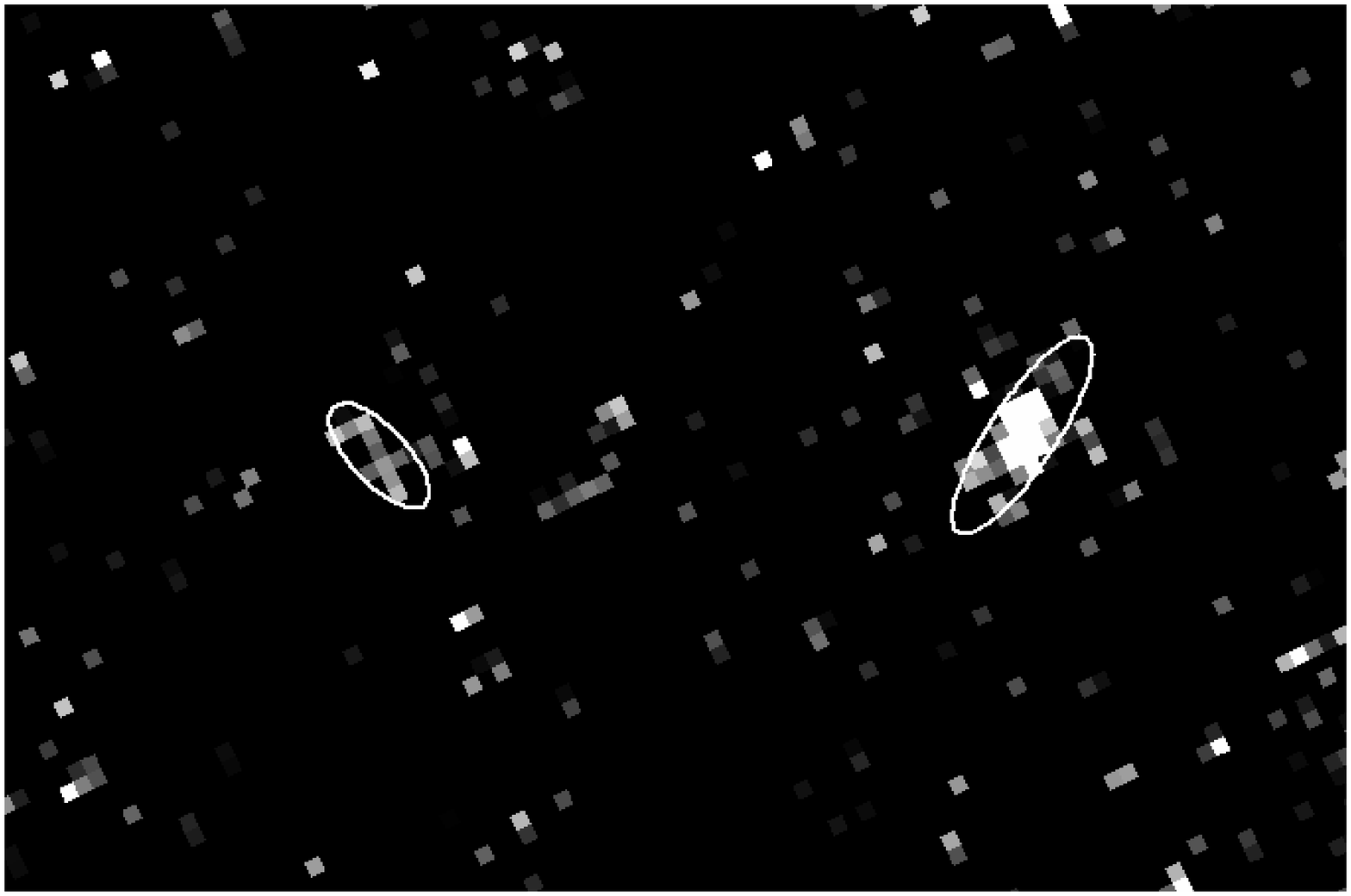}\medskip
\end{minipage}
\medskip
\begin{minipage}[t]{0.22\linewidth}
\centering\includegraphics[width=\linewidth]{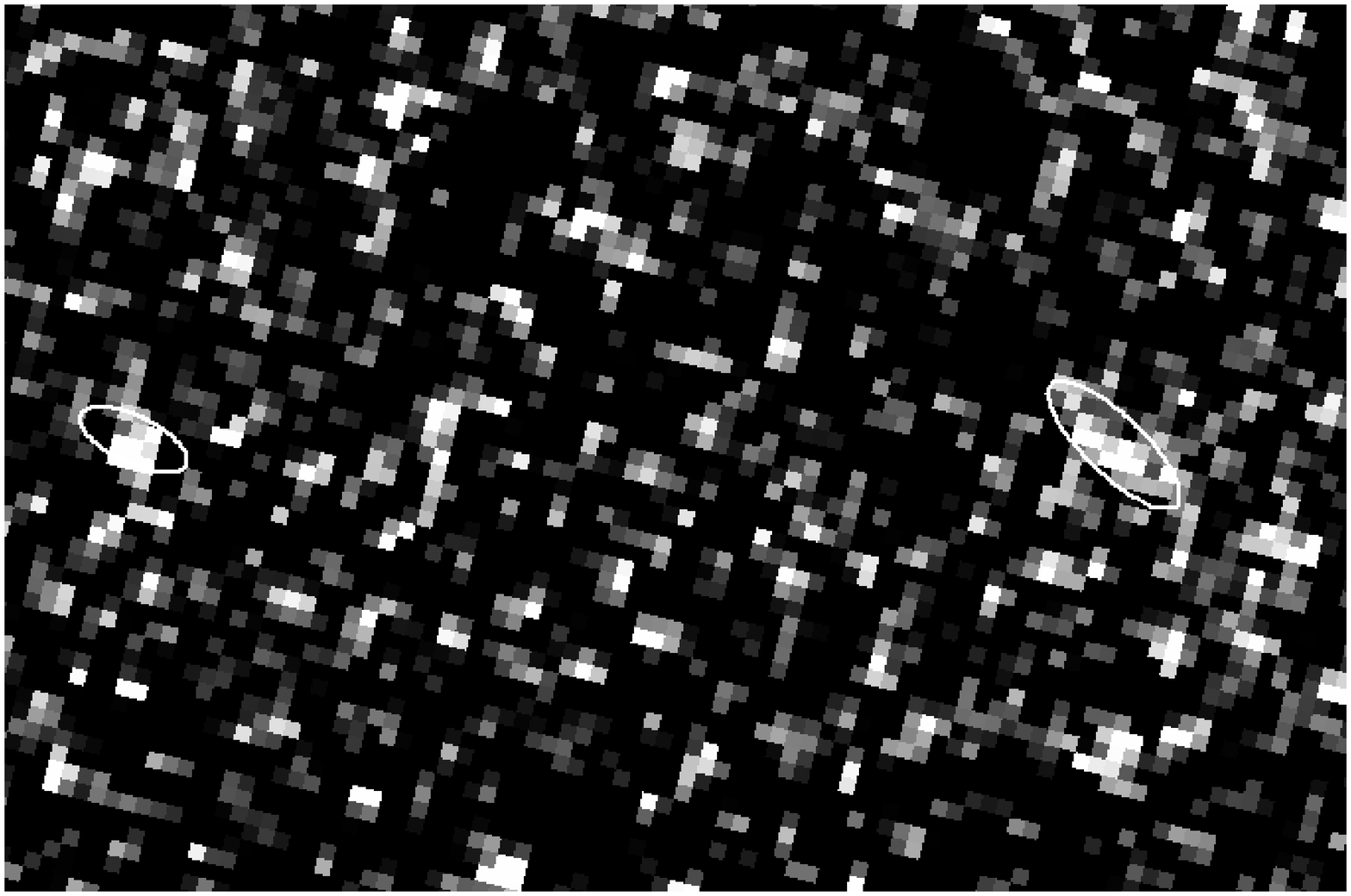}\medskip
\end{minipage}
\medskip
\begin{minipage}[t]{0.295\linewidth}
\centering\includegraphics[width=\linewidth]{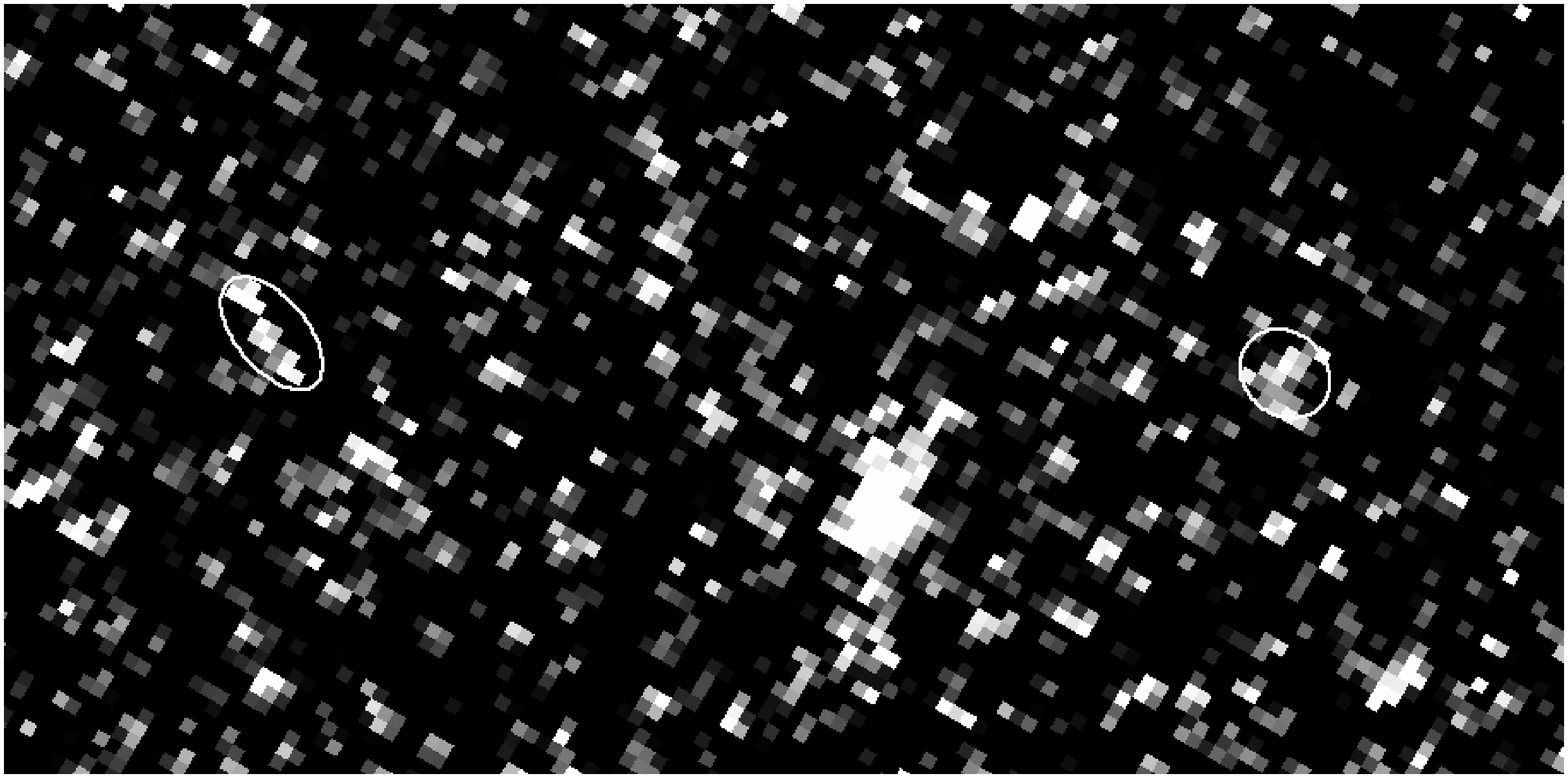}\medskip
\end{minipage}
\medskip
\begin{minipage}[t]{0.22\linewidth}
\centering\includegraphics[width=\linewidth]{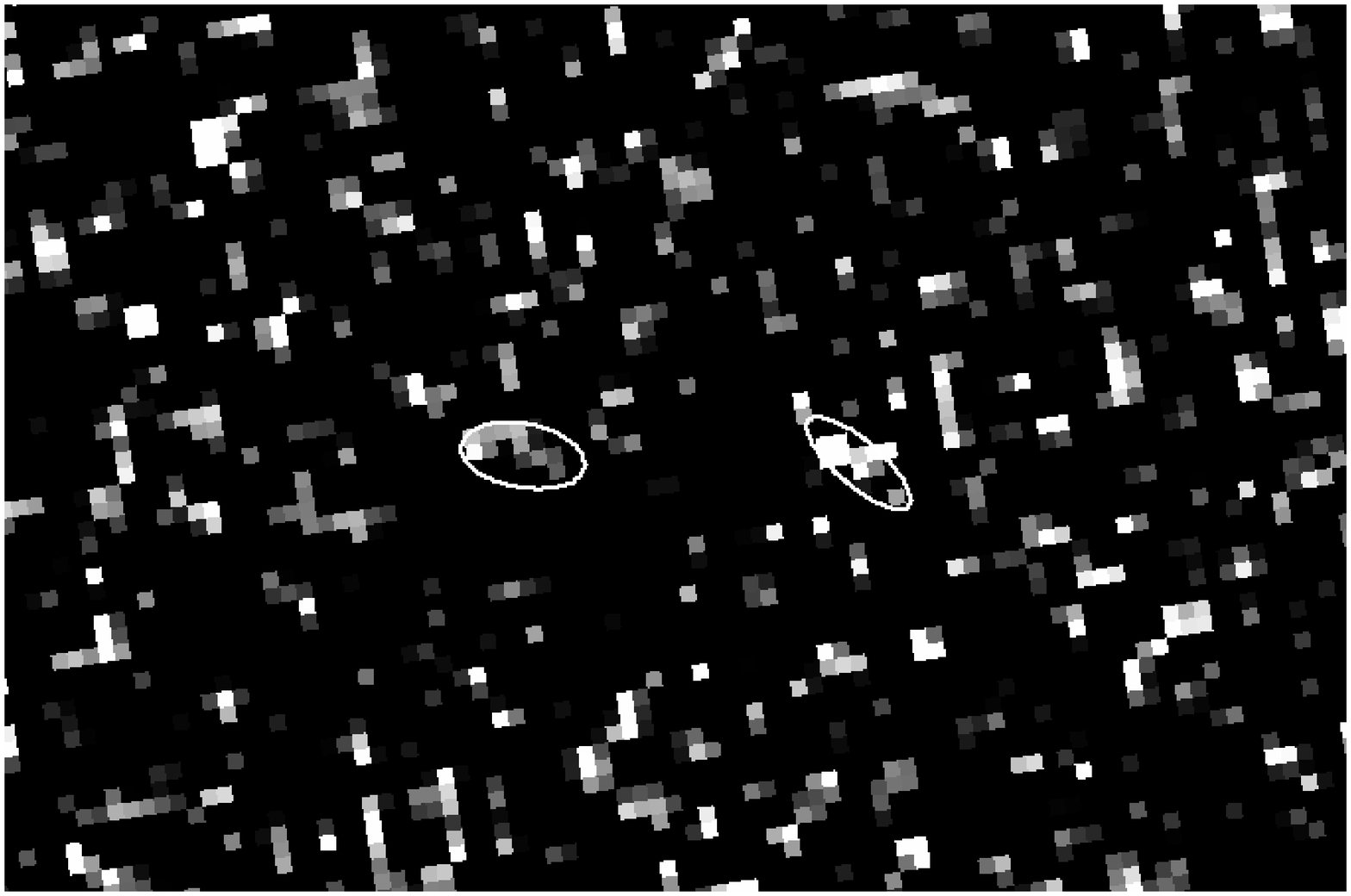}\medskip
\end{minipage}
\medskip
\begin{minipage}[t]{0.22\linewidth}
\centering\includegraphics[width=\linewidth]{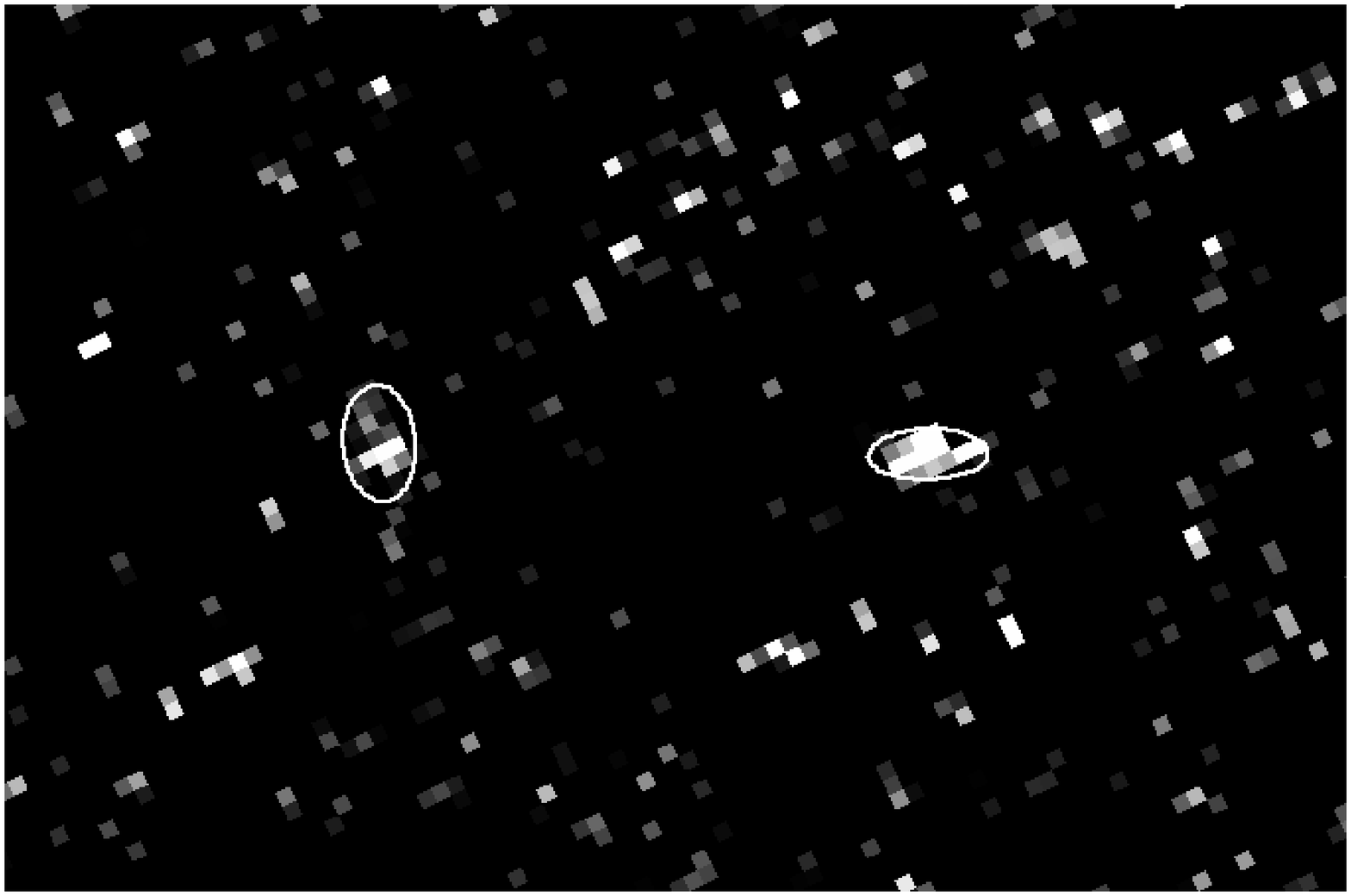}\medskip
\end{minipage}
\medskip
\begin{minipage}[t]{0.22\linewidth}
\centering\includegraphics[width=\linewidth]{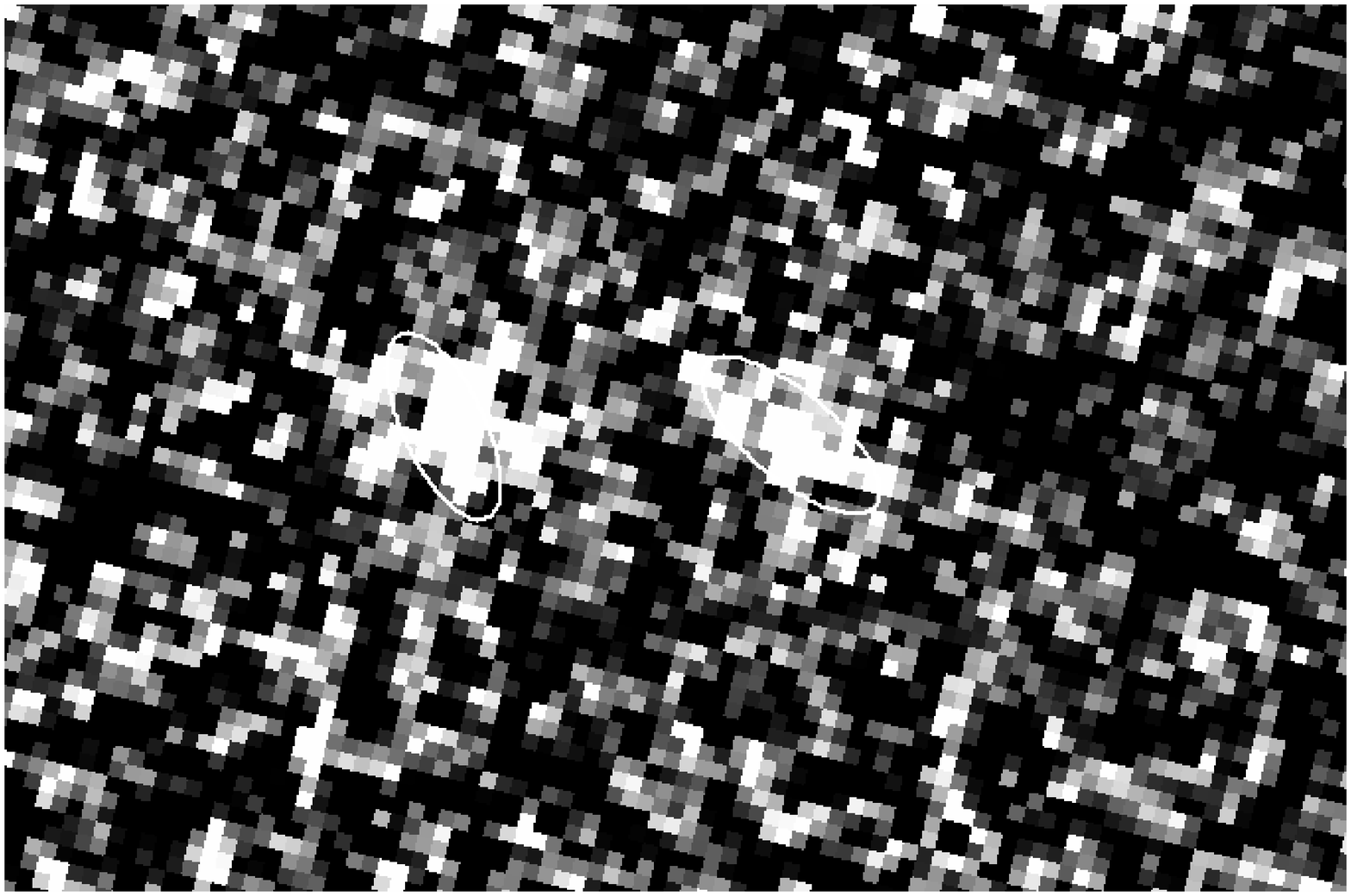}\medskip
\end{minipage}
\medskip
\begin{minipage}[t]{0.295\linewidth}
\centering\includegraphics[width=\linewidth]{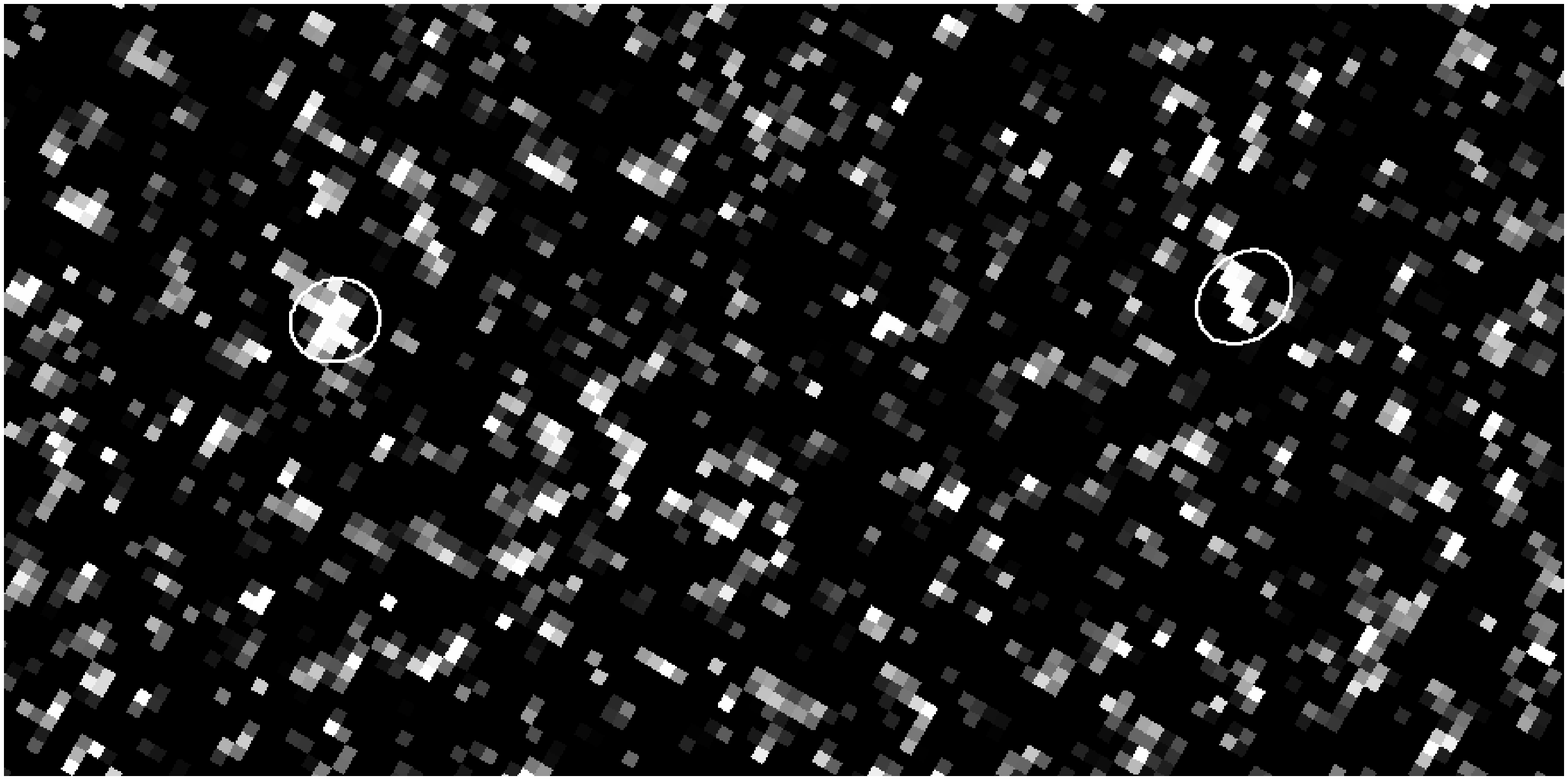}\medskip
\end{minipage}
\medskip
\begin{minipage}[t]{0.22\linewidth}
\includegraphics[width=\linewidth]{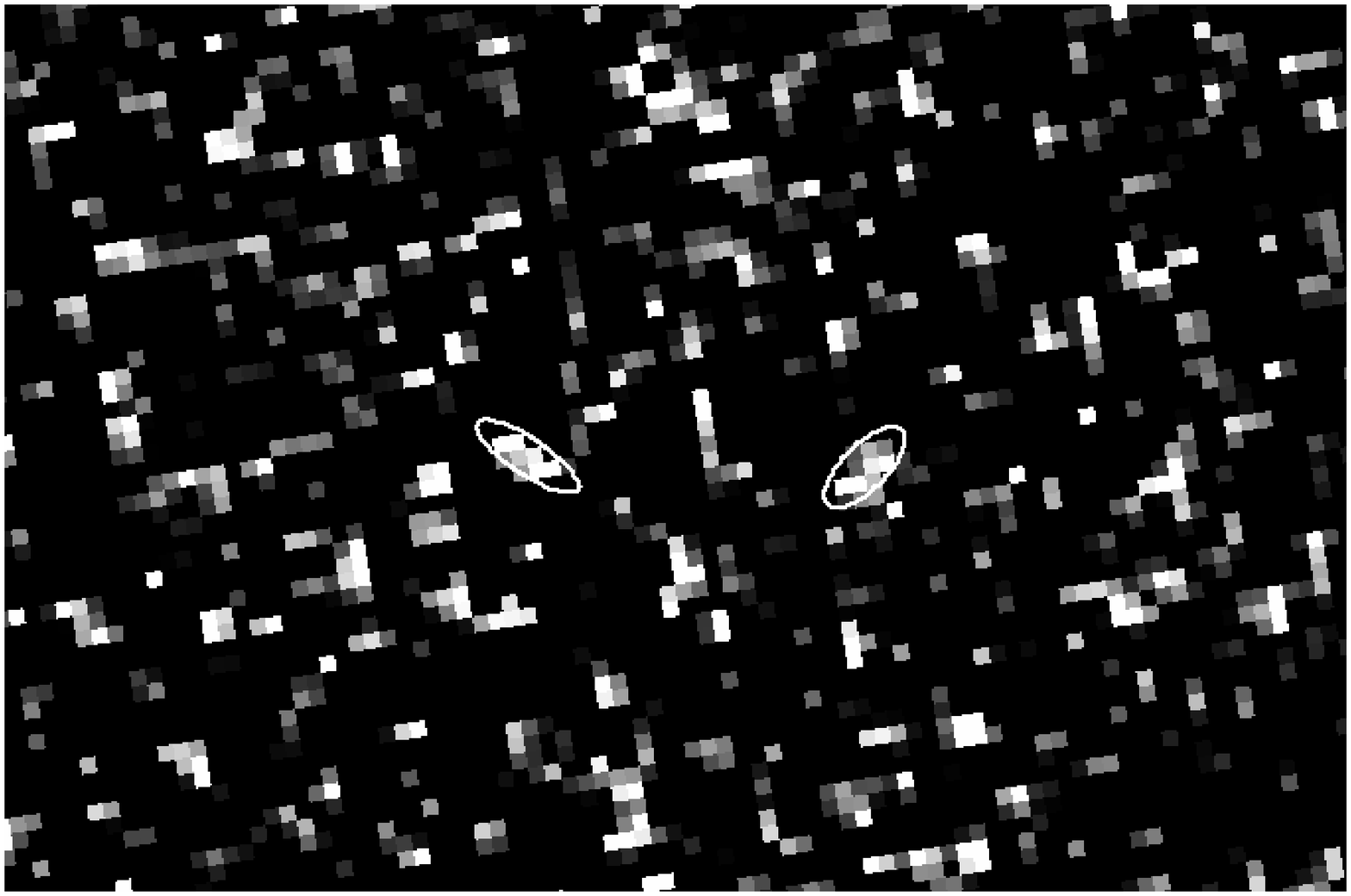}
\end{minipage}
\end{flushleft}
\caption{Example string-lens image pairs from field ACSJ195021-405350 (left), ACSJ181424+411240 
(left-centre), ACSJ042115+193600 (right-centre) and ACSJ144146-095150 (right). Each image is 
rotated such that the most likely string is vertical. The ellipses show the ellipticity and 
orientation of each object, as measured with \sex. The quality of these images was, on average, 
fairly poor, and they had to be scaled differently from other data images so that sources we visible above
the noise. This accounts for the different appearance.}
\label{fig:realpairs}
\end{figure*}

We are left with four potential strings from the single band GO fields (see
Table~\ref{tab:candidates} for more information). We emphasise that there is nothing
particularly suggestive about these remaining candidates, but only a lack of
data that prevents us from eliminating them. We display the constituent pairs of
each string candidate in Fig \ref{fig:realpairs}. These string candidates come from  
some of our noisiest data, and the images we display here are scaled differently from other 
data. Even qualitative analysis of these images suggests that these pairs are not exact copies.  
We show in this section that statistical analysis strongly suggests that these string candidates 
are indeed false detections. But we also show the limits on $\G$ that each of these string candidates 
would imply if it were the result of a real cosmic string.

\begin{table*}
\begin{center}
\begin{tabular}{cccccccccc}
Field & $\theta_{\rm{max}}$ & Right Ascension & Declination & Slope & $N_{\rm{pairs}}$ & Threshold & Score & $\log_{10}(\G)$ & $\log_{10}(\Os)$  \\
\hline
ACSJ195021-405350        & 1.0 & 297.59409 & -40.88897 & 0.1717   & 4 & 9.43  & 10.25 & $-6.19 \pm 0.27$   & $-5.11 \pm 0.29$ \\
ACSJ181424+411240        & 1.4 & 273.60678 & 41.20152  & 0.6494   & 3 & 6.58  & 7.88  & $-6.00 \pm 0.26$   & $-4.96 \pm 0.28$ \\
ACSJ042115+193600        & 2.0 &  65.31551 & 19.59700  & 0.1949   & 3 & 7.09  & 9.11  & $-5.81 \pm 0.21$   & $-4.81 \pm 0.23$ \\
ACSJ144146-095150        & 3.0 & 220.44917 & -9.86378  & 0.8195   & 3 & 5.55  & 5.70  & $-5.628 \pm 0.051$ & $-4.65 \pm 0.10$ \\
\hline
\end{tabular}
\end{center}
\caption{The four candidates which passed two rounds of automated analysis, were not part of large surveys and lacked multifilter data. These candidates are all in the single filter GO survey (GO-S).}\label{tab:candidates}
\end{table*}

The rejection of the 46 potential string candidates which met the same criteria as
the remaining four allows us to estimate the probability that these remaining
candidates are false detections. If we start with the prior assumption that
the fraction of string candidates which are false detections, $f$, is a random
variable chosen uniformly between 0 and 1 (a generous assumption) and label
the condition of having 46 false detections ``$46\ \rm{false}$'', then Bayes'
Theorem gives us a distribution on f:
\bea
P(f | 46\ \rm{false}) &=& \frac{P(f) P(46\ \rm{false} | f) }{P( 46\ \rm{false})}\\
&=& \frac{1 \times f^{46}}{\int_0^1f'^{46} df'}\nonumber\\
&=& 47 f^{46}\nonumber
\eea
And in turn, the probability of the remaining three detections being false is:
\be
P(4\ \rm{false}) = \int_0^1 47 f'^{46} f'^4 df' = \frac {47}{51} = 0.92
\ee

This 0.92 is a minimum probability that makes use of a generous prior and does
not take note of the fact that the scores and numbers of pairs found in the four
remaining string candidates are atypically small for cosmic string detections. 
We thus
believe that these detections are spurious, although we cannot prove it with
current data.

\begin{figure*}
\begin{minipage}[t]{0.49\linewidth}
\centering\includegraphics[width=\linewidth]{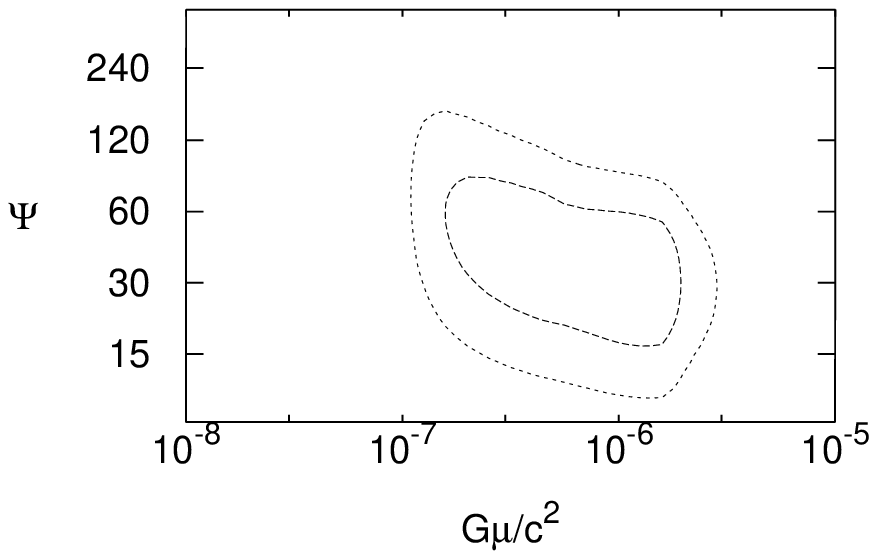}
\end{minipage}
\begin{minipage}[t]{0.49\linewidth}
\centering\includegraphics[width=\linewidth]{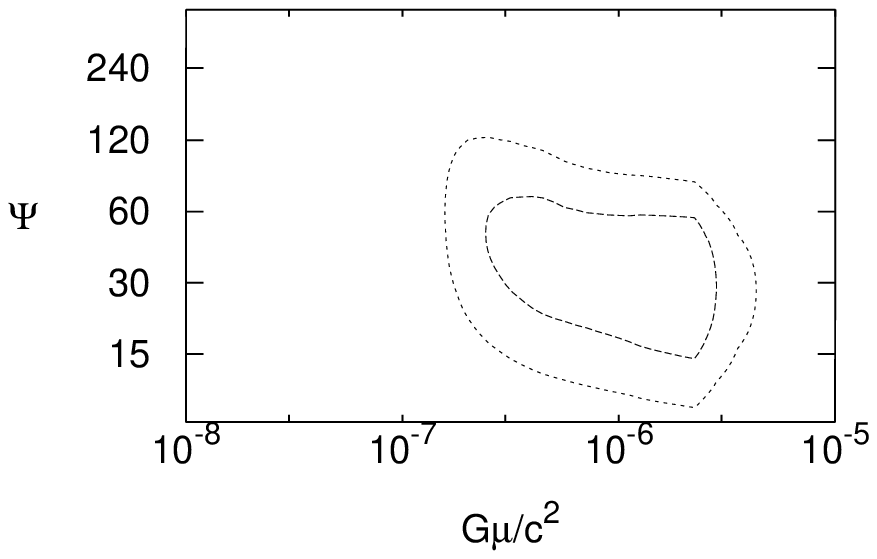}
\end{minipage}
\begin{minipage}[t]{0.49\linewidth}
\centering\includegraphics[width=\linewidth]{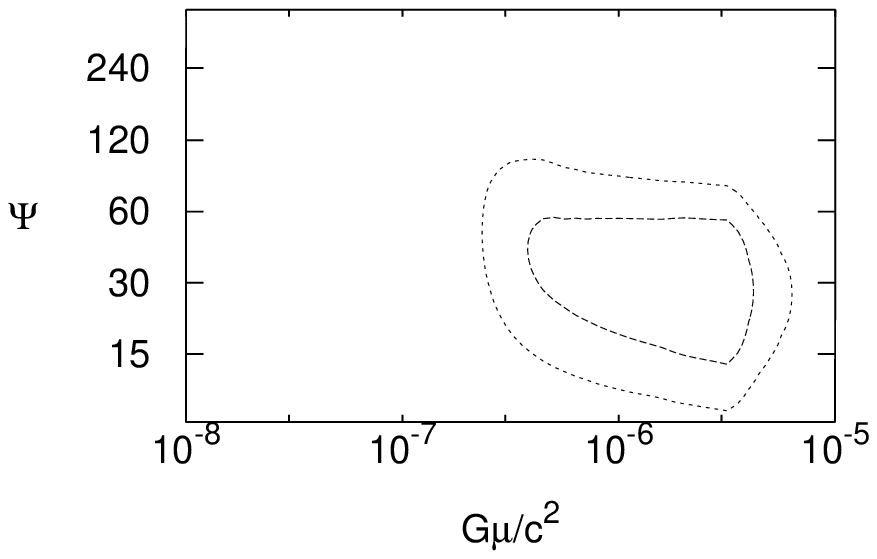}
\end{minipage}
\begin{minipage}[t]{0.49\linewidth}
\centering\includegraphics[width=\linewidth]{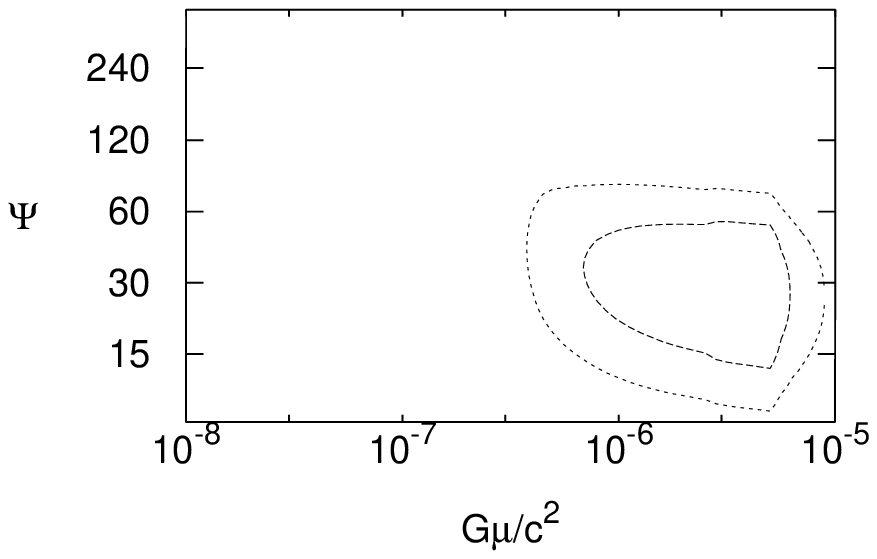}
\end{minipage}
\caption{The $\G$-$\Psi$ limits if we assume that our string
detections in ACSJ195021-405350 (top left), ACSJ181424+411240 (top right). 
ACSJ042115+193600 (bottom left) or ACSJ144146-095150 (bottom right) is genuine. 
The inner contour is the 68\% confidence limits and the outer contour is the 95\% 
confidence limit. Our priors are described in Section~\ref{sect:theory:survey}.
}
\label{fig:detectlimits}
\end{figure*}

If we accept these detections as genuine, we can calculate the implied 
tension and density of cosmic strings as shown in Table~\ref{tab:candidates}. We show the 
limits (as calculated in Section~\ref{sect:theory:survey} for potential strings in Fig.~\ref{fig:detectlimits}.

From Table~\ref{tab:candidates}, these four strings would imply a most likely $\G$ of $6.5 \times 10^{-7}$,
$1.0 \times 10^{-6}$, $1.5 \times 10^{-6}$ and $2.4 \times 10^{-6}$ respectively and a most likely
$\Os$ of $7.7 \times 10^{-6}$, $1.1 \times 10^{-5}$, $1.5 \times 10^{-5}$ and $2.2 \times 10^{-5}$, 
respectively. Any of these strings would be consistent with limits from other direct
detection methods, but inconsistent with the stricter indirect CMB detection
limits (see Section~\ref{sect:limits}).

%%%%%%%%%%%%%%%%%%%%%%%%%%
%%%%%%%%%%%%%%%%%%%%%%%%%%

Following from the results of the previous section, it would be useful to
image these four string candidates with comparably high resolution
imaging in a second filter. In the absence of such data, we will present our 
limits on string concentration and tension in the next section with the multi-band GO (for
which we have no string candidates) and total GO as separate limits.

%-------------------------------------------------------------------------------

\section{A new direct detection upper limit on the string tension}
\label{sect:obslimits}

In this section we use the results from our direct detection programme to set
direct detection limits on the cosmic string tension, $\G$, and string 
density, $\Os$, which vastly improve upon previous optical direct  detection
limits and are competitive with indirect detection methods such as  CMB power
spectrum analysis. This improvement is largely due to our using the  many
scattered GO fields rather than just one or two contiguous surveys.  Our
limits become much tighter if we assume that the four detections in the GO 
single band images are false detections.

\begin{figure*}
\begin{minipage}{\linewidth}
\centering\includegraphics[width=0.49\linewidth]{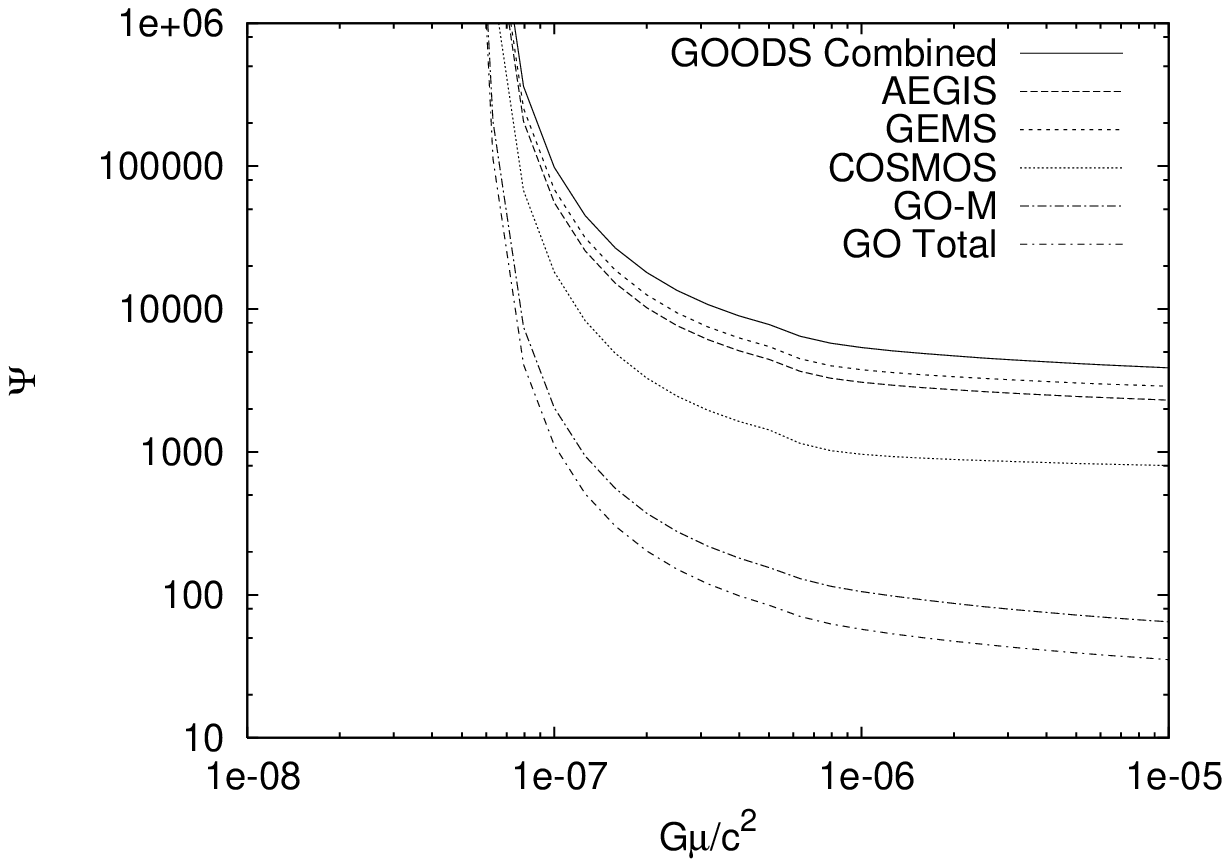}\medskip
\centering\includegraphics[width=0.49\linewidth]{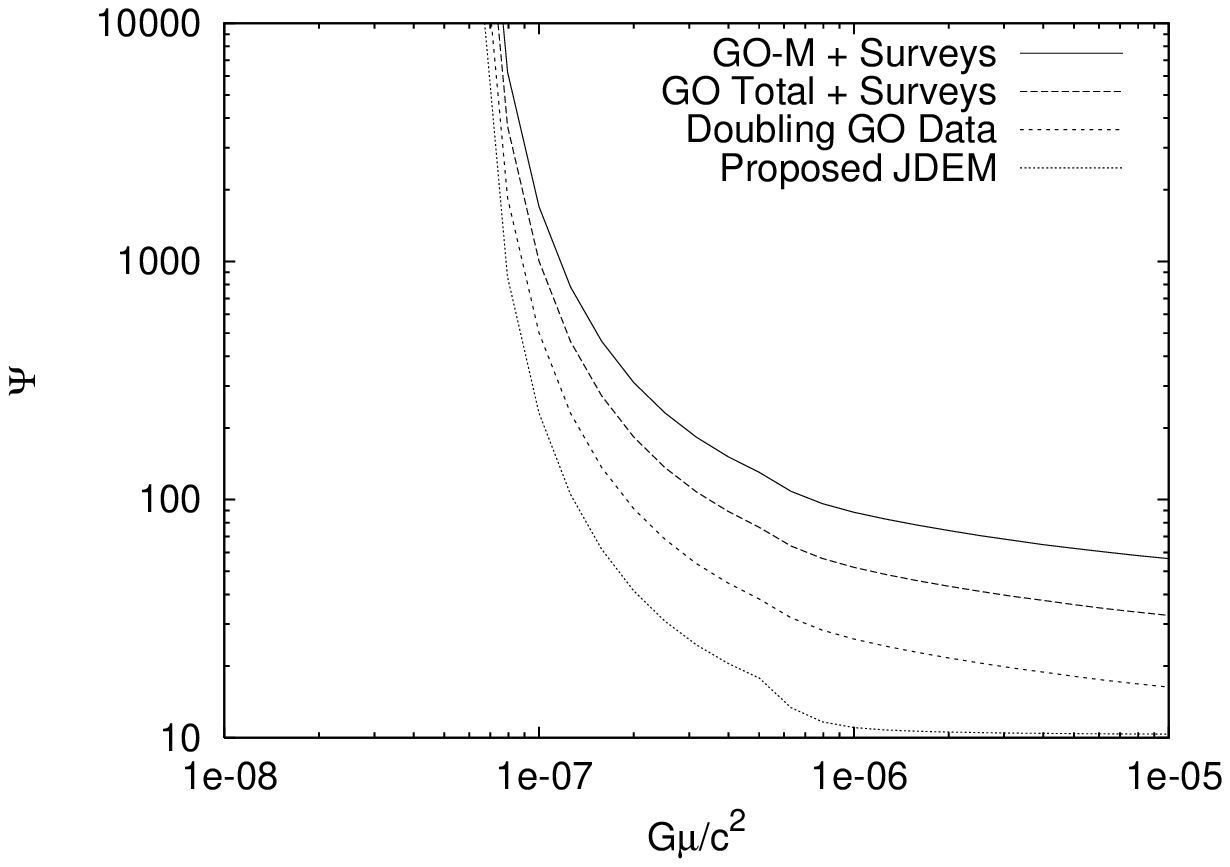}\medskip
\end{minipage}
\caption{The $\G$-$\Psi$ 95% detection contours in the
for GOODS, AEGIS, GEMS, COSMOS, GO multifilter and GO total surveys (left).
We show these contours for all surveys+multiband GO, all surveys+GO, doubling 
the data from all surveys+GO and a proposed JDEM weak lensing survey (right).}
\label{fig:surveylimits}
\end{figure*}

In Figure~\ref{fig:surveylimits}, we show the contours in $\G-\Psi$ space
for which we would detect a string with 95\% probability, derived from each of a range
of HST/ACS surveys of increasing scale. We use the formulae from
Section~\ref{sect:limits}. We also show the contours for our combined dataset,
the projected contour if we doubled its size, and the contour that would be
obtained by a putative Joint Dark Energy Mission (JDEM) weak lensing
survey covering 10000 square to an AB limiting magnitude of 26 \citep{SNAP}. These 
contours are not confidence intervals and make no prior assumption on the distribution of
$\G$ or $\Psi$.

%%%%%%%%%%%%%%%%%%%%%%%%%%%%%%%%
%%%%%%%%%%%%%%%%%%%%%%%%%%%%%%%%

Figure~\ref{fig:surveylimits} illustrates the sharp dependence of our
ability to probe $\G$ on the number of fields being searched through.
This dependence is due to the requirement that a moderate redshift string 
cross through at least one field. At the theoretically-predicted 
values of $\Psi = 60$, strings at observable redshifts would
cross fewer than 1\% of the ACS fields. 
Very high $\G$ strings would be observed nearly any time
they cross a field, but moderate redshift strings with smaller $\G$ 
become undetectable if their projection along the line of sight, $\sin i$,
is small. When we add the 318 single filter GO fields to our analysis, we are
exponentially decreasing the chance that every cosmic string ``missed'' every
one of our fields. This leads to a marked decrease in our limits on $\G$.

We produce proper confidence limits using the methods in 
Section~\ref{sect:theory:survey}. As a reminder, the prior on $\G$ is 
log-uniform from $10^{-8}$ to $10^{-5}$, and the prior on $\Psi$ is 
Gaussian in logspace centererd around $\log(\Psi) = \log(60)$ and with width 
$\log(2)/2$. The limits for each GO field were calculated independently,
while the limits for each survey (with GOODS North and South as separate
surveys) were calculated as single large fields. Limits for individual
GO fields and surveys were then combined.

%%%%%%%%%%%%%%%%%%%%%%%%%%%
\begin{figure*}
\begin{minipage}{\linewidth}
\centering\includegraphics[width=0.49\linewidth]{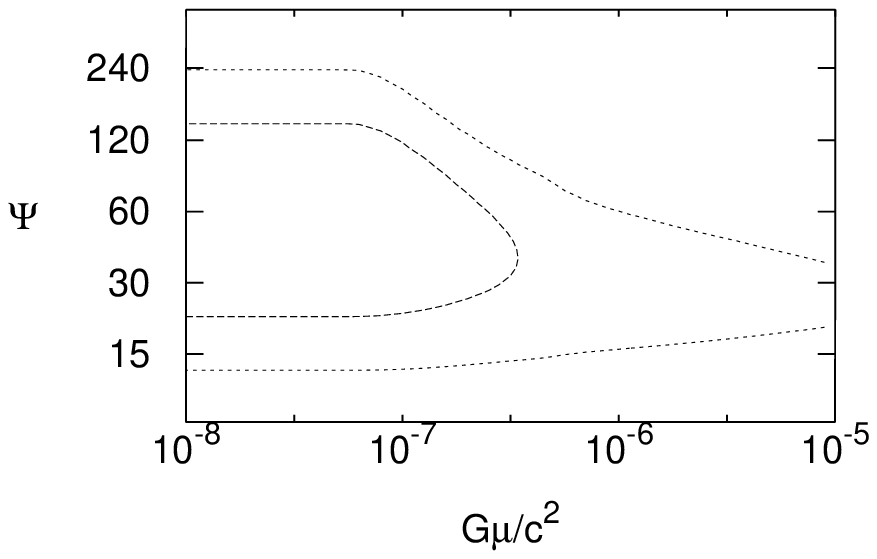}\medskip
\centering\includegraphics[width=0.49\linewidth]{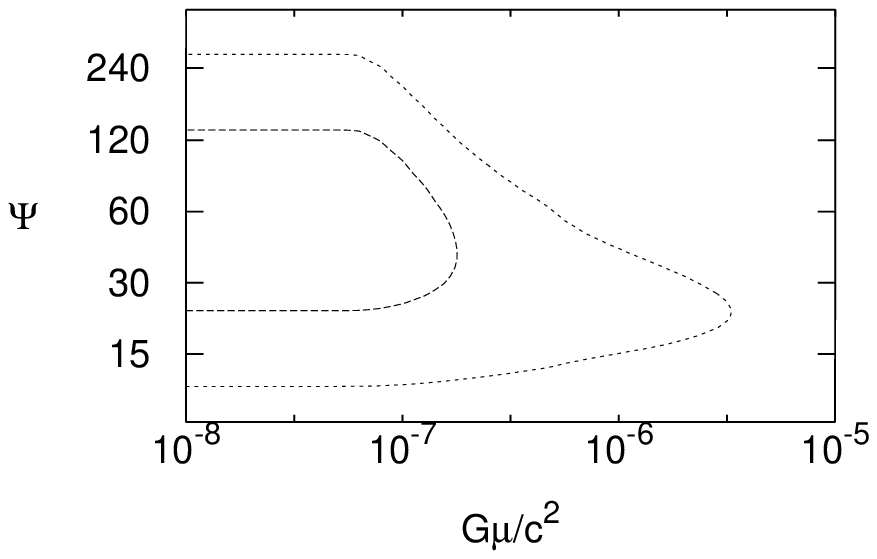}\medskip
\end{minipage}
\caption{The 68\% and 95\% confidence limits for all surveys+GO-M (left) 
and all surveys+GO (right). Our priors are described in 
Section~\ref{sect:theory:survey}.}\label{fig:nulllimits}
\end{figure*}
%%%%%%%%%%%%%%%%%%%%%%%%%%%

In Figure~\ref{fig:nulllimits}, we show the 68\% and 95\% confidence contours
in  string $\G-\Psi$ space using all surveys and the multifilter GO fields. We
also show the analogous contours calculated using surveys and all GO fields.
For surveys and multifilter data, our 95\% confidence limits are $\G <
2.3 \times10^{-6}$, $\log_{10} \Psi = 1.72 \pm 0.41$ and $\Os < 2.1 \times 10^{-5}$. Using all of our data, 
our 95\% confidence limits are $\G < 6.5 \times10^{-7}$, $\log_{10} \Psi = 1.50 \pm 0.30$ and $\Os < 7.3
\times 10^{-6}$. Our limits on $\Psi$ are particularly sensitive on the prior we take in
Eq. \ref{eq:pofl}, but we have lowered the expected value of $\Psi$ from 60 (the prior) to 52 (using GO-M)
or 32 (using all data). 

%-------------------------------------------------------------------------------

\section{Discussion}
\label{sect:disc}

% PJM: I disagree - you have elevated us into the realm of 
% diminishing returns!
% The exponential dependence of our limits on the amount of data we can search
% through points to a bright future for this type of string search. 

Our string
constraints could be lowered somewhat
by new data that will soon be delivered by several
different projects.  
The
recently deployed HST Wide Field Camera 3 will provide 7 square arcminute
exposures of high resolution data with similar limiting magnitudes to the
Advanced Camera for Surveys \citep{Kim++06}: this new GO data, combined with the
growing ACS imaging archive should provide at least a factor of two increase
in sky coverage.
Looking ahead, the James Webb Space Telescope (JWST) is scheduled to launch in
2014, and will provide more  images of similar depth and resolution to HST in
the infrared \citep{Gar++06},  
% PJM: new caveats: 
albeit over smaller ($\simeq
4$ square arcmin) fields of view (and which may already have been observed in
the optical with HST). 
% PJM:
These data would certainly enable the constraint projected in the ``doubled
dataset'' curve in the right-hand panel of Figure~\ref{fig:surveylimits}.

Can we envision extending this study to much larger surveys?
As we showed in \PI, 
large ground-based projects like the Dark Energy Survey
\citep{DES05}, the Panoramic Survey Telescope and Rapid Response System
(Pan-STARRS) \citep{Jed++07} and the Large Synoptic Survey Telescope
(LSST)\citep{Ive++08} would provide sufficient area and source counts for
string searches, but lack the angular resolution to further constrain $\G$.
Weak lensing surveys have been proposed for the
Joint Dark Energy Mission (JDEM) that cover 10000 square degrees
to an AB limiting magnitude of 26 \citep{SNAP}, although the exact
details of the JDEM mission are still in flux \citep{G+A09}. 

Searching through more data will indeed lower our detection limits. If
we were to double the number of HST fields we searched through, our 
95\% confidence limits would fall by a factor of 2--3
to $\G < 2.5\times10^{-7}$ and $\Os < 3.1 \times 10^{-6}$. The proposed 
JDEM survey would lower these confidence limits by a further factor of just 1.7
to $\G < 1.5\times10^{-7}$ and 
$\Os < 1.8 \times 10^-6$: a marginal improvement over the doubled HST limits,
given the 3 orders of magnitude increase in survey area!
At this point, resolution, and not lack of data would be
the limiting factor. 
% PJM: check this out:
In fact, we argue that with this work we have within sight of
the realm of diminishing returns, with only the next factor of two increase 
in survey scale gaining us a factor of two in parameter constraints.

The computational and human requirements for this work are modest and 
could be easily scaled up to analyse the forthcoming
increased HST archive dataset. We did not use any special 
computational tools other than parallel batch processing, and all 
computation
times quoted below are for modest desktop processors. We produced 6000 
full HST-size simulated string images and 120000 10'' wide string images (see 
Section~\ref{sect:sims}) automatically in one month of processing time. The 
first round of automatic analysis (see Section~\ref{sect:method:assess1}) was
performed on simulated data in one week and on all HST data in three days. The
manual filtering described in Section~\ref{sect:method:assess2} was performed 
on automatically-generated webpages for each field and required an average of
twenty seconds per potential string for a total of ten hours of human work. 
In 
the future, the extraction of cosmic rays and other false sources would need 
to be
automated for JDEM-scale datasets; the greater uniformity of such a survey's
images would make this process easier. The final
automated phase described in Section~\ref{sect:method:assess2} took
essentially no computer time. The analysis of the remaining candidates (see
Section~\ref{sect:results:inspection}) took several hours and would have to be
automated for larger projects.

The technique for long string detection developed here will continue to be
applicable to wider searches for lower tension, smaller image separation, and
more sparsely distributed strings.
Previous direct detection methods
relied on finding single distinctive string lensing events \citep{Saz++07} or
searched for a large excess of similar pairs \citep{Ch++08}. The former method
requires bright sources at large separation and thus does not make full use of
survey depth or resolution. This prevents it from probing the smallest $\G$.
In addition, it is difficult to characterise a single pair as being caused by
a string lensing event. The latter method does not efficiently extract string
lensing events from the background and is only effective if the field is
crowded with strings. By searching for several pairs along straight lines, we
probe to smaller separations (which imply smaller $\G$), examine fainter
sources (whose exponentially large number allow us to probe smaller $\Psi$)
and limit the possibility that a real string will be obscured by background
pairs. 

However, 
there are several straightforward improvements one could make to our string
searching methods that could improve both the final limits and the searching
efficiency in future projects. Firstly, our pair characterisation, based on
magnitude, ellipticity and the alignment of a pair with the proposed string,
does not use all the information available, and is unable to recognise a very
incompletely copied pair. One could instead match the images of each pair
pixel by pixel \citep[as performed by][]{AHP06}, and allow incomplete copying
to account for galaxy images ``cut'' by the string.
We did not not implement this method
on the grounds that our 
search focuses on barely resolved pairs for which there is little
information, and because it would vastly increase computational time.
One could imagine a hybrid survey whereby candidate image pairs were fed to an
automated pixel-matching routine; this would both improve the purity of the
candidate samples and also reduce the human inspection time of the search.

Using exclusively multifilter data would also make searching more efficient.
None of our automatic string detection algorithms use color information
because both COSMOS and half of the HAGGLeS GO fields are single filter. But
we saw in Section~\ref{sect:results} that checking for color consistency
across pairs effectively eliminates potential string candidates. It may be
impossible to guarantee multifilter data for large, diverse archival datasets
like the GO fields, but the kind of homogeneous 
survey like JDEM would automatically
provide color information which could be profitably exploited.

The main reason why our limits on string concentration and tension are much
stricter than those of previous works is our use of a survey comprised of many
scattered small fields. 
Because the probability of a string crossing a field scales as the
linear size of the field and not its area, searching for cosmic strings in
randomly scattered fields is more effective than searching many contiguous
fields in a survey. This method also has the practical advantage that it does
not require a dedicated survey and naturally thrives on archival data.
Essentially any high resolution extragalactic images without overwhelmingly
large objects in them can contribute to our limits on cosmic strings. 
However, the relatively small size of the ACS field means that at the low
lensing cross-sections of interest we only expect to see a small number of
lensed image pairs per field. This may be a problem for string searches based
on the smaller WFC3 and JWST fields of view.

%-------------------------------------------------------------------------------
% PJM: CONCLUSIONS MUST BE BRIEF!

\section{Conclusions}
\label{sect:concl}

% PJM: We should aim to be the last word on this topic!
% We have presented significant direct detection limits on the tension and density of
% cosmic strings. 
% Several near-term projects will produce data that will allow
% us to lower these limits significantly. 
% We employed a variety of techniques
% that should be applied to future searches, but depending on the source of
% future data, other techniques should be used to improve on our methods and set
% even stricter limits.
%

From an exhaustive search of some 4.5 square degrees of archival HST/ACS imaging
data, we have derived significant direct detection limits on the tension and
density of cosmic strings, employing a variety of techniques that could be
applied to future searches. We draw the following conclusions:

\begin{itemize}

\item Using multi-filter data from the HAGGLeS, COSMOS, GEMS, AEGIS and
GOODS surveys, we are able to constrain the dimensionless 
string tension
$\G$ to be below $2.3\times10^{-6}$ and the energy density in long strings
$\Os$ to be below
$2.1 \times 10^{-5}$ with 95\% confidence. 

\item Extending the search to single filter GO imaging, we found 4 string
candidates that we were unable to reject formally on the basis of the data in
hand. Applying our experience with the multi-band data, we suggest that these
detections are false positives: under this assumption
we find stronger upper limits of  
$\G < 6.5\times10^{-7}$ and $\Os < 7.3 \times 10^{-6}$. 

% PJM: maybe a conclusion based on figure 16? If we assume the concentration 
% expected by theory...

\item Unlike previous optical imaging searches, we cover sufficient sky area
to make these limits universal: if there were 
cosmic strings present in the 
redshift~$\lesssim 1$ universe with tension and density greater than our
limits, we would have detected them.

\item Color information was found to be important in rejecting string
candidates: high resolution imaging in multiple filters would be the most
efficient way of ruling out the candidates presented here, and any future
single-filter detections.

\item Using the technology developed in this study, the upcoming 
factor of two increase in sky area imaged at comparably high resolution
expected from the refurbished HST should enable these upper limits on string
tension and density to be further reduced, by slightly more than a factor of two.

\end{itemize}

% PJM: Hmm - usually the referee says this sort of thing, not the authors!
% This work represents a major step forward in the field of direct cosmic string
% detection. Our limits on $\G$ differ from previous direct detection searches,
% because they do not assume that a string crosses through a given field, and
% they effectively constrain $\G$ on a universal level. 

% PJM: read carefully and see what you think, esp CMB part!
This work represents the the first
direct detection limits on cosmic strings 
that are both competitive with and
complimentary to other indirect methods. Our limit on the string tension is only a
factor of 2--3 higher than the indirect upper 
limit of $\G = 3 \times 10^{-7}$ derived from the CMB power spectrum. 
We have shown that 
direct detection of strings by their gravitational lensing effect
has the
advantage of constraining both $\G$ and $\Psi$, 
and does not require that
strings emit gravitational radiation, as the pulsar timing
constraints require. Perhaps most importantly, 
direct detection is the only
method currently available that would provide the precise location of a
cosmic string for future study. 
%
% PJM - last word?
% Given that cosmic string limits can be lowered
% using essentially any high resolution, extragalactic data, extending this work
% in the near and longterm future will be profitable.
%
Prospects for using near future data to improve upon these limits are
bright; however, we anticipate that this will bring us into the era where
large increases in survey area would bring relatively small changes in the
limits. The next significant advance in this field maybe the increase in
angular resolution promised by future radio surveys.

%-----------------------------------------------------------------------

\section*{Acknowledgments} 

We thank Joe Polchinski for inspiring this work with a Blackboard Lunch Talk
at the Kavli Institute for Theoretical Physics (KITP), and for numerous useful
suggestions.  We are also very grateful for input to the early part of this
project on the theoretical side from Alice Gasparini and Florian Dubath. We
acknowledge Chris Fassnacht's efforts in testing the HAGGLeS image processing
pipeline and thank David Hogg for useful discussions about automated
string detection.
PJM received support from the TABASGO foundation in the form of
a research fellowship.  TT acknowledges support from the NSF through CAREER
award NSF-0642621, and from the Sloan Foundation through a Sloan Research
Fellowship.
Support for this work was provided by NASA through grant number HST-AR-10676
(the HAGGLeS project) from the Space Telescope Science Institute, which is
operated by AURA, Inc., under NASA contract NAS 5-26 555.
This work was supported by the NSF under award AST05-07732 and in part by the U.S Department of Energy under contract number DE-AC02-76SF00515.

%-------------------------------------------------------------------------------
\section*{Appendix: The Probability of a 1 Radian String Crossing a Field}

\begin{figure}
\centering\includegraphics[width=\linewidth]{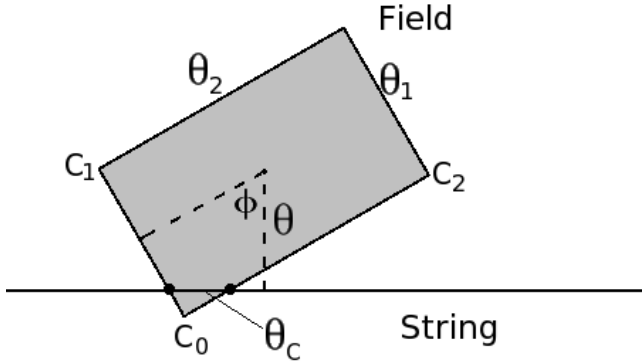}
\caption{A string crossing a $\theta_1 \times \theta_2$ field. The distance from the centre of the field to the string and orientation of the field are $\theta$ and $\phi$, respectively. The length of string crossing the field is $\theta_c$. The lowest corner is $C_0$. The adjacent corners separated from $C_0$ by $\theta_1$ and $\theta_2$ are $C_1$ and $C_2$ respectively.}\label{fig:stringcross}
\end{figure}

When calculating the probability of a rectangular, $\theta_1\times\theta_2$
field lying on a projected $1$ radian string such that $\theta_c$ string
crosses the field, we start by making useful approximations and other
simplifications. We use the fact that our fields are much smaller than a
radian so that the sky can be approximated as a rectangle with area $4\pi$
steradians. We ignore the end effects and curvature of the string. As shown in 
Fig. \ref{fig:stringcross},  we imagine
that the centre of our field is a distance $\theta$ from our string, and our
field is oriented with the angle $\phi$ where $\phi = 0$ implies that the long
sides of our field (with length $\theta_2$) are perpendicular to the string.
We need only consider the cases where $\theta > 0$ and $0 < \phi < \pi/2$. The
rest of parameter space is proportional by symmetry. The probability of a 
string crossing our field with
toal overlap $\theta_c$ is then:
\be\label{eq:peqdef}
\Prad(\theta_c) = \frac{1}{\pi^2} \int_0^1 d\theta \int_0^{\pi/2}d\phi\ \delta(\theta_c'(\theta,\phi)-\theta_c)
\ee
%Where the $4 \pi$ in the denominator normalises for a $1$ radian string in a
%$4\pi$ steradian sky. The $4/\pi$ normalises for the quarter rotation that we
%integrate our field over and the symmetry for positive and negative $\theta$.
%The overall normalisation factor is then $1/\pi^2$.

From Fig.~\ref{fig:stringcross}, we can see that the vertical displacement 
between each of the three corners and the string is: 

\bea
c_0 &=& \frac{\theta_0}{2} \cos(\phi-\phi_0)\\
c_1 &=& \frac{\theta_0}{2} \cos(\phi+\phi_0)\label{eq:thetac1}\\
c_2 &=& -\frac{\theta_0}{2} \cos(\phi+\phi_0)\label{eq:thetac2}
\eea
Where $\theta_0 = \sqrt{\theta_1^2+\theta_2^2}$ and $\phi_0 = \arctan(\theta_1/\theta_2)$.

A string can cross our field in three different ways: along the short axis 
($\theta_1$), along the long axis ($\theta_2$) or cutting a corner. If the 
string crosses our field along the short axis (entering and exiting via 
the two long sides), it's crossing length is:

\be
\theta_c = \frac{\theta_1}{\cos(\phi)},\ \rm{for}\ \phi < \pi/2-\phi_0\ \rm{and}\ \theta < c_1\label{eq:thetacs}
\ee
Using Eq. \ref{eq:peqdef} we can convert this into the component of $\Prad$
that represents the string crossing the field along the short axis, $P_{\rm{short}}$:
\bea\label{thetac:3}
P_{\rm{short}}(\theta_{c}) &=& \frac{1}{\pi^2}\int_0^{c_1}d\theta\int_0^{\pi/2-\phi_0}d\phi\ \delta(\frac{\theta_1}{\cos(\phi)}-\theta_c),\\
&&\rm{for}\ \theta_1 <\theta_c < \theta_0\nonumber\\
 &=& \frac{1}{\pi^2}\cos(\arccos(\theta_1/\theta_c)+\phi_0)\frac{\theta_1}{\theta_c\sqrt{\theta_c^2-\theta_1^2}},\\
&&\rm{for}\ \theta_1 <\theta_c < \theta_0\nonumber\\
&=& \frac{\theta_1^2}{2\pi^2\theta_c^2}\left(\frac{\theta_2}{\sqrt{\theta_c^2-\theta_1^2}}-1\right) ,\\
&& \rm{for}\ \theta_1 <\theta_c < \theta_0\nonumber
\eea

If a string crosses along the long side of the field:
\be
\theta_c = \frac{\theta_2}{\sin(\phi)},\ \rm{for}\ \phi > \pi/2-\phi_0\ \rm{and}\ \theta < c_2
\ee
And:
\be
P_{\rm{long}} = \frac{\theta_2^2}{2\pi^2\theta_c^2}\left(\frac{\theta_1}{\sqrt{\theta_c^2-\theta_2^2}}-1\right) ,\ \rm{for}\ \theta_2 <\theta_c < \theta_0
\ee

When the string ``cuts the corner'', entering and exiting via adjacent sides, $\theta_c$ is:
\bea
\theta_c &=& \frac{\theta_0 \cos(\phi-2\phi_0)-2\theta}{\sin(2\phi)},\label{eq:thetac3}\\
&&\rm{for}\ \phi < \pi/2-\phi_0,\ c_1 < \theta < c_0,\nonumber\\
&&\rm{and}\  \phi > \pi/2-\phi_0, c_2 < \theta < c_0\nonumber
\eea

Again using Eq. \ref{eq:peqdef} we can find the component of $\Prad$ that
comes from the string crossing the field via adjacent sides, $P_2$:
\bea
P_{\rm{corner}}(\theta_c)& =&\frac{1}{\pi^2} \int_{\phi_{\rm{min}}}^{\phi_{\rm{max}}}d\phi\\
               &&\int_0^{c_1} d\theta\ \delta\left(\frac{\theta_0 \cos(\phi-\phi_0)-2\theta}{\sin(2\phi)}-\theta_c\right)\nonumber\\
&=&\frac{1}{2\pi^2}\int_{\phi_{\rm{min}}}^{\phi_{\rm{max}}}d\phi\ \sin(2\phi)
\eea
$\phi_{\rm{min}}$ and $\phi_{\rm{max}}$ mark the points when a second corner rotates across the 
string so that $\theta_C$ reverts to either $\theta_1$ or $\theta_2$. We can derive them from Eq.
\ref{eq:thetac1} and Eq. \ref{eq:thetac2} and obtain:
\bea
P_{\rm{corner}}(\theta_c)& =& \frac{1}{\pi^2}\int_{0}^{\pi/2}d\phi\ \sin(2\phi),\ \rm{for}\ \theta_c < \theta_1\\
& =& \frac{1}{\pi^2}\int_{\arccos(\theta_1/\theta_c)}^{\pi/2}d\phi\ \sin(2\phi),\nonumber\\ 
&&\rm{for}\ \theta_1 < \theta_c < \theta_2\\
& =& \frac{1}{\pi^2}\int_{\arccos(\theta_1/\theta_c)}^{\arcsin(\theta_2/\theta_c)}d\phi\ \sin(2\phi),\nonumber\\
&&\rm{for}\ \theta_2 < \theta_c < \theta_0\\
P_{\rm{corner}}(\theta_c)& =& \frac{1}{2\pi^2},\ \rm{for}\ \theta_c < \theta_1\\
& =& \frac{1}{2\pi^2}\frac{\theta_1^2}{\theta_c^2},\ \rm{for}\ \theta_1 < \theta_c < \theta_2\\
& =& \frac{1}{2\pi^2}\left(\frac{\theta_1^2+\theta_2^2}{\theta_c^2}-1\right),\ \rm{for}\ \theta_2 < \theta_c < \theta_0
\eea

Finally, we add $P_{\rm{short}}$, $P_{\rm{long}}$ and $P_{\rm{corner}}$ to obtain:

\bea
\Prad(\theta_c) &=& \frac{1}{2\pi^2},\ \rm{for}\ \theta_C < \theta_1\label{eq:peq}\\
                  &=& \frac{1}{2\pi^2}\left(\frac{\theta_1}{\theta_c}\right)^2\frac{\theta_2}{\sqrt{\theta_c^2-\theta_1^2}},\ \rm{for}\ \theta_1 < \theta_c < \theta_2\nonumber\\
                  &=& \frac{1}{2\pi^2}\left(\frac{\theta_1}{\theta_c}\right)^2\frac{\theta_2}{\sqrt{\theta_c^2-\theta_1^2}}+\nonumber\\
&&\frac{1}{2\pi^2}\left(\left(\frac{\theta_2}{\theta_c}\right)^2\frac{\theta_1}{\sqrt{\theta_c^2-\theta_2^2}}-1\right), \nonumber\\
                  & & \rm{for}\ \theta_2 < \theta_c < \sqrt{\theta_1^2+\theta_2^2}\nonumber
\eea

\label{lastpage}
%\bibliography{references}
% MNRAS can be tricked into accepting bibtex but I forget how...  I used bubble
% to make a bbl file from our bib file and then edited it to just have the
% papers we want. In principle we should comment out the non-used refs from the
% bib file...

%\bsp
%-------------------------------------------------------------------------------
\end{document}